%% file: 0-arXiv-fairglucose.tex
\documentclass{article}

\PassOptionsToPackage{numbers, sort&compress}{natbib}

\usepackage[preprint]{arxiv}

\input{0-preamble-shared}

\title{FairGlucose: A CGM Fairness Benchmark Reveals Subgroup Disparities Hidden in Population-Level Validation}

\author{%
  Junjie Luo\thanks{Corresponding author: \texttt{jluo41@jhu.edu}} \\
  Carey Business School \\
  Johns Hopkins University \\
  Baltimore, MD, USA \\
  \And
  Xuzhe Zhi \\
  Carey Business School \\
  Johns Hopkins University \\
  Baltimore, MD, USA \\
  \And
  Rui Han \\
  Welldoc, Inc. \\
  Columbia, MD, USA \\
  \AND
  Abhimanyu Kumbara \\
  Welldoc, Inc. \\
  Columbia, MD, USA \\
  \And
  Anand K. Iyer \\
  Welldoc, Inc. \\
  Columbia, MD, USA \\
  \And
  Mansur E. Shomali \\
  Welldoc, Inc. \\
  Columbia, MD, USA \\
  \AND
  Ritu Agarwal \\
  Carey Business School \\
  Johns Hopkins University \\
  Baltimore, MD, USA \\
  \And
  Guodong Gordon Gao \\
  Carey Business School \\
  Johns Hopkins University \\
  Baltimore, MD, USA \\
}

\date{}

\begin{document}
\maketitle
\thispagestyle{empty}
\pagestyle{plain}

\input{0-sections/00_abstract}

\keywords{continuous glucose monitoring \and fairness \and benchmark \and time-series forecasting \and digital health \and diabetes}

\input{0-sections/01_introduction}
\input{0-sections/02_results}
\input{0-sections/03_discussion}
\input{0-sections/04_methods}

\section*{Data availability}
\input{0-sections/94_data_availability}

\section*{Code availability}
\input{0-sections/95_code_availability}

\section*{Acknowledgements}
\input{0-sections/90_acknowledgements}

\section*{Author contributions}
\input{0-sections/96_author_contributions}

\section*{Competing interests}
\input{0-sections/92_competing_interests}

\bibliographystyle{naturemag}
\bibliography{0-fairglucose-preprint}

\clearpage
\appendix
\input{0-sections/A_dataset_statistics}
\input{0-sections/B_model_evaluation}
\input{0-sections/C_experiment_details}

\end{document}

%% file: 0-preamble-shared.tex
\usepackage[utf8]{inputenc}
\usepackage[T1]{fontenc}
\usepackage{hyperref}
\usepackage{url}
\usepackage{booktabs}
\usepackage{amsmath,amssymb,amsfonts,mathtools,amsthm}
\usepackage{nicefrac}
\usepackage{microtype}
\usepackage{xcolor}
\usepackage{cleveref}
\usepackage{graphicx}
\usepackage{subcaption}
\usepackage{multirow}
\usepackage{tcolorbox}
\usepackage{algorithm}
\usepackage{algorithmicx}
\usepackage{algpseudocode}
\usepackage{listings}
\usepackage[title]{appendix}
\usepackage{doi}

\theoremstyle{plain}

\theoremstyle{definition}

\theoremstyle{remark}

%% file: 0-sections/00_abstract.tex

\begin{abstract}


As CGM-based AI tools approach clinical deployment, whether their accuracy is equitable across patient demographics remains insufficiently tested. To enable this evaluation, we constructed FairGlucose, a 300-patient CGM cohort balanced across 12 demographic strata (age $\times$ gender $\times$ type~1/type~2 diabetes), with 132{,}480 forecasting samples and 3{,}945 unique behavioral events (meals, exercise, medication) logged by 81 patients. Benchmarking 33 models across four families on 2-hour glucose forecasting, we find that population-level external validation can conceal substantial subgroup disparities. Aggregate out-of-distribution metrics appear stable ($\approx$1.0), yet subgroup-level ratios range from 0.8 to 1.4, with T1D patients showing 6\,mg/dL higher prediction error than T2D ($p < 0.001$). This disparity persists across \emph{all} 33 models, suggesting a property of the prediction task rather than any single architecture. Further analysis shows that subgroup performance gaps align with the proportion of clinically hard cases, and that input-length sensitivity varies across demographics, motivating personalized configurations. Frontier LLMs underperform specialized neural models by 1--6\,mg/dL; behavioral events contribute negligibly ($\sim$0.1\,mg/dL) even under oracle event access. These findings establish that population-level validation alone is insufficient for equity assessment of digital health AI, motivating subgroup-disaggregated reporting as a default standard.

\end{abstract}

%% file: 0-sections/01_introduction.tex
\section{Introduction}

%
Continuous glucose monitoring (CGM) is now embedded in a growing class of digital diabetes interventions, from real-time alerting and automated insulin delivery to mobile-app coaching, each relying on predictive models of near-term glucose trajectories \cite{elsayed20237, contreras2018artificial, kim2020lessons}.
%
As these AI-driven tools scale from research prototypes to clinical deployment, an unresolved question is whether their predictive accuracy is \emph{equitable} across the demographic subgroups they will serve, or whether population-level performance metrics mask systematic subgroup-level failures \cite{armandpour2021deep, luo2025large, mirshekarian2019lstms}.

%
This concern is not hypothetical. In clinical AI more broadly, recent work has shown that models with high average performance can underperform significantly for specific patient subgroups \cite{subbaswamy2024framework}, and a scoping review of 467 AI fairness studies identified substantial evidence gaps across medical specialties \cite{liu2025scoping}. In medical imaging, systematic demographic disparities have been documented \cite{xu2024fairness}. Yet in CGM-based forecasting, fairness evaluation remains virtually absent.

%
A major obstacle is the data itself. Existing CGM benchmarks are demographically imbalanced: GlucoBench \cite{sergazinov2024glucobench} consolidates five public datasets totaling $\sim$461 subjects but does not enforce demographic balance or include fairness metrics. OhioT1DM \cite{marling2020ohiot1dm} provides only 12 type~1 diabetes patients, precluding generalization. GluFormer \cite{gluformer2025}, a recent foundation model trained on 10 million CGM measurements, acknowledges that its training data is geographically and demographically limited but does not perform subgroup-level evaluation. Additionally, behavioral event annotations (meals, exercise, medication) are typically missing from these resources \cite{dubosson2018open, hall2018glucotypes}, preventing analysis of how glucose-altering events interact with prediction accuracy across demographics. The open question is therefore not whether CGM forecasting models are accurate on average, but whether population-level validation metrics provide a reliable guarantee of equitable performance across the diverse patient populations these tools will serve.

%
To answer this question, we constructed \textbf{FairGlucose}, a CGM cohort specifically designed for fairness-aware evaluation: 300 patients balanced across 12 demographic strata (age $\times$ gender $\times$ type~1/type~2 diabetes), with 132{,}480 standardized forecasting samples, 3{,}945 unique timestamped behavioral events (meals, exercise, medication) logged by 81 of the 300 patients, and uniform data quality across all subgroups. Each sample contains a 24-hour CGM input and an 8-hour output window; the main benchmark reports the first 2 hours of that output window, with additional horizons reported in the Supplementary Information. Using this balanced infrastructure, we benchmarked 33 models spanning statistical, machine learning, neural time-series, and frontier LLM families on 2-hour glucose forecasting, with evaluation protocols that disaggregate performance by demographic subgroup \cite{obermeyer2019dissecting}.

%
We formulate two clinically meaningful evaluation tasks for CGM modeling: (1) prediction accuracy measured by root mean square error (rMSE) for 2-hour ahead glucose forecasting supporting real-time glycemic control,
and (2) clinical safety assessed via Clarke Error Grid Analysis, reporting the percentage of predictions in the most clinically accurate zone (Clarke A), representing predictions with no clinical errors (within 20\% or both $<$70 mg/dL) \cite{clarke1987evaluating}.
%
To support rigorous and equitable evaluation, we introduce a fairness-aware benchmarking protocol that integrates overall prediction error with subgroup-disaggregated performance reporting and event-aware metrics.
%
We benchmark 33 models across four families: statistical methods (3), machine learning (4), neural time-series architectures (20, including the zero-shot foundation model Chronos2), and frontier LLMs and foundation models (6).
%
Among neural models, attention-based architectures dominate the top ranks: \textbf{NS-Transformer} achieves the best performance (rMSE=25.6\,mg/dL, Clarke A=76.4\%), followed by Transformer (25.7), TimeXer (25.7), and PatchTST \cite{nie2022time} (25.8). \textbf{LightGBM} \cite{ke2017lightgbm} remains competitive as the best ML model (rMSE=26.0\,mg/dL, Clarke A=76.8\%), demonstrating that tree-based methods rival neural architectures for glucose forecasting.
%
Foundation and LLM models show mixed results: TimeGPT \cite{garza2023timegpt} achieves competitive prediction error (rMSE=26.6), while general-purpose LLMs (GPT-5.1 \cite{openai_gpt51_model_docs}, GPT-5 Mini \cite{openai_gpt5mini_model_docs}, Claude 4.5 \cite{anthropic_claude45_whats_new}, Gemini 3 \cite{google_gemini3_dev_guide}) lag by 1--6\,mg/dL, suggesting time-series-specific models adapt better for physiological forecasting.

%
Baseline evaluation reveals key patterns.
%
First, substantial subgroup disparities exist even with balanced data: T1D patients show consistently higher prediction error than T2D patients, and performance varies significantly across age-gender intersections.
%
Although population-level generalization appears strong (OD/ID ratio $\approx$ 1.0), subgroup-level ratios range from 0.8 to 1.4, demonstrating that aggregate metrics can mask heterogeneity.
%
Second, input-length sensitivity is subgroup-dependent: older T2D patients require longer CGM histories for stable prediction, while younger T1D patients tolerate short input windows with minimal accuracy loss, supporting personalized input-length strategies rather than one-size-fits-all configurations.
%
Third, instance-level difficulty analysis reveals that subgroup performance disparities align with the \emph{proportion} of clinically hard cases (rapid glucose excursions, post-meal spikes); top models concentrate gains on easy cases yet offer minimal improvement on hard cases, precisely the scenarios where accurate prediction is most clinically valuable.
%
Fourth, explicit event features provide small but consistent improvements: incorporating glucose-altering events (meals, exercise, medication) as model features yields minor performance gains ($\sim$0.1 mg/dL) even under oracle event access (including post-observation events), with benefits consistent across demographic subgroups.
%
These findings demonstrate that equitable glucose prediction requires careful attention to subgroup composition, input configuration, difficulty-aware evaluation, and event handling.

%
Our work makes three contributions.

\textbf{(1) Clinical finding.}~~Population-level external-validation metrics ($\approx$1.0) systematically mask substantial subgroup-level variation (0.8--1.4 range) in CGM forecasting accuracy. This masking persists across all 33 evaluated models spanning four paradigms, suggesting that the disparity is a task-level property rather than an architecture-level artifact. Subgroup performance gaps align with the proportion of clinically hard cases, and input-length sensitivity varies by demographic, motivating personalized configurations. Together these findings demonstrate that population-level validation alone is \emph{insufficient} for equity assessment of digital health AI tools.

\textbf{(2) Resource.}~~FairGlucose is a 300-patient demographically balanced CGM cohort with 132{,}480 standardized forecasting samples and 3{,}945 unique timestamped behavioral events logged by 81 patients, to be released with held-in/held-out splits, an evaluation server, and a 33-model reference leaderboard, under a data-use agreement.

\textbf{(3) Model benchmarking.}~~Frontier LLMs (GPT-5.1, Claude~4.5, Gemini~3) underperform specialized neural models by 1--6\,mg/dL, while behavioral events contribute negligibly ($\sim$0.1\,mg/dL) to top models even under oracle event access, suggesting modern forecasters implicitly capture event-driven dynamics from CGM signal alone.

\begin{figure*}[t]
    \centering
    \includegraphics[width=0.7\textwidth]{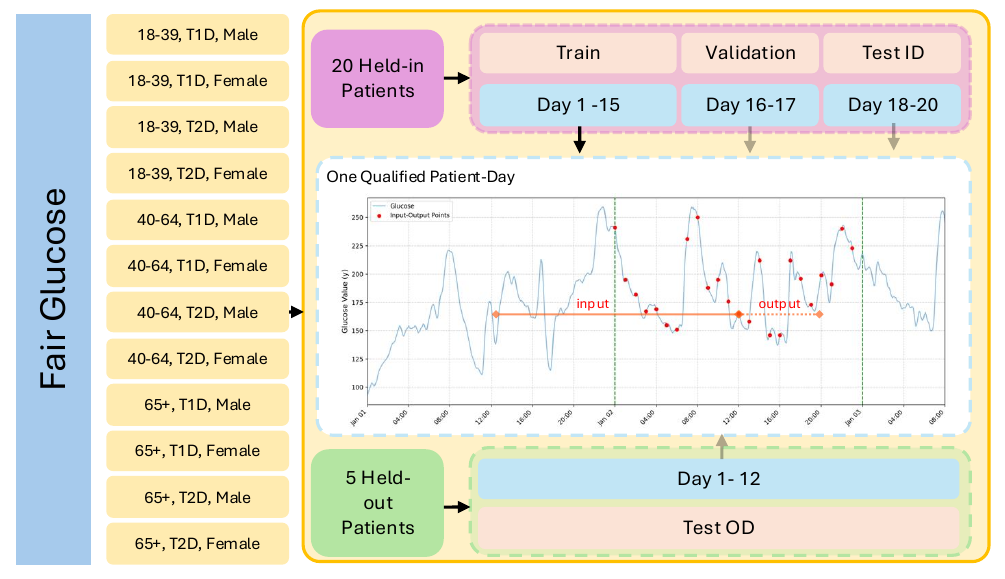}
    \caption{FairGlucose Benchmark Curation Process. 300 patients stratified across 12 balanced subgroups (age: 18–39, 40--64, $\geq$65 $\times$ gender: M/F $\times$ diabetes type: T1D/T2D), 25 per subgroup. Each subgroup has 20 held-in patients (train/valid/test-id: 15/2/3 days) and 5 held-out patients (test-od: 12 days). Each day yields 24 prediction pairs (24h input + 8h output). Includes 3{,}945 unique event annotations (meals, exercise, medication) logged by 81 patients across 132{,}480 prediction pairs.}
\label{fig:figure1}
\end{figure*}

%% file: 0-sections/02_results.tex

\section{Results}\label{sec:results}

\noindent\textbf{Study design and dataset overview.}~~FairGlucose comprises 300 patients sampled from a US-based mobile health platform, evenly stratified across 12 demographic subgroups defined by age (18--39 / 40--64 / $\geq$65), gender (male / female), and diabetes type (T1D / T2D), with 25 patients per subgroup (Figure~\ref{fig:figure1}; see Methods for full cohort and protocol details). Each patient contributes multiple 24-hour CGM traces sampled at 5-minute intervals; from these we extract 132{,}480 input--output prediction pairs comprising a 24-hour input window and an 8-hour output window, alongside 3{,}945 unique timestamped behavioral events (meals, exercise, medication) logged by 81 patients. The main benchmark evaluates the first 2 hours of the output window. We benchmark 33 baseline models spanning four families: statistical (Na\"ive, ARIMA, AutoETS), machine learning (LightGBM, XGBoost, CatBoost, Linear), neural time-series (20 architectures including PatchTST, iTransformer, TimeXer, Crossformer, TFT, N-HiTS, N-BEATS, DLinear, TimeMixer, TimesNet, Chronos2, and others; see Methods), and frontier LLMs and foundation models (TimeGPT, GPT-5.1, GPT-5~Mini, Claude~4.5~Sonnet/Haiku, Gemini~3~Flash), on 2-hour-ahead forecasting. Predictive accuracy is reported as root-mean-square error (rMSE) and Clarke Error Grid Zone~A percentage (Clarke~A); equity is quantified as cross-subgroup disparity in rMSE.

\noindent\textbf{Glycemic characteristics of the cohort.}~~Glycemic profiles differ measurably across the 12 demographic strata even after balancing for sample size (Table~\ref{tab:cohort_char}). T1D patients exhibit higher coefficient of variation (CV $\sim$0.29) and lower time-in-range (TIR $\sim$65\%) than T2D patients (CV $\sim$0.23, TIR $\sim$69\%), consistent with the clinical literature on subtype-level glycemic variability. Within-subtype, younger T1D males show the highest variability (mean glucose 174.6\,mg/dL, CV 0.30). These differences motivate subgroup-disaggregated evaluation and provide the substrate against which model performance is later compared.

\input{0-display/Table/table1_cohort_characteristics.tex}

\input{0-display/Table/table1}


\noindent\textbf{Behavioral event annotations.}~~A defining feature of FairGlucose is comprehensive timestamped behavioral events that accompany the CGM signal. Of the 300 patients, \textbf{81 (27.0\%)} logged at least one event via the mobile health platform, producing \textbf{3{,}945 unique behavioral events}: medication events (2{,}809; 71.2\%), meal events with nutritional details (648; 16.4\%), and exercise events with duration and intensity (488; 12.4\%). Logging patients recorded a median of 38 events (mean 48.7, range 1--237) over their observation period. Because the hourly sliding-window design produces overlapping 48-hour observation windows, each unique event appears in approximately 30 prediction pairs; consequently 22{,}239 of 132{,}480 samples (16.8\%) contain at least one event within their window. Event density varies systematically across demographic subgroups: patients aged 65+ log 22\% fewer events than younger groups, while T2D patients log 12\% more events than T1D, reflecting differences in app engagement and disease-management style. Event timing is highly asymmetric: 33.8\% of event occurrences fall in the 24-hour input window, 3.0\% in the evaluated 2-hour forecast window, and 63.2\% in the 2--24 hours after the prediction time. Event-aware results should therefore be interpreted as retrospective event-aware analyses or upper-bound estimates unless restricted to events available at prediction time.

\subsection{Overall Results}

%
\input{0-display/Table/table2_rmse_clarkeA}

\noindent\textbf{Model Family Performance.}~~Table~\ref{tab:main_performance} shows neural time-series models achieve the best performance. Attention-based architectures dominate the top ranks: \textbf{NS-Transformer} ranks first (rMSE=25.6\,mg/dL, Clarke A=76.4\%), followed closely by Transformer (25.7), TimeXer (25.7), and PatchTST (25.8). \textbf{LightGBM} achieves competitive performance as the best ML model (rMSE=26.0\,mg/dL, Clarke A=76.8\%), demonstrating that tree-based methods remain viable alternatives to neural architectures for glucose forecasting.

\begin{figure}[t!]
    \centering
    \includegraphics[width=0.95\linewidth]{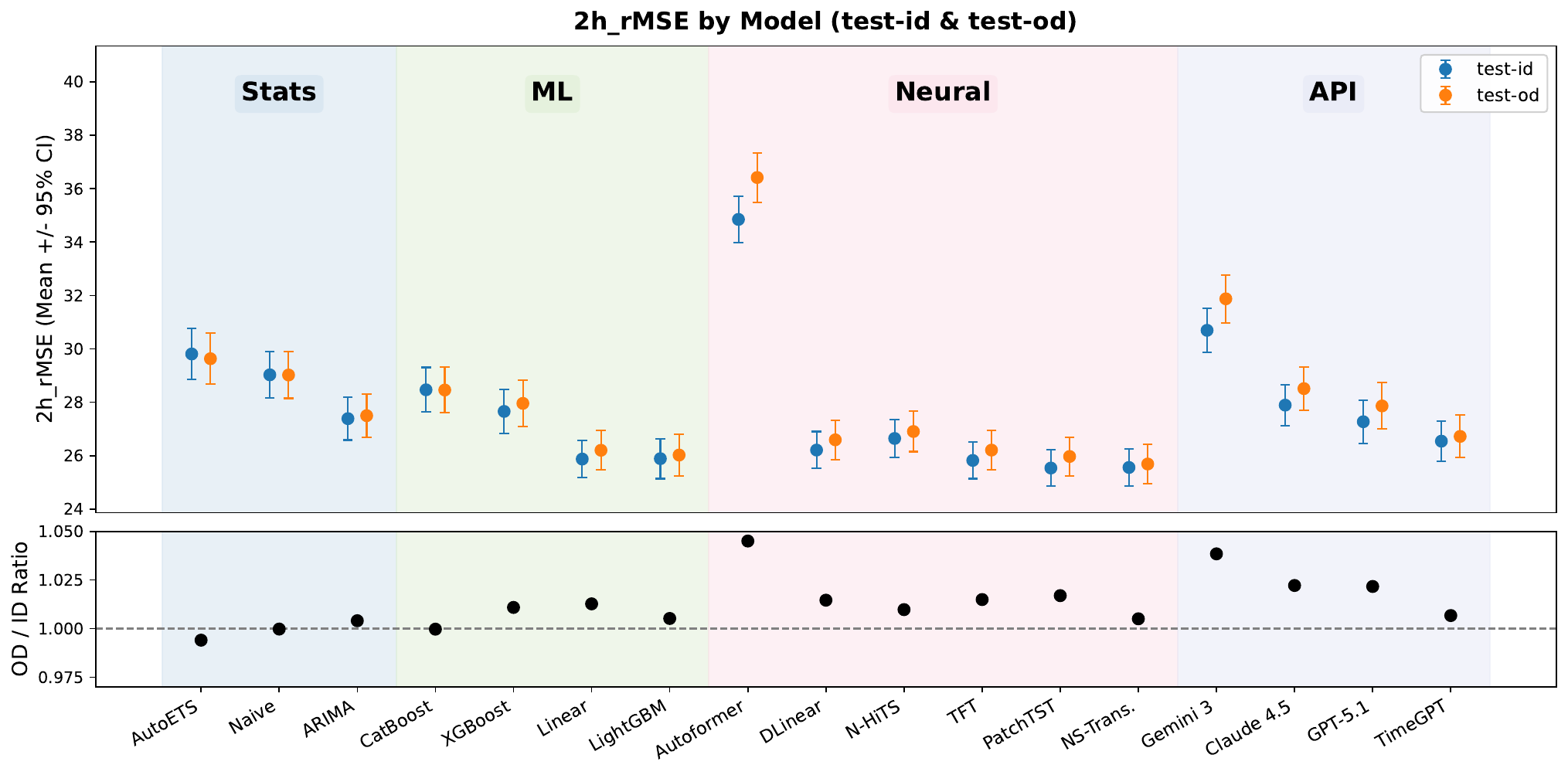}
    \caption{
        Generalization performance across models. Top: 2-hour rMSE on test-ID and test-OD with 95\% CIs. Bottom: OD/ID ratio. Figure created by the authors using Matplotlib.}
    \label{fig:testid-od}
\end{figure}

%
ML models achieve competitive performance, with LightGBM within 0.2\,mg/dL of PatchTST on prediction error (25.96 vs 25.75) and in fact ahead of it on clinical accuracy (Clarke~A 76.8\% vs 76.0\%).
%
ARIMA remains competitive as a statistical baseline, demonstrating that classical time series methods provide strong performance for well-structured physiological data.

%
Foundation and LLM models show mixed performance. TimeGPT achieves competitive prediction error (rMSE=26.6\,mg/dL, within 1.0\,mg/dL of top neural performers), while Chronos2 as a zero-shot foundation model lags substantially (rMSE=39.7).
%
Frontier LLMs (GPT-5.1, GPT-5 Mini, Claude 4.5, Gemini 3) show larger gaps (rMSE: 1--6\,mg/dL behind top neural models; Clarke A: 70--75\%), suggesting that general-purpose LLMs are not yet competitive with specialized time-series architectures for physiological forecasting \cite{luo2025large}.

%
\textbf{Clinical Accuracy Profile.}
Clarke Error Grid analysis shows that top-performing models achieve strong clinical accuracy, with \textbf{LightGBM} reaching 76.8\% of predictions in Zone A (clinically accurate), followed by NS-Transformer and Transformer (both 76.4\%), and Informer and TimeXer (76.1\%) (Table~\ref{tab:main_performance}).
%
The remaining 23--24\% of predictions fall into zones B through E, with the majority in Zone B (benign errors requiring no treatment adjustment).
%
Across the 20 neural architectures, Clarke A rates range from 54.2\% (Chronos2, zero-shot) to 76.4\% (NS-Transformer), demonstrating that architecture selection substantially impacts clinical accuracy.
%

\begin{figure}[t!]
    \centering
    \includegraphics[width=0.99\linewidth]{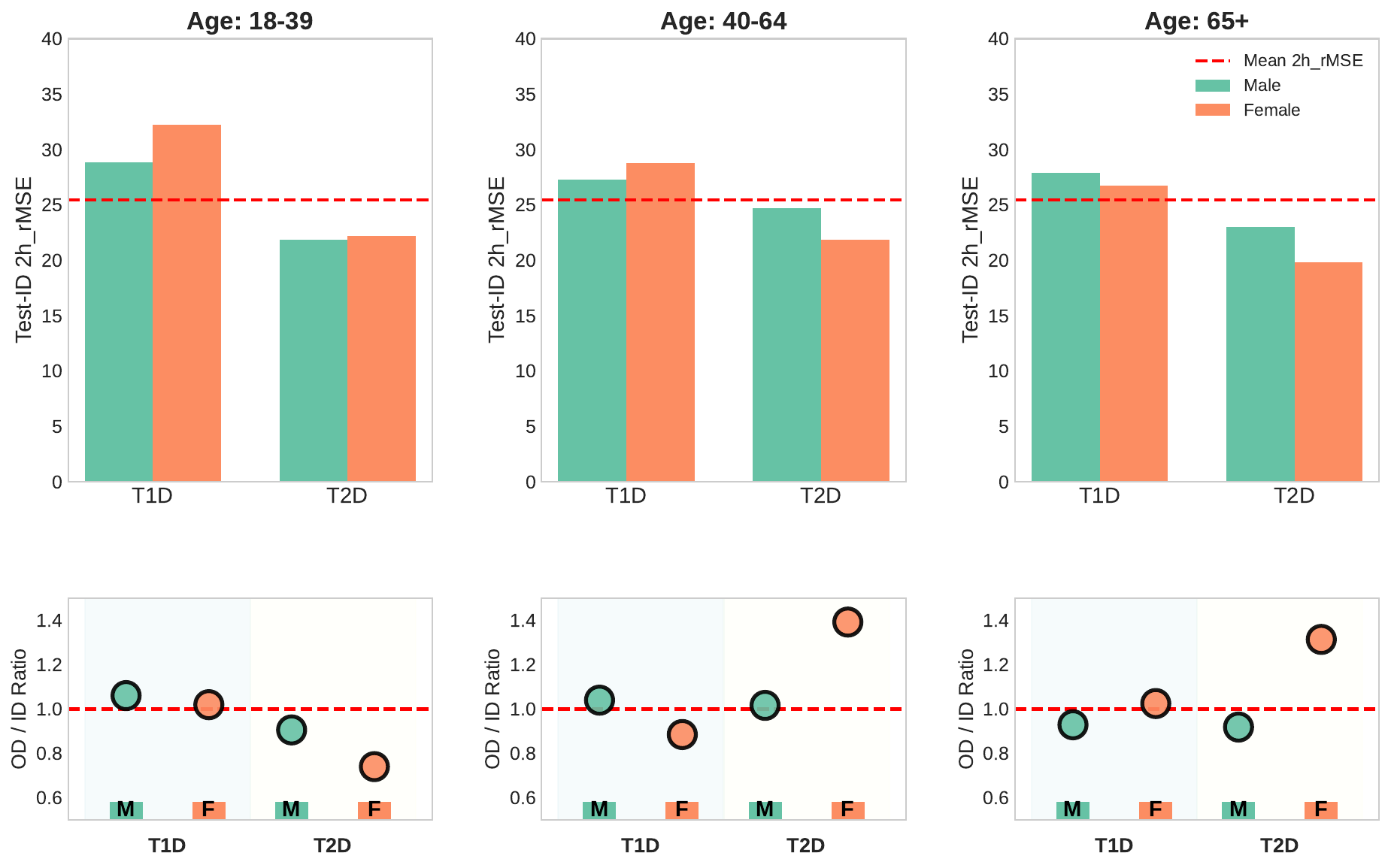}
    \caption{Intersectional subgroup accuracy and generalization across 12 demographic strata. Figure created by the authors using Matplotlib.}
    \label{fig:12groups-idod}
\end{figure}

%
\noindent\textbf{Performance Across Subgroups.}~~Subgroup patterns are consistent across models. T1D patients show consistently higher errors than T2D patients (rMSE difference 6.3\,mg/dL; 95\% bootstrap CI [4.3, 8.2]; permutation $p < 0.001$), reflecting the greater glucose variability and insulin dynamics characteristic of type~1 diabetes. This gap persists on held-out patients (test-od: $\Delta = 5.6$\,mg/dL, 95\% CI [1.0, 9.6], $p = 0.011$).
%
T2D patients also achieve notably higher clinical accuracy (Clarke A: 80.1\% vs 72.2\% for T1D among top-5 models), further demonstrating the prediction challenges posed by type~1 diabetes.
%
At the marginal level, gender differences are small and not significant ($\Delta_{\text{M}-\text{F}} = -0.19$\,mg/dL, 95\% CI $[-2.12, 1.76]$, $p = 0.86$), and younger adults (18--39) show higher error than the oldest group without reaching significance ($\Delta_{\text{18--39 vs 65+}} = 1.85$\,mg/dL, 95\% CI $[-0.63, 4.66]$, $p = 0.16$).
%
This patient-level marginal test should not be read against the per-model gender columns of Table~\ref{tab:main_performance}, which are sample-weighted and show a consistent female-minus-male gap near $+0.8$\,mg/dL; the marginal difference is small precisely because the gender effect changes sign across diabetes types.
%
Our balanced sampling protocol confirms these disparities reflect true physiological heterogeneity rather than sample size artifacts.

%
\noindent\textbf{Generalization Across Distributions.}~~Figure~\ref{fig:testid-od} shows overlapping confidence intervals between test-id and test-od across all models, with OD/ID ratios near 1.0, indicating robust generalization to out-of-distribution patients.
%
While population-level generalization appears strong, subgroup-level analysis reveals hidden heterogeneity in both accuracy and generalization behavior, motivating deeper fairness evaluation.

\subsection{Fairness Analysis}

%
\noindent\textbf{Intersectional Subgroup Disparities.}~~Figure~\ref{fig:12groups-idod} reveals intersectional patterns across age, gender, and diabetes type.
%
The gender pattern is conditional on diabetes type rather than uniform: among T1D patients males show lower error than females at ages 18--39 ($\Delta_{\text{F}-\text{M}} = +3.3$\,mg/dL) and 40--64 ($+2.1$), while among T2D patients females show lower error at 40--64 ($-3.4$) and 65+ ($-2.4$), and the T1D male advantage has disappeared by age 65+ ($-0.5$).
%
Critically, OD/ID ratios range from 0.8 to 1.4 across subgroups despite the population-level average near 1.0, demonstrating that aggregate metrics mask subgroup vulnerabilities.
%
For example, T2D males aged 65+ generalize robustly (OD/ID $=$ 0.90), while T2D females aged 40--64 degrade substantially (OD/ID $=$ 1.42), the widest gap in the cohort.
%
These findings underscore the necessity of intersectional fairness evaluation rather than single-axis demographic analysis.

\begin{figure}[t!]
    \centering
    \includegraphics[width=0.99\linewidth]{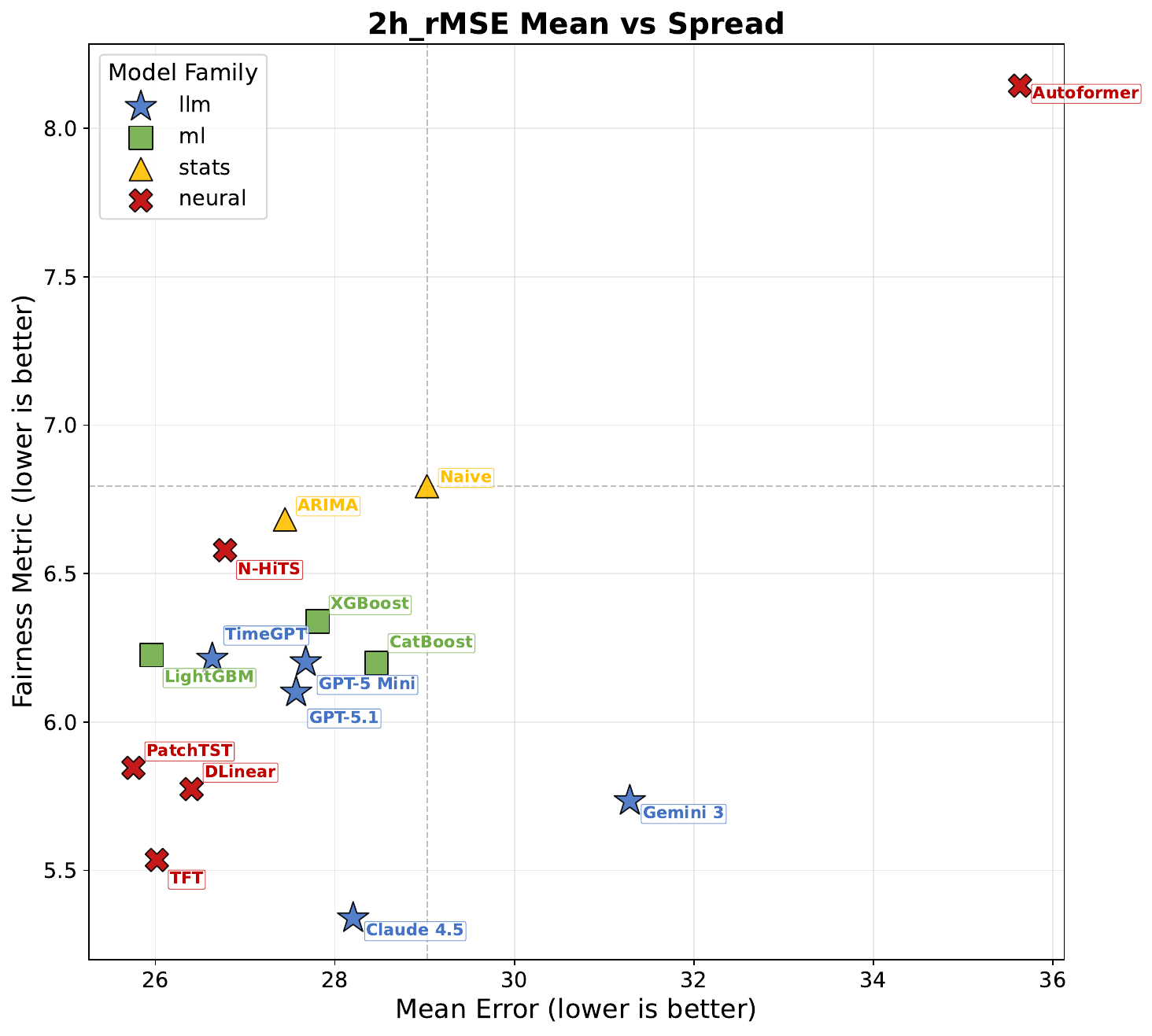}
    \caption{Model accuracy vs. subgroup disparity. Models near bottom-left (low error, low disparity) are preferred. PatchTST and TFT achieve strong performance on both axes. Figure created by the authors using Matplotlib.}
    \label{fig:model-fair}
\end{figure}

%
\noindent\textbf{Accuracy-Fairness Tradeoffs.}~~To assess both accuracy and fairness jointly, Figure~\ref{fig:model-fair} plots each model's mean 2-hour rMSE (x-axis) against subgroup performance disparity (y-axis), measured as rMSE standard deviation across the 12 demographic strata.
%
Models in the bottom-left quadrant (low error, low disparity) are preferred for equitable deployment.
%
Neural models PatchTST and TFT achieve this ideal combination, demonstrating that state-of-the-art accuracy does not require sacrificing fairness.
%
Gradient-boosted models sit mid-range on disparity (LightGBM 6.23, XGBoost 6.34) despite LightGBM's competitive mean error, and they are not the least equitable family: the neural model N-HiTS (6.58) is worse than both.
%
The statistical baselines are in fact the least equitable of the reasonable performers: ARIMA (6.68) and Na\"ive (6.79) rank 13th and 14th of 15 models on disparity, with Autoformer the outlier on both axes (error 35.6\,mg/dL, disparity 8.14).
%
The frontier LLMs separate on the two axes rather than clustering: they carry high mean error (Gemini~3 31.3\,mg/dL, Claude~4.5 28.2\,mg/dL) yet some of the lowest cross-subgroup disparity in the panel (Claude~4.5 at 5.34 is the lowest of all 15 models; Gemini~3 5.73 is third lowest).
%
Low disparity here reflects uniformly mediocre zero-shot prediction rather than equitable competence, so it is not evidence of deployment readiness; it does show that error magnitude and error equity are separable properties that a single leaderboard column would hide.
%
TimeGPT, a time-series-specific foundation model, sits closer to the neural models than to the general-purpose LLMs on mean error (26.6\,mg/dL) with mid-range disparity (6.21).


\subsection{Input Length Sensitivity}

%
\noindent\textbf{Subgroup-Level Patterns.}~~To assess whether longer input histories benefit all patient subgroups equally, we computed the temporal ratio (TR), defined as the rMSE at a given input length $L$ divided by the rMSE at the longest available input (288 steps, 24 hours), for each subgroup (see Methods for definition). A TR near 1.0 indicates that shorter inputs preserve full performance; higher values indicate degradation.

\begin{figure*}[t]
    \centering
    \begin{subfigure}[t]{0.48\linewidth}
        \centering
        \includegraphics[width=\linewidth]{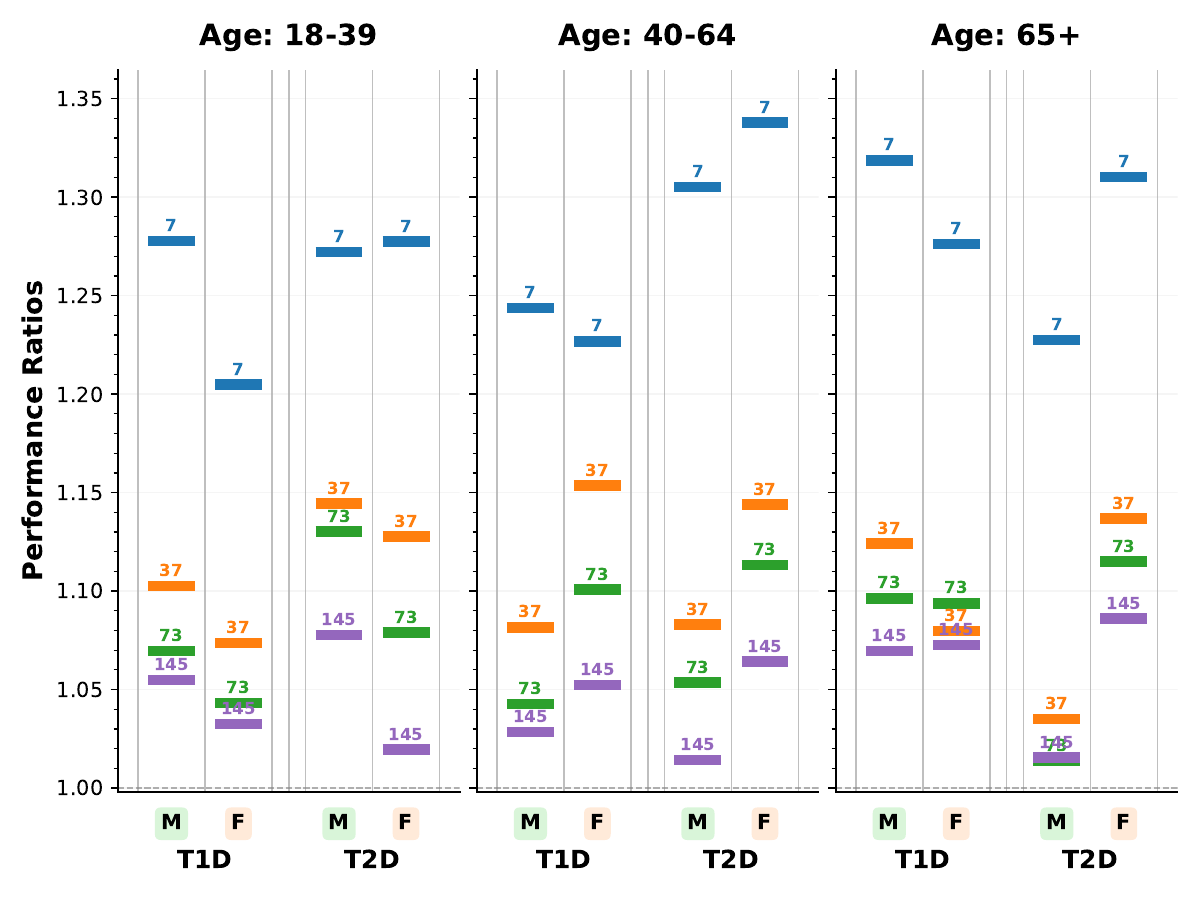}
        \caption{Subgroup-level sensitivity}
    \end{subfigure}
    \hfill
    \begin{subfigure}[t]{0.48\linewidth}
        \centering
        \includegraphics[width=\linewidth]{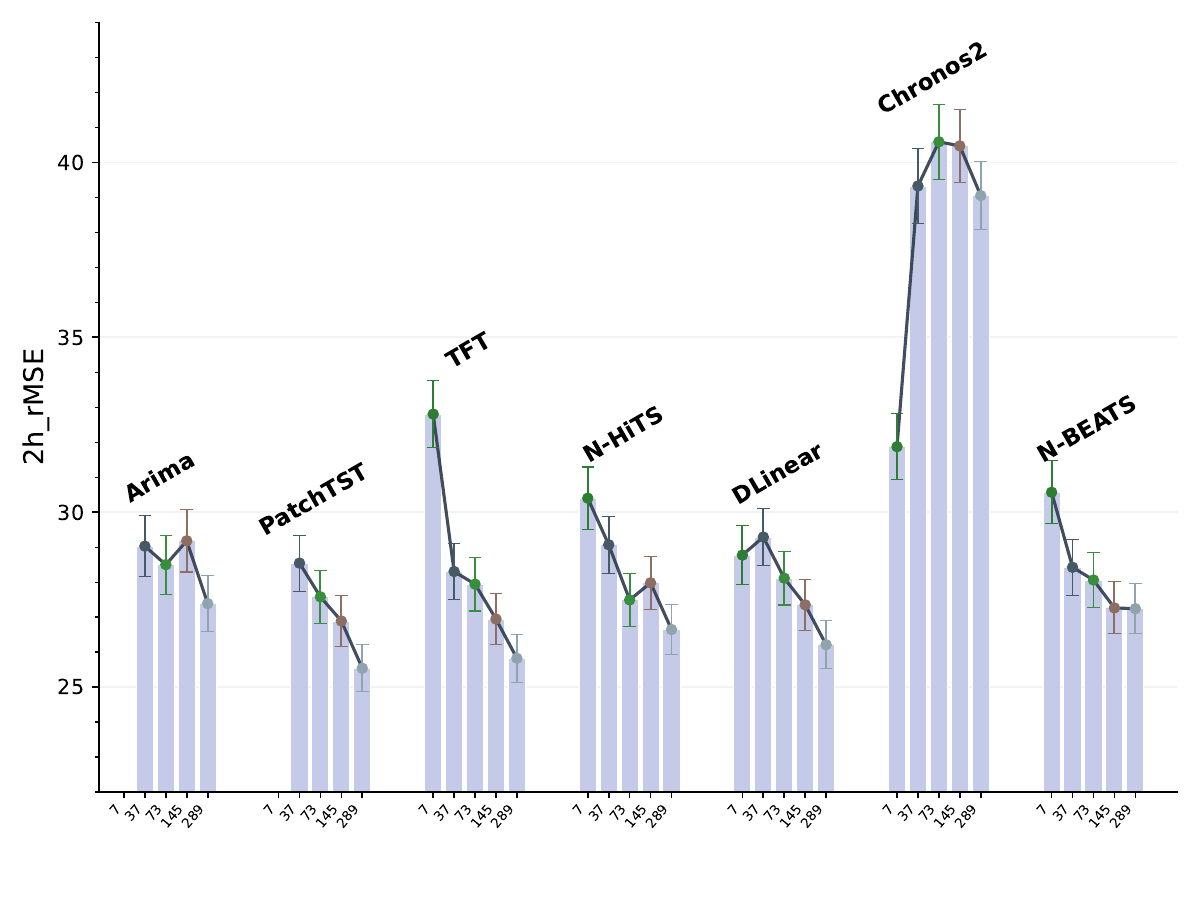}
        \caption{Model-level sensitivity}
    \end{subfigure}
    \caption{Input-length sensitivity analysis. (a) Subgroup-level temporal ratio (TR) across 12 demographic strata: shorter input windows degrade performance more for older T2D patients. (b) Model-level 2-hour rMSE across input lengths: neural architectures benefit most from extended input windows. Figure created by the authors using Matplotlib.}
    \label{fig:input-length-combined}
\end{figure*}

%
Figure~\ref{fig:input-length-combined} (left) reveals that the relative utility of input length varies substantially across demographic and clinical subgroups. Among younger patients (ages 18--39), particularly those with T1D, shorter input windows (7 or 37 steps) yield performance close to the 288-step baseline, suggesting that recent glucose history is sufficient for accurate forecasting in these groups. In contrast, older patients (ages 65+), especially those with T2D, experience larger performance degradation when input length is reduced, implying greater dependence on longer historical context.
%
Gender-specific trends also emerge: male and female patients within the same age and disease category respond differently to input window truncation, indicating complex interactions between physiology, disease progression, and temporal information. These findings demonstrate that the optimal input history length is not uniform across populations, highlighting the potential for subgroup-adaptive input-length strategies.

%
\noindent\textbf{Model-Level Patterns.}~~Figure~\ref{fig:input-length-combined} (right) investigates how varying input lengths affect model performance across families. Neural models, particularly PatchTST, TFT, and N-HiTS, exhibit the largest gains as input length increases, especially from short (7 or 37 steps) to long (145 or 288 steps) windows, indicating their capacity to capture long-range temporal dependencies in glucose dynamics. In contrast, linear models show marginal improvement, and foundation models (TimeGPT, LLM-based) show modest gains, suggesting limitations in their zero-shot temporal modeling capabilities.
%
These findings challenge prior claims that extended CGM input length provides limited benefit \cite{pikulin2024enhanced, sun2023adaptive}, showing instead that the benefit is model-dependent: architectures equipped to leverage longer histories gain substantially, with implications for both model selection and personalized input configuration.


\subsection{Instance-Level Difficulty}

%
\noindent\textbf{Difficulty Distribution Across Subgroups.}~~To investigate model robustness across prediction samples of varying difficulty, we define high-difficulty (hard) and low-difficulty (easy) instances using a majority-vote approach across models (see Methods for protocol). Instances where most models produce large errors are labeled hard; those where most produce small errors are labeled easy.

\begin{figure*}[t]
    \centering
    \begin{subfigure}[t]{0.48\linewidth}
        \centering
        \includegraphics[width=\linewidth]{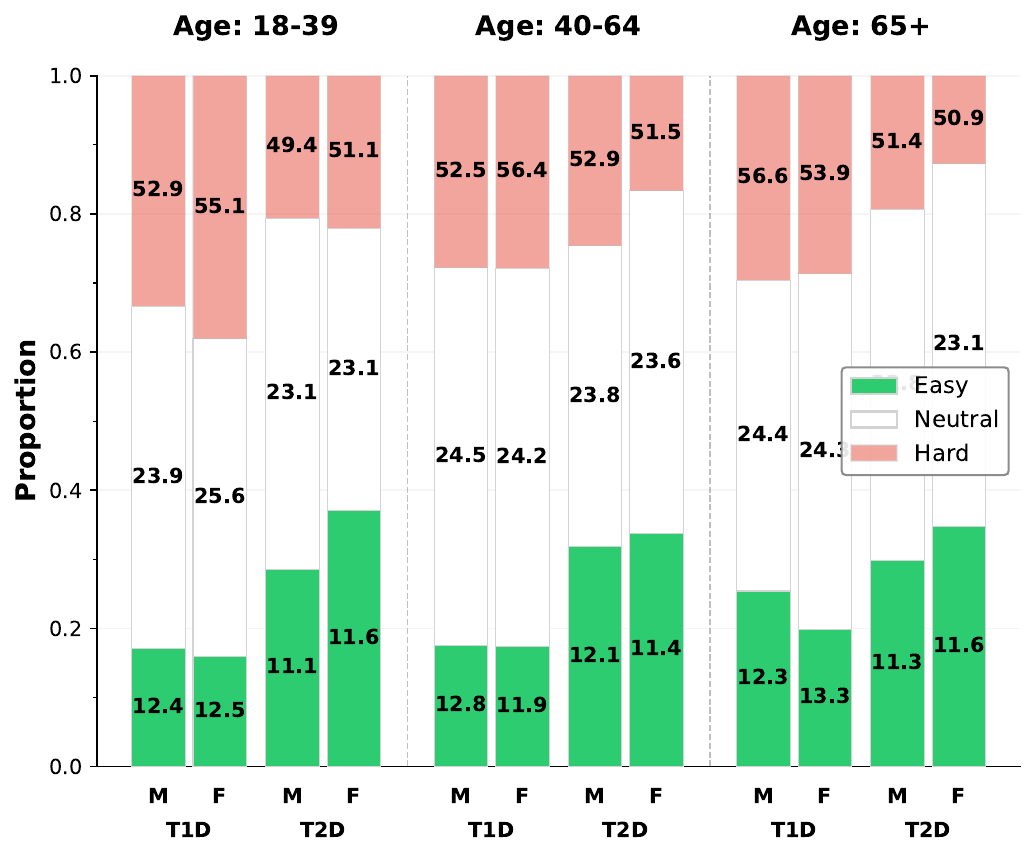}
        \caption{Difficulty distribution by subgroup}
    \end{subfigure}
    \hfill
    \begin{subfigure}[t]{0.48\linewidth}
        \centering
        \includegraphics[width=\linewidth]{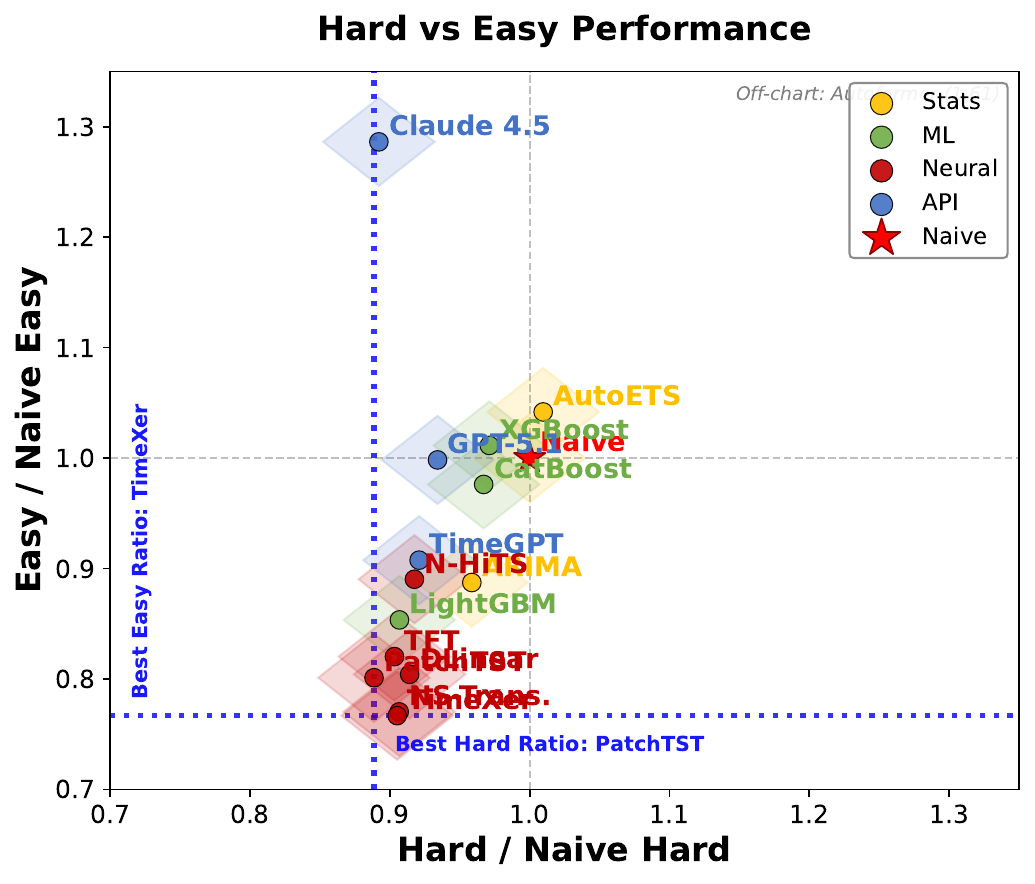}
        \caption{Hard vs.\ easy performance by model}
    \end{subfigure}
    \caption{Instance-level difficulty analysis. (a) Difficulty distribution across 12 demographic strata: hard cases concentrate in T1D subgroups. (b) Hard vs.\ easy case performance by model (normalized by Naive): top models concentrate gains on easy cases with limited improvement on hard cases. Figure created by the authors using Matplotlib.}
    \label{fig:difficulty-combined}
\end{figure*}

%
Figure~\ref{fig:difficulty-combined} (left) shows that the distribution of hard and easy cases varies substantially across demographic strata. Young male T1D and middle-aged female T1D patients have the highest proportion of hard cases, while older T2D males show the highest proportion of easy cases. These patterns indicate that prediction difficulty reflects underlying physiological and behavioral complexity rather than random variation.
%
While the share of hard and easy cases differs across subgroups, the average rMSE within each difficulty level is relatively stable: hard cases cluster around 63 mg/dL and easy cases around 10 mg/dL. This consistency suggests that instance-level difficulty, not subgroup identity alone, primarily drives prediction error. Subgroup disparities in overall performance therefore largely reflect differences in the \emph{proportion} of hard cases.
%
Hard cases often correspond to clinically critical periods such as rapid glucose excursions, post-meal spikes, or exercise-induced drops, scenarios where accurate prediction is most valuable. Evaluation metrics focused on average performance may underweight these high-risk moments, masking potential real-world performance gaps.

%
\noindent\textbf{Model Robustness Across Difficulty.}~~Figure~\ref{fig:difficulty-combined} (right) compares model performance on easy and hard instances, normalized by the Naive baseline. Top neural models improve on both easy cases (easy ratios 0.77--0.89) and hard cases (hard ratios 0.89--0.92), but the gains are asymmetric: models achieve larger relative improvements on easy cases than on hard cases. PatchTST achieves the best hard ratio (0.89), while TimeXer achieves the best easy ratio (0.77).
%
This asymmetry reveals that top-performing models concentrate their aggregate accuracy gains on easy cases, with more limited improvement on hard cases, precisely the clinically critical scenarios where accurate prediction is most needed. Statistical baselines (ARIMA, AutoETS) maintain ratios near 1.0 on both easy and hard cases, while frontier LLMs show variable performance. Future work should prioritize difficulty-stratified evaluation and hard-case-targeted training to improve both clinical robustness and fairness.


%
Beyond demographic fairness, input-length sensitivity, and instance difficulty, we examine how behavioral events (meals, exercise, medication) impact prediction difficulty and whether event handling varies equitably across patient subgroups, adding a temporal-behavioral dimension to fairness evaluation.

\input{0-display/Table/table3_event_vs_noevent.tex}

\subsection{Event Impact on Performance}

%
\noindent\textbf{Event Prevalence and Demographics.}~~Event annotations enable stratified evaluation of prediction difficulty under glucose-altering conditions.
%
Among the 16.8\% of samples containing events (22,239 of 132,480 total), medication events are most prevalent (72.8\% of window-level event occurrences), followed by diet (15.9\%) and exercise (11.3\%).
%
Event density varies across demographics: patients aged 65+ log 22\% fewer events than younger groups, while T2D patients log 12\% more events than T1D patients, reflecting differences in disease management patterns and app engagement.

%
\noindent\textbf{Event-Stratified Prediction Difficulty.}~~To quantify how event features affect prediction accuracy, we compare models with event features enabled (``Event'') versus disabled (``NoEvent'') (Table~\ref{tab:event_vs_noevent}). Because event features include post-observation events not necessarily available at prediction time, these results represent an upper-bound estimate of prospective event-aware forecasting performance.

%
Table~\ref{tab:event_vs_noevent} reveals a surprising finding: events barely affect accuracy for top-performing models. PatchTST and TFT show differences of only $-0.11$ and $-0.08$ mg/dL respectively (negative values indicating event model performs slightly \textit{better}), far below clinical significance thresholds ($\sim$5 mg/dL).
%
This pattern holds across all subgroups: T1D patients show $-0.15$ to $-0.25$ mg/dL differences, age groups span $+0.16$ to $-0.35$\,mg/dL, and both genders exhibit differences near $-0.10$ mg/dL.

%
In contrast, Autoformer shows larger event-related improvements ($-0.77$ mg/dL), while VanillaTransformer shows minimal difference ($-0.20$ mg/dL).
%
Across all 32 comparisons (4 models $\times$ 8 subgroups), models with event features consistently match or outperform those without; in fact, 31 of 32 show the event model performing better or equivalently.
%
This suggests that explicit event features provide small but consistent gains, though modern neural forecasters may implicitly learn event-driven glucose dynamics from 24-hour temporal patterns even without explicit event annotations.

%
\noindent\textbf{Event-Related Fairness Analysis.}~~To assess fairness, we examine whether event features benefit all demographic subgroups equally. Table~\ref{tab:event_vs_noevent} shows consistent patterns across 32 model-subgroup comparisons: 22 show event models performing better ($\Delta < -0.1$ mg/dL), 9 show negligible differences ($|\Delta| < 0.1$), and only 1 shows the noevent model slightly better (PatchTST for 65+: $\Delta = +0.16$ mg/dL).

%
The consistency across subgroups indicates event features do not create performance disparities. For top-performing models (PatchTST, TFT), event benefits typically range from $-0.08$ to $-0.23$ mg/dL for most subgroups; one exception (PatchTST for 65+) shows $+0.16$ mg/dL, though this difference remains clinically negligible.
%
T1D and T2D patients show similar patterns, males and females benefit comparably, and age groups show only minor variation.
%
This uniformity suggests that the evaluated neural forecasters handle available behavioral event features similarly across diverse patient profiles, supporting further evaluation of event-aware forecasting without immediate evidence of demographic-specific event-feature harm.

%% file: 0-display/Table/table1_cohort_characteristics.tex
\begin{table}[t]
    \centering
    \caption{Glycemic characteristics of the FairGlucose cohort by demographic stratum (all 25 patients per stratum, held-in and held-out pooled; split-wise breakdowns in Supplementary Note~A). Each stratum contains 25 patients (20 held-in for training/validation/test-ID; 5 held-out for test-OD). Glucose values in mg/dL; CV is the coefficient of variation; TIR is time-in-range (70--180~mg/dL); MAGE is the mean amplitude of glycemic excursions.}
    \label{tab:cohort_char}
    \small
    \begin{tabular}{l|c|cccc}
    \hline
    \textbf{Stratum} & \textbf{N} & \textbf{Mean glucose} & \textbf{CV} & \textbf{TIR (\%)} & \textbf{MAGE} \\
    \hline
    \multicolumn{6}{l}{\textit{Overall and marginal subgroups}} \\
    \hline
    Overall                   & 300 & 162.7 & 0.26 & 67.1 & 4.06 \\
    Age 18--39                & 100 & 163.8 & 0.26 & 65.9 & 4.40 \\
    Age 40--64                & 100 & 165.9 & 0.26 & 64.9 & 4.23 \\
    Age $\geq$65              & 100 & 158.5 & 0.25 & 70.6 & 3.57 \\
    Female                    & 150 & 162.9 & 0.25 & 67.9 & 4.11 \\
    Male                      & 150 & 162.6 & 0.26 & 66.3 & 4.02 \\
    Type 1 (T1D)              & 150 & 163.7 & 0.29 & 65.3 & 4.20 \\
    Type 2 (T2D)              & 150 & 161.8 & 0.23 & 68.9 & 3.92 \\
    \hline
    \multicolumn{6}{l}{\textit{12-way intersectional strata (age $\times$ gender $\times$ type)}} \\
    \hline
    18--39 F T1D              & 25  & 165.7 & 0.29 & 64.2 & 4.67 \\
    18--39 F T2D              & 25  & 160.6 & 0.22 & 69.0 & 4.02 \\
    18--39 M T1D              & 25  & 174.6 & 0.30 & 59.0 & 4.67 \\
    18--39 M T2D              & 25  & 154.1 & 0.23 & 71.3 & 4.24 \\
    40--64 F T1D              & 25  & 163.1 & 0.27 & 65.0 & 4.12 \\
    40--64 F T2D              & 25  & 169.3 & 0.23 & 66.1 & 4.32 \\
    40--64 M T1D              & 25  & 161.0 & 0.30 & 65.5 & 4.32 \\
    40--64 M T2D              & 25  & 170.4 & 0.22 & 62.9 & 4.15 \\
    65+ F T1D                 & 25  & 157.5 & 0.27 & 72.3 & 3.89 \\
    65+ F T2D                 & 25  & 161.0 & 0.22 & 70.9 & 3.62 \\
    65+ M T1D                 & 25  & 160.5 & 0.29 & 65.8 & 3.55 \\
    65+ M T2D                 & 25  & 155.1 & 0.24 & 73.2 & 3.20 \\
    \hline
    \end{tabular}
\end{table}

%% file: 0-display/Table/table1.tex
\begin{table}[t!]
    \centering
    \caption{Dataset partition overview. Each cell shows the per-subgroup value (total across 12 subgroups in parentheses).}
    \label{tab:split}
    \small
    \begin{tabular}{cc|ccc}
    \hline
    \multicolumn{2}{c|}{\textbf{Split}} & \textbf{Patients} & \textbf{Patient-Days} & \textbf{I/O Pairs} \\ \hline
    \multirow{3}{*}{\textbf{Held-in}}
        & Train   & 20 (240) & 300 (3,600)  & 7,200 (86,400)      \\
        & Valid   & 20 (240) & 40 (480)     & 960 (11,520)        \\
        & Test-ID & 20 (240) & 60 (720)     & 1,440 (17,280)      \\ \hline
    \textbf{Held-out}
        & Test-OD & 5 (60)   & 60 (720)     & 1,440 (17,280)      \\ \hline
    \multicolumn{2}{c|}{\textbf{Total}}
                 & 25 (300) & 460 (5,520)   & 11,040 (132,480)    \\ \hline
    \end{tabular}
\end{table}

%% file: 0-display/Table/table2_rmse_clarkeA.tex
\begin{table*}[t]
\centering
\resizebox{\textwidth}{!}{%
\begin{tabular}{l|llllllll|llllllll}
\hline
\multirow{2}{*}{\textbf{Model}} & \multicolumn{8}{c|}{\textbf{rMSE (mg/dL)}} & \multicolumn{8}{c}{\textbf{Clarke A (\%)}} \\ \cline{2-17}
 & \textbf{All} & \textbf{T1D} & \textbf{T2D} & \textbf{Female} & \textbf{Male} & \textbf{18--39} & \textbf{40--64} & \textbf{65+} & \textbf{All} & \textbf{T1D} & \textbf{T2D} & \textbf{Female} & \textbf{Male} & \textbf{18--39} & \textbf{40--64} & \textbf{65+} \\ \hline

\textbf{ARIMA} & \underline{27.44} & \underline{30.78} & \underline{24.10} & \underline{27.89} & \underline{26.99} & \underline{27.76} & \underline{28.05} & \underline{26.51} & \underline{74.67} & \underline{70.26} & \underline{79.07} & \underline{74.14} & \underline{75.19} & \underline{74.11} & \underline{74.55} & \underline{75.38} \\
\textbf{Naive} & 29.03 & 32.42 & 25.63 & 29.54 & 28.51 & 29.41 & 29.78 & 27.85 & 72.86 & 68.61 & 77.11 & 72.45 & 73.27 & 71.98 & 72.90 & 73.76 \\
\textbf{AutoETS} & 29.72 & 33.35 & 26.09 & 30.44 & 29.00 & 30.27 & 30.53 & 28.31 & 72.96 & 68.63 & 77.29 & 72.41 & 73.51 & 71.53 & 73.03 & 74.46 \\
\hline
\textbf{LightGBM} & \underline{25.96} & 29.07 & \underline{22.84} & \underline{26.34} & 25.58 & \underline{26.59} & 26.58 & \underline{24.63} & \textbf{76.77*} & \underline{72.45} & \textbf{81.10*} & \textbf{76.30*} & \textbf{77.24*} & \textbf{75.25*} & \textbf{77.16*} & \textbf{78.08*} \\
\textbf{Linear} & 26.04 & \underline{28.92} & 23.15 & 26.51 & \underline{25.56} & 26.80 & \underline{26.50} & 24.75 & 75.57 & 71.64 & 79.50 & 74.97 & 76.17 & 73.87 & 76.22 & 76.74 \\
\textbf{XGBoost} & 27.81 & 30.98 & 24.64 & 28.41 & 27.20 & 28.07 & 28.74 & 26.56 & 75.15 & 70.62 & 79.68 & 74.57 & 75.73 & 74.18 & 74.94 & 76.49 \\
\textbf{CatBoost} & 28.46 & 31.56 & 25.36 & 29.06 & 27.87 & 29.10 & 29.07 & 27.19 & 73.81 & 69.94 & 77.68 & 73.23 & 74.38 & 72.37 & 74.12 & 75.03 \\
\hline
\textbf{NS-Transformer} & \textbf{25.62*} & \textbf{28.46*} & \textbf{22.79} & \textbf{26.02*} & \textbf{25.23*} & 25.99 & \textbf{26.39*} & \textbf{24.44} & \textbf{76.45} & \textbf{72.70*} & 80.20 & 75.86 & \textbf{77.03} & \textbf{75.22} & \textbf{76.61} & 77.61 \\
\textbf{Transformer} & \textbf{25.65} & 28.56 & \textbf{22.74*} & \textbf{26.02} & \textbf{25.28} & 26.10 & \textbf{26.46} & \textbf{24.34*} & 76.43 & \textbf{72.45} & \textbf{80.41} & \textbf{75.87} & 77.00 & 75.16 & 76.24 & \textbf{78.02} \\
\textbf{TimeXer} & 25.75 & \textbf{28.53} & 22.96 & 26.12 & 25.38 & \textbf{25.97*} & 26.47 & 24.76 & 76.06 & 72.07 & 80.05 & 75.32 & 76.80 & 74.93 & 76.33 & 77.02 \\
\textbf{PatchTST} & 25.75 & 28.68 & 22.83 & 26.08 & 25.43 & \textbf{25.98} & 26.57 & 24.66 & 75.98 & 72.02 & 79.93 & 75.20 & 76.75 & 75.07 & 75.99 & 76.97 \\
\textbf{Crossformer} & 25.85 & 28.70 & 23.00 & 26.29 & 25.41 & 26.06 & 26.60 & 24.86 & 75.79 & 71.86 & 79.73 & 75.05 & 76.53 & 74.62 & 76.02 & 76.82 \\
\textbf{Informer} & 25.94 & 28.79 & 23.09 & 26.29 & 25.59 & 26.12 & 26.65 & 25.01 & 76.13 & 72.18 & 80.09 & 75.54 & 76.73 & 75.00 & 76.45 & 77.05 \\
\textbf{TFT} & 26.02 & 28.78 & 23.25 & 26.40 & 25.64 & 26.21 & 26.75 & 25.06 & 75.52 & 71.68 & 79.37 & 74.85 & 76.20 & 74.72 & 75.71 & 76.20 \\
\textbf{MICN} & 26.11 & 28.96 & 23.26 & 26.45 & 25.77 & 26.36 & 26.88 & 25.04 & 75.57 & 71.66 & 79.48 & 75.02 & 76.12 & 74.47 & 75.81 & 76.50 \\
\textbf{TimesNet} & 26.40 & 29.45 & 23.35 & 26.75 & 26.05 & 26.68 & 27.01 & 25.46 & 75.61 & 71.29 & 79.92 & 75.11 & 76.10 & 74.33 & 76.15 & 76.45 \\
\textbf{DLinear} & 26.40 & 29.29 & 23.52 & 26.65 & 26.16 & 26.63 & 27.03 & 25.51 & 75.15 & 71.04 & 79.26 & 74.74 & 75.57 & 74.07 & 75.62 & 75.83 \\
\textbf{iTransformer} & 26.49 & 29.49 & 23.50 & 26.74 & 26.25 & 26.78 & 27.19 & 25.44 & 75.07 & 70.99 & 79.15 & 74.50 & 75.64 & 73.72 & 75.35 & 76.29 \\
\textbf{SCINet} & 26.50 & 29.32 & 23.68 & 26.83 & 26.17 & 26.80 & 27.09 & 25.56 & 75.12 & 71.20 & 79.03 & 74.70 & 75.53 & 73.92 & 75.59 & 75.91 \\
\textbf{N-HiTS} & 26.78 & 30.07 & 23.49 & 26.96 & 26.60 & 27.09 & 27.30 & 25.89 & 74.70 & 70.24 & 79.15 & 74.28 & 75.12 & 73.55 & 75.08 & 75.55 \\
\textbf{TiDE} & 26.80 & 29.71 & 23.88 & 27.05 & 26.54 & 26.93 & 27.42 & 26.01 & 74.50 & 70.43 & 78.58 & 74.07 & 74.94 & 73.65 & 74.80 & 75.13 \\
\textbf{TSMixer} & 26.80 & 29.85 & 23.75 & 27.04 & 26.55 & 26.94 & 27.53 & 25.87 & 74.26 & 69.97 & 78.55 & 73.80 & 74.72 & 73.33 & 74.64 & 74.87 \\
\textbf{TimeMixer} & 26.93 & 29.94 & 23.92 & 27.21 & 26.65 & 27.21 & 27.64 & 25.89 & 74.34 & 70.09 & 78.59 & 73.98 & 74.70 & 73.08 & 74.72 & 75.31 \\
\textbf{N-BEATS} & 27.42 & 30.57 & 24.28 & 27.65 & 27.20 & 27.65 & 28.00 & 26.61 & 73.33 & 68.85 & 77.81 & 72.98 & 73.67 & 72.31 & 73.78 & 73.93 \\
\textbf{FEDformer} & 31.17 & 34.67 & 27.67 & 31.58 & 30.76 & 31.02 & 31.67 & 30.82 & 66.98 & 61.77 & 72.19 & 66.38 & 67.58 & 66.85 & 67.34 & 66.78 \\
\textbf{Autoformer} & 35.63 & 39.71 & 31.56 & 36.34 & 34.93 & 34.85 & 36.03 & 36.12 & 60.68 & 54.50 & 66.85 & 59.73 & 61.62 & 61.39 & 61.78 & 58.72 \\
\textbf{Chronos2} & 39.70 & 44.21 & 35.18 & 40.14 & 39.25 & 38.86 & 39.71 & 40.63 & 54.22 & 47.92 & 60.52 & 53.92 & 54.52 & 55.25 & 55.84 & 51.38 \\
\hline
\textbf{TimeGPT} & \underline{26.63} & \underline{29.74} & \underline{23.53} & \underline{27.29} & \underline{25.98} & \underline{26.96} & \underline{27.39} & \underline{25.52} & \underline{75.55} & \underline{71.40} & \underline{79.70} & \underline{74.72} & \underline{76.38} & \underline{74.79} & \underline{75.32} & \underline{76.62} \\
\textbf{GPT-5.1} & 27.57 & 30.62 & 24.52 & 28.03 & 27.10 & 27.89 & 28.83 & 25.90 & 74.91 & 70.60 & 79.22 & 74.39 & 75.43 & 74.07 & 74.23 & 76.58 \\
\textbf{GPT-5 Mini} & 27.67 & 30.78 & 24.57 & 28.26 & 27.09 & 27.91 & 28.84 & 26.23 & 74.81 & 70.48 & 79.15 & 74.08 & 75.54 & 73.91 & 74.34 & 76.31 \\
\textbf{Claude 4.5 Sonnet} & 28.20 & 30.87 & 25.53 & 28.87 & 27.54 & 28.50 & 29.71 & 26.30 & 73.17 & 69.64 & 76.70 & 72.22 & 74.11 & 71.79 & 72.67 & 75.20 \\
\textbf{Claude 4.5 Haiku} & 28.77 & 31.78 & 25.76 & 29.31 & 28.23 & 29.15 & 29.84 & 27.26 & 73.33 & 69.19 & 77.47 & 72.51 & 74.16 & 71.95 & 73.35 & 74.87 \\
\textbf{Gemini 3 Flash} & 31.29 & 34.15 & 28.42 & 31.88 & 30.69 & 31.27 & 32.69 & 29.85 & 69.92 & 65.63 & 74.21 & 69.13 & 70.72 & 69.43 & 69.40 & 71.08 \\
\hline

\end{tabular}%
}
\caption{
Model performance on rMSE (mg/dL) and Clarke Error Grid Zone A (\%) at the 2-hour prediction horizon, averaged across test-ID and test-OD splits, across all subgroups defined by diabetes type (T1D, T2D), gender (Female, Male), and age (18--39, 40--64, 65+).
Clarke A represents clinically accurate predictions. Higher Clarke A percentage indicates better clinical accuracy.
\underline{Underlined} values indicate the best performance within each model family. 
\textbf{Bolded} values highlight the top-2 overall performers per subgroup: 
\textbf{*} denotes the best performance and bold without asterisk denotes the second best.
}
\label{tab:main_performance}
\end{table*}

%% file: 0-display/Table/table3_event_vs_noevent.tex
\definecolor{darkgreen}{RGB}{0,100,0}
\definecolor{darkred}{RGB}{139,0,0}
\begin{table*}[!htbp]
\centering
\caption{
Comparison of event vs noevent model performance at 2-hour prediction horizon across subgroups.
Event features include both historical (pre-observation) and future (post-observation) events; these results represent an upper-bound estimate of prospective event-aware forecasting.
$\Delta$ = Event - NoEvent: \textcolor{darkgreen}{Negative values (dark green, $\Delta < -0.1$)} indicate event model is better,
\textcolor{darkred}{positive values (dark red, $\Delta \geq 0.1$)} indicate noevent model is better.
Differences with $|\Delta| < 0.1$ mg/dL are shown in black (negligible). Note that the models in this table are the Nixtla/\texttt{neuralforecast} implementations used for the event ablation, whereas Table~\ref{tab:main_performance} reports the TSLib implementations; \textbf{VanillaTrans.}\ here is therefore not the same code path as \textbf{Transformer} there, and the two should not be read across. Only the Event vs.\ NoEvent contrast within a column is meaningful.
}
\label{tab:event_vs_noevent}
\small
\resizebox{\textwidth}{!}{%
\begin{tabular}{l|cccc|cccc|cccc}
\hline
\multirow{2}{*}{\textbf{Subgroup}} & \multicolumn{4}{c|}{\textbf{Event rMSE (mg/dL)}} & \multicolumn{4}{c|}{\textbf{NoEvent rMSE (mg/dL)}} & \multicolumn{4}{c}{\textbf{$\Delta$ (Event - NoEvent)}} \\ \cline{2-13}
 & \textbf{PatchTST} & \textbf{TFT} & \textbf{Autoformer} & \textbf{VanillaTrans.} & \textbf{PatchTST} & \textbf{TFT} & \textbf{Autoformer} & \textbf{VanillaTrans.} & \textbf{PatchTST} & \textbf{TFT} & \textbf{Autoformer} & \textbf{VanillaTrans.} \\ \hline
\textbf{All} & 25.37 & 26.00 & 37.95 & 47.52 & 25.48 & 26.08 & 38.71 & 47.72 & \textcolor{darkgreen}{-0.11} & -0.08 & \textcolor{darkgreen}{-0.77} & \textcolor{darkgreen}{-0.20} \\
\textbf{T1D} & 28.21 & 28.94 & 42.30 & 48.66 & 28.36 & 28.97 & 43.04 & 48.90 & \textcolor{darkgreen}{-0.15} & -0.04 & \textcolor{darkgreen}{-0.74} & \textcolor{darkgreen}{-0.25} \\
\textbf{T2D} & 22.53 & 23.06 & 33.59 & 46.37 & 22.60 & 23.18 & 34.39 & 46.53 & -0.07 & \textcolor{darkgreen}{-0.12} & \textcolor{darkgreen}{-0.80} & \textcolor{darkgreen}{-0.16} \\
\textbf{Female} & 25.78 & 26.39 & 38.55 & 49.07 & 25.91 & 26.51 & 39.26 & 49.24 & \textcolor{darkgreen}{-0.13} & \textcolor{darkgreen}{-0.13} & \textcolor{darkgreen}{-0.71} & \textcolor{darkgreen}{-0.16} \\
\textbf{Male} & 24.96 & 25.61 & 37.34 & 45.96 & 25.05 & 25.64 & 38.17 & 46.20 & -0.09 & -0.03 & \textcolor{darkgreen}{-0.83} & \textcolor{darkgreen}{-0.24} \\
\textbf{18--39} & 25.69 & 26.23 & 37.08 & 47.35 & 25.92 & 26.33 & 37.83 & 47.36 & \textcolor{darkgreen}{-0.23} & -0.10 & \textcolor{darkgreen}{-0.74} & -0.00 \\
\textbf{40--64} & 26.10 & 26.74 & 38.14 & 46.93 & 26.33 & 26.79 & 38.92 & 47.28 & \textcolor{darkgreen}{-0.23} & -0.05 & \textcolor{darkgreen}{-0.77} & \textcolor{darkgreen}{-0.35} \\
\textbf{65+} & 24.28 & 24.99 & 38.67 & 48.40 & 24.12 & 25.07 & 39.46 & 48.65 & \textcolor{darkred}{+0.16} & -0.08 & \textcolor{darkgreen}{-0.79} & \textcolor{darkgreen}{-0.25} \\
\hline
\end{tabular}%
}
\end{table*}

%% file: 0-sections/03_discussion.tex

\section{Discussion}\label{sec:discussion}

%
Our principal finding is that population-level external-validation metrics can create false reassurance about CGM forecasting equity. Aggregate out-of-distribution performance appears stable ($\approx$1.0), yet subgroup-level ratios range from 0.8 to 1.4 across 12 demographic strata. This gap is not limited to any single model: it persists across all 33 evaluated models spanning four paradigms (statistical, machine-learning, neural time-series, and frontier LLM/foundation), suggesting that the disparity reflects the prediction task itself. Because the FairGlucose cohort is demographically balanced by construction, these disparities cannot be attributed to unequal sample sizes.

%
This has direct implications for how CGM-based AI tools should be qualified before clinical deployment. If evaluation relies solely on population-level metrics, subgroups such as young T1D females, who carry the highest in-distribution error in this cohort (31.9\,mg/dL against 19.9\,mg/dL for the best-served subgroup), may be systematically underserved; notably the worst-generalizing subgroup is a different one again (T2D females aged 40--64, OD/ID $=$ 1.42), so accuracy and generalization single out different groups and both need reporting. 
Subbaswamy et al. \cite{subbaswamy2024framework} demonstrated this problem in general clinical AI; our results show that analogous concerns arise in CGM forecasting across 33 models. Recent reviews of AI fairness in healthcare \cite{liu2025scoping, xu2024fairness} have identified evidence gaps across medical specialties; FairGlucose addresses this gap for CGM-based digital diabetes tools. We recommend that subgroup-disaggregated reporting become a default standard for digital health AI evaluation, analogous to how clinical trials report outcomes by demographic subgroup.

%
Compared to existing CGM benchmarks, FairGlucose addresses several key limitations. GlucoBench \cite{sergazinov2024glucobench} aggregates five public datasets ($\sim$461 subjects) but does not enforce demographic balance, making subgroup-level fairness evaluation unreliable. OhioT1DM \cite{marling2020ohiot1dm} provides only 12 type~1 diabetes patients. GluFormer \cite{gluformer2025}, trained on 10 million CGM measurements, acknowledges demographic limitations but does not perform subgroup evaluation. FairGlucose provides 300 patients across 12 balanced strata with behavioral event annotations, enabling a systematic fairness evaluation of CGM forecasting models. Cohort metadata, splits, code, and the reference leaderboard will be made available at \url{https://github.com/JHU-CDHAI/FairGlucose} upon acceptance; raw CGM data can be requested under a data-use agreement with Welldoc, Inc.

%
Two additional dimensions reinforce this conclusion. Input-length sensitivity varies across subgroups: older T2D patients require longer CGM histories for stable prediction, while younger T1D patients tolerate short input windows with minimal accuracy loss. This motivates personalized or subgroup-adaptive input-length strategies rather than fixed-length configurations. Instance-level difficulty analysis further shows that subgroup performance disparities align with the proportion of clinically hard cases (rapid glucose excursions, post-meal spikes, exercise-induced drops). Top-performing neural and ML models concentrate their aggregate accuracy gains on easy cases, yet offer minimal improvement on these clinically critical hard cases compared to simple baselines like ARIMA. This asymmetry suggests that average-case evaluation metrics can underweight high-risk scenarios, and that difficulty-stratified reporting should accompany subgroup-disaggregated evaluation for comprehensive equity assessment.

%
Several important limitations warrant attention.
%
First, event annotations are limited to app-logged data (meals, exercise, medication); passive sensing modalities (sleep, stress, physical activity patterns) are not captured.
%
Second, data originates from a US-based mobile health platform; generalization to other populations, geographic regions, and CGM devices should be validated.
%
Third, the 12 subgroup stratification (age, gender, diabetes type) does not capture all fairness-relevant factors such as race, socioeconomic status, or comorbidities.
%
Fourth, there is risk of overfitting to specific dataset properties; we encourage model development on multiple datasets alongside FairGlucose.

%
Future directions for the community include:
(1) expanding to multi-site cohorts with broader demographic and device coverage;
(2) developing event-aware and personalized forecasting approaches;
(3) investigating fairness-aware model architectures using the subgroup performance metrics; and
(4) incorporating hybrid or mechanistic models for interpretability.
%
The standardized evaluation protocols can inform regulatory approval and clinical deployment decisions.
%
FairGlucose provides the infrastructure needed for reproducible, fairness-focused CGM forecasting development.


%
FairGlucose provides a demographically balanced CGM benchmark with behavioral event annotations and a 33-model reference leaderboard, enabling reproducible fairness research in glucose forecasting. We hope this infrastructure accelerates development of fairness-aware CGM forecasting and supports more equitable diabetes care.


%% file: 0-sections/04_methods.tex

\section{Methods}\label{sec:methods}

This section details the construction of the FairGlucose benchmark and the protocol used for baseline evaluation. We describe (i) cohort construction and dataset partitioning, (ii) input-output sample construction and event annotation, (iii) the predictive-accuracy and fairness metrics used for evaluation, and (iv) the baseline models and training setup.

\subsection{Dataset Partitioning}

%
The dataset is constructed from de-identified CGM records collected via a mobile health platform.
%
%
Data were collected from existing users of a US-based mobile health digital platform for diabetes management, reflecting naturalistic self-management behaviors rather than controlled study conditions.
%
Participants used either Dexcom G6 or Abbott FreeStyle Libre continuous glucose monitors, both factory-calibrated devices providing readings every 5 minutes with manufacturer-stated accuracy of approximately 9--10\% mean absolute relative difference (MARD).
%
The mixed-device cohort ensures model robustness across CGM platforms, reflecting real-world deployment scenarios where patients may use different devices. Incomplete CGM traces are excluded via the qualified patient-day criterion described below; no imputation is applied.
%
As shown in Figure~\ref{fig:figure1}, to capture physiological and behavioral heterogeneity, we stratify patients across three axes: age (18–39, 40–64, $\geq$65), gender (male, female), and diabetes type (Type 1, Type 2), yielding 12 disjoint subgroups.
%
No additional personal information beyond the reported stratification variables is used for modeling or evaluation.
%
We sample 25 patients from each subgroup, ensuring sufficient CGM records for robust evaluation.
%
Each subgroup is partitioned into 20 held-in and 5 held-out patients. Hold-in patients contribute to training, validation, and in-distribution testing.
%
Specifically, we extract 20 qualified days per held-in patient and assign 15 days to training set, the next 2 to validation set, and final 3 days to the in-distribution (\texttt{test-id}) set.
%
Held-out patients are entirely excluded from model development and each provides 12 qualified days for the out-of-distribution (\texttt{test-od}) set.

\subsection{Input-Output Construction}

%
To ensure data quality and consistency across subgroups, we define a qualified patient-day as one with a complete 24-hour CGM trace (288 values) and adjacent continuity: the 24 hours prior and 8 hours following must also be complete. Each such day yields 24 hourly prediction input-output pairs, each comprising a 24-hour input and an 8-hour output segment.
%
Each input-output pair must meet strict quality criteria: the input contains all 288 values, the output all 96; fewer than 40\% of readings are constant (to exclude flatline readings); no values fall below 20 mg/dL; and fewer than 20\% exceed 400 mg/dL. If any pair within a patient-day fails these checks, the entire day is discarded. This rigorous filtering ensures that every retained patient-day yields 24 high-quality input–output pairs, providing a reliable foundation for both classical and neural forecasting models.

%
From each qualified day, we can extract 24 hourly prediction input-output pairs.
%
Each pair consists of a 288-length input sequence (representing the prior 24 hours) and a 96-length output sequence (representing the subsequent 8 hours).
%
The full output window supports evaluation up to 8 hours. The main text focuses on the first 2 hours of this window (24 steps), while results for 30-minute and 1-hour horizons are presented in Supplementary Note B.
%
We also append static covariates (age group, gender, diabetes type) and dynamic indicators (time-of-day) to each instance.
%
Table~\ref{tab:split} shows statistics: there are non-overlapping sets of \texttt{train}, \texttt{valid}, \texttt{test-id}, and \texttt{test-od}, supporting evaluation across 12 balanced subgroups, with total 300 patients, and 5520 patient-days, and 132,480 input-output pairs.
%
By enforcing uniform data quality and matched sample sizes, the partitioning scheme enables both robust generalization and fairness assessment.
%
Detailed data descriptions, analysis, and accessibility information are provided in Supplementary Note A.


\subsection{Event Characterization}
%
FairGlucose includes timestamped behavioral event annotations logged via the mobile health platform: (1) meal events with nutritional details (carbs, calories, protein, fat, fiber), (2) exercise events with duration and intensity, and (3) medication events with dosage and timing.
%
Of the 300 patients, 81 (27.0\%) logged at least one event, producing 3{,}945 unique behavioral events: medication events (2{,}809; 71.2\%), meal events (648; 16.4\%), and exercise events (488; 12.4\%). Logging patients recorded a median of 38 events (mean 48.7, range 1--237) over their observation period.
%
Because the hourly sliding-window design produces overlapping 48-hour observation windows, each unique event appears in approximately 30 prediction pairs; consequently 22{,}239 of 132{,}480 samples (16.8\%) contain at least one event within their window, yielding 117{,}979 event occurrences at the window level.
%
Event timing is highly asymmetric: 33.8\% of event occurrences fall in the 24-hour input window, 3.0\% in the evaluated 2-hour forecast window, and 63.2\% in the 2--24 hours after the prediction time. Because after-observation events are not necessarily known at deployment time, event-aware experiments quantify retrospective event-aware performance or an upper-bound estimate unless restricted to events available at prediction time.
%
Demographic variations reveal age-related differences (patients 65+ log 22\% fewer events than younger groups) and disease-type differences (T2D patients log 12\% more events than T1D).
%
This distinguishes FairGlucose from existing CGM benchmarks (which provide only glucose traces) and enables event-aware model development and subgroup-specific event impact analysis.

\subsection{Predictive Accuracy}

%
\textbf{Root Mean Square Error (rMSE)}. rMSE measures the average magnitude of the prediction error across time steps and is sensitive to large deviations:
\begin{equation}
\text{rMSE} = \sqrt{\frac{1}{T} \sum_{t=1}^{T} (\hat{y}_t - y_t)^2}
\end{equation}
%
Despite the right-skewed nature of rMSE, we report the mean to capture difficult cases and summarize overall model robustness \cite{armandpour2021deep}.
%
Median values and other metrics are provided in Supplementary Note B for a fuller assessment.

%
\textbf{Clarke Error Grid Analysis (Clarke A)}: To assess clinical safety, we use the Clarke Error Grid \cite{clarke1987evaluating}, which categorizes predictions into five zones.
%
Zone A represents the most clinically accurate predictions (within 20\% or both $<$70 mg/dL) with no clinical errors, while Zone B captures benign errors with no treatment changes.
%
Zones C, D, and E represent progressively dangerous errors: overcorrection (C), failure to detect hypo/hyperglycemia (D), and confusion between hypoglycemia and hyperglycemia (E).
%
We report the \textbf{Clarke A rate}, representing predictions with perfect clinical accuracy.
%
Critically, Zones D and E carry asymmetric clinical costs—hypoglycemia misses can lead to seizures or loss of consciousness within minutes, while hyperglycemia overcorrection may cause acute hypoglycemic events. This strict metric ensures models meet the highest clinical standards for glucose prediction validation \cite{battelino2019clinical, clarke1987evaluating}.

\subsection{Statistical Uncertainty}

Because overlapping sliding windows produce multiple prediction pairs per patient, sample-level confidence intervals underestimate uncertainty. We therefore compute \textbf{patient-level bootstrap} confidence intervals: in each of $B = 1{,}000$ resamples, we draw patients with replacement, aggregate each patient's per-sample rMSE to a patient-level mean, and compute the group statistic of interest (e.g., subgroup mean rMSE or between-group difference). The 95\% CI is taken as the 2.5th and 97.5th percentiles of the bootstrap distribution.
%
For subgroup comparisons (e.g., T1D vs.\ T2D, event vs.\ no-event), we additionally report two-sided \textbf{permutation tests} ($N = 10{,}000$ permutations) on the patient-level mean rMSE difference. All reported $p$-values and confidence intervals in figures and tables use these patient-level procedures unless noted otherwise.


\subsection{Fairness Evaluation Frameworks}

%
To evaluate equity and robustness in glucose prediction models, we propose a fairness evaluation framework spanning five dimensions.

%
\noindent\textbf{Subgroup Fairness.}~~Subgroup fairness examines whether predictive accuracy is consistent across patient demographics and clinical characteristics \cite{mehrabi2021survey}. We partition the population into \( K \) disjoint subgroups (e.g., by age, sex, or diabetes type) and compute the average root mean squared error ($\text{rMSE}_i$) within each subgroup $i$.
%
We quantify fairness disparities using two measures. First, the \textbf{rMSE Spread} captures the absolute difference between the best and worst performing subgroups.
%
It is defined as
\begin{equation}
\text{rMSE-Spread} = \max_i(\text{rMSE}_i) - \min_i(\text{rMSE}_i)
\end{equation}
%
The second is the \textbf{Gini Index} \cite{dixon2018measuring} over the set \(\{\text{rMSE}_1, \dots, \text{rMSE}_K\}\), which measures the distributional inequality in performance across subgroups.
%
A higher Gini value indicates greater inequality in prediction error across subgroups, while a value closer to 0 indicates more equitable performance. Gini results are reported in Supplementary Note B.

%
\noindent\textbf{Input Length Sensitivity.}~~Input length sensitivity assesses whether model performance improves as the input window length increases, and whether such gains are consistent across subgroups.
%
Let $L \in \{7, 37, 73, 145, 288\}$ denote the number of CGM input steps. For each length $L$ and subgroup $i$, we define the \textbf{temporal ratio} (TR) as:
\begin{equation}
\text{TR}_i^{(L)} = \text{rMSE}_i^{(L)} \;/\; \text{rMSE}_i^{(288)}
\end{equation}
where $\text{rMSE}_i^{(288)}$ is the performance under the longest (24-hour) input. A TR near 1.0 indicates consistent performance across input lengths; higher values indicate degradation with shorter input.
%
Substantial variation in TR across subgroups at the same input length signals inequitable sensitivity to data availability; some subgroups may lose disproportionately more accuracy when historical data is limited.

%
\noindent\textbf{Instance-Level Difficulty.}~~Instance-level difficulty investigates model robustness across prediction samples of varying difficulty. We define hard and easy instances using a vote-based approach across all $M$ evaluated models \cite{swayamdipta2020dataset}. For each model, we rank all samples by per-sample rMSE; samples in the top 25\% (highest error) receive a hard vote, and samples in the bottom 25\% (lowest error) receive an easy vote. We then compute vote counts $\text{Vote}_j^{\text{hard}}$ and $\text{Vote}_j^{\text{easy}}$ for each sample $j$. Samples with $\text{Vote}_j^{\text{hard}}$ in the top 25th percentile of all vote counts are labeled \emph{hard}; those with $\text{Vote}_j^{\text{easy}}$ in the top 25th percentile are labeled \emph{easy}; the remainder are \emph{neutral}.
%
For each model, we compute rMSE over the hard and easy subsets separately. To assess robustness, we introduce normalized difficulty ratios using the Naive model as reference:
\begin{equation}\text{Hard Ratio} = \frac{\text{rMSE}^{\text{hard}}_{\text{model}}}{\text{rMSE}^{\text{hard}}_{\text{naive}}}, \quad
\text{Easy Ratio} = \frac{\text{rMSE}^{\text{easy}}_{\text{model}}}{\text{rMSE}^{\text{easy}}_{\text{naive}}}
\end{equation}
These ratios control for the inherent difficulty of the samples and emphasize how much a model improves upon a naive baseline in different difficulty regimes. A fair and robust model should yield reductions in both ratios, narrowing the performance gap between easy and hard cases.

%
\noindent\textbf{Event-Aware Evaluation.}~~We compare models trained with event features (meals, exercise, medication) versus without these features. Event features are evaluated under the availability caveat described above, because some events occur after prediction time.
%
Computing $\text{rMSE}^{\text{Event}}$ and $\text{rMSE}^{\text{NoEvent}}$ on the same test set, negative differences ($\Delta < 0$) indicate event features improve accuracy.
%
We evaluate across demographic subgroups to assess whether benefits are consistent and identify event-fairness disparities.

\subsection{Models}

%
To benchmark performance, we compare 33 models spanning four families. For \textbf{statistical baselines} (3 models), the Naive method predicts the most recent value, ARIMA \cite{yang2018arima} captures temporal dependencies, and AutoETS fits exponential smoothing; all are implemented via Nixtla StatsForecast. For \textbf{machine learning} (4 models), we evaluate Linear Regression and tree-based models: LightGBM \cite{ke2017lightgbm}, CatBoost \cite{prokhorenkova2018catboost}, and XGBoost \cite{chen2016xgboost}, implemented via Nixtla MLForecast.
%
For \textbf{neural time-series models} (20 architectures), we use the Time-Series-Library (TSLib) framework, which provides a unified training pipeline across diverse architectures. These include attention-based models: PatchTST \cite{nie2022time}, iTransformer, TimeXer, Crossformer, Transformer \cite{vaswani2017attention}, Informer \cite{zhou2021informer}, Non-stationary Transformer, Autoformer \cite{wu2021autoformer}, FEDformer, and TFT \cite{lim2021temporal}; MLP/linear models: DLinear \cite{zeng2023transformers}, TiDE, TSMixer, TimeMixer, N-BEATS, N-HiTS \cite{challu2023nhits}, and MICN; convolutional models: TimesNet and SCINet; and the zero-shot foundation model Chronos2.
%
For \textbf{frontier LLMs and foundation models} (6 models), we include TimeGPT \cite{garza2023timegpt}, a time-series-specific foundation model; GPT-5.1 \cite{openai_gpt51_model_docs} and GPT-5 Mini \cite{openai_gpt5mini_model_docs}; Claude 4.5 Sonnet and Claude 4.5 Haiku \cite{anthropic_claude45_whats_new}; and Gemini 3 Flash \cite{google_gemini3_dev_guide}. These are evaluated via API-based inference.

\subsection{Experiment Details}
%
To ensure reproducibility, we trained and tested each model on predefined splits (train, validation, test-id, and test-od) with consistent input and output lengths.
%
We focus on out-of-the-box performance to establish strong and fair baselines, deliberately avoiding model-specific enhancements such as pretraining, auxiliary losses, data augmentation, knowledge distillation, or learning rate schedules.
%
Statistical and ML models use the Nixtla library \cite{olivares2022library_neuralforecast} with Bayesian hyperparameter optimization via Optuna \cite{akiba2019optuna} (50 trials per model). Neural time-series models use the TSLib framework with default hyperparameters and up to 20-epoch training with early stopping (patience~5); see Supplementary Note~C for search spaces.
%
For LLMs, we set the temperature to zero for deterministic outputs with the same prompts.
%
Experiment details are reported in Supplementary Note C.
%
Our evaluation prioritizes general-purpose time-series forecasters over glucose-specific architectures \cite{sergazinov2023gluformer, zhu2022personalized} to assess out-of-the-box performance and facilitate broader adoption.
%
Our framework establishes fairness-aware baselines that future work can extend with domain-specialized methods under the FairGlucose protocol.

\subsection{Ethics approval and consent to participate}
\input{0-sections/93_ethics}

%% file: 0-sections/93_ethics.tex
This research was performed in accordance with the Declaration of Helsinki. The secondary
analysis of de-identified Welldoc patient data was reviewed and approved by the Johns
Hopkins University Institutional Review Board (approval reference IRB00447704). No other
ethics committee or institutional review board reviewed this work. Informed consent for
the collection and research use of the underlying data was obtained from participants at
the time of original data collection under Welldoc's standard consent procedures. The
Institutional Review Board waived the requirement for additional informed consent for the
present secondary analysis because all records were fully de-identified before they were
made available to the research team, so that participants could not be identified directly
or through linked identifiers.

%% file: 0-sections/94_data_availability.tex
The FairGlucose cohort metadata, demographic strata definitions, evaluation splits, and the 33-model reference leaderboard will be made available at \url{https://github.com/JHU-CDHAI/FairGlucose} upon acceptance for publication. The underlying CGM and behavioral event records from the Welldoc cohort are not publicly available due to patient privacy; access can be requested under a data use agreement with Welldoc, Inc.

%% file: 0-sections/95_code_availability.tex
All code for cohort construction, baseline training, evaluation, and figure generation will be made available at \url{https://github.com/JHU-CDHAI/FairGlucose} upon acceptance for publication. Pretrained model checkpoints used for the 33 baselines will be available on request from the corresponding author.

%% file: 0-sections/90_acknowledgements.tex
The authors thank the Welldoc team for data access and clinical guidance. No specific
funding was received for this work. Welldoc, Inc.\ provided access to the de-identified
cohort and computing infrastructure but had no role in the study design, the analysis and
interpretation of the results, the decision to publish, or the preparation of the
manuscript.

%% file: 0-sections/96_author_contributions.tex
J.L.\ and X.Z.\ led the cohort construction, data curation, software development, formal analysis, investigation, and visualization, and prepared the original draft of the manuscript. R.H.\ contributed to baseline-model implementation, data infrastructure, investigation, and manuscript review. A.K., A.K.I., and M.E.S.\ provided the Welldoc cohort and infrastructure access, contributed clinical and product expertise (validation), and reviewed and edited the manuscript. R.A.\ and G.G.G.\ contributed to conceptualization, supervision, and project administration, and reviewed and edited the manuscript. All authors reviewed and approved the final version of the manuscript.

%% file: 0-sections/92_competing_interests.tex
R.H., A.K., A.K.I., and M.E.S. are employees of Welldoc, Inc.\ and may hold equity in the company. The remaining authors declare no competing interests.

%% file: 0-sections/A_dataset_statistics.tex
\section{Dataset Statistics and Details}

\subsection{FairGlucose Dataset Statistics}

For each subgroup in the dataset (e.g., by age, gender, diabetes type), we computed key signal metrics such as mean, median, standard deviation, minimum, maximum, coefficient of variation (CV), interquartile range (IQR), and selected percentiles (5th, 25th, 75th, 95th). Additionally, we calculated clinically relevant measures including TIR\% (Time in Range), MAGE (Mean Amplitude of Glycemic Excursions), and dynamic features such as the average rate of change, number of peaks/valleys, average trend length, and entropy of the signal distribution. The implementation used \texttt{scipy.stats.entropy} for entropy computation and \texttt{scipy.signal.find\_peaks} to detect peaks and valleys \cite{scikit-learn}. The results were aggregated to provide an overview of the variability and distribution of glucose values across subsets.

For all calculations, each metric was first computed at the individual sequence level using a window of \textbf{288 CGM} (corresponding to a full-day CGM trace). We then aggregated these results by taking the mean across all sequences within the subgroup to obtain the representative metric value for that subset. Formally, the metrics are defined as follows:

\begin{itemize}
    \item \textbf{Interquartile Range (IQR)}:
    \begin{equation}    \text{IQR} = Q_{75} - Q_{25}
    \end{equation}
    where \(Q_p\) denotes the \(p\)-th percentile.
    
    \item \textbf{Selected Percentiles}: 
    Values at 5th, 25th, 75th, and 95th percentiles of the sequence.
    
    \item \textbf{Time in Range (TIR\%)}:
    \begin{equation}    \text{TIR\%} = \frac{\#\{x_i \mid L \le x_i \le U\}}{n} \times 100
    \end{equation}
    where \(L\) and \(U\) are clinically defined lower and upper bounds (typically 70 and 180 mg/dL).
    
    \item \textbf{Mean Amplitude of Glycemic Excursions (MAGE)}:
    The average of all significant excursions exceeding one standard deviation from the mean:
    \begin{equation}    \text{MAGE} = \frac{\sum_{k=1}^{m} |e_k|}{m}, \quad e_k > \sigma
    \end{equation}
    where \(e_k\) are individual excursions and \(m\) is the number of qualifying excursions.
    
    \item \textbf{Average Rate of Change}:
    \begin{equation}    \text{Rate of Change} = \frac{1}{n-1} \sum_{i=1}^{n-1} \frac{|x_{i+1} - x_i|}{\Delta t}
    \end{equation}
    where \(\Delta t\) is the time interval between measurements.
    
    \item \textbf{Number of Peaks / Valleys}:
    The count of local maxima and minima in the sequence, detected using \texttt{scipy.signal.find\_peaks}.
    
    \item \textbf{Average Trend Length}:
    The mean duration of monotonic segments between peaks and valleys.
    
    \item \textbf{Entropy}:
    Shannon entropy of the empirical distribution:
    \begin{equation}    H = - \sum_{j=1}^{k} p_j \log(p_j)
    \end{equation}
    where \(p_j\) is the probability of values falling in bin \(j\), computed using \texttt{scipy.stats.entropy}.
\end{itemize}

\subsection{Cohort Representation in Feature Space}

\begin{figure}[t]
    \centering
    \begin{subfigure}[t]{0.65\linewidth}
        \centering
        \includegraphics[width=\linewidth]{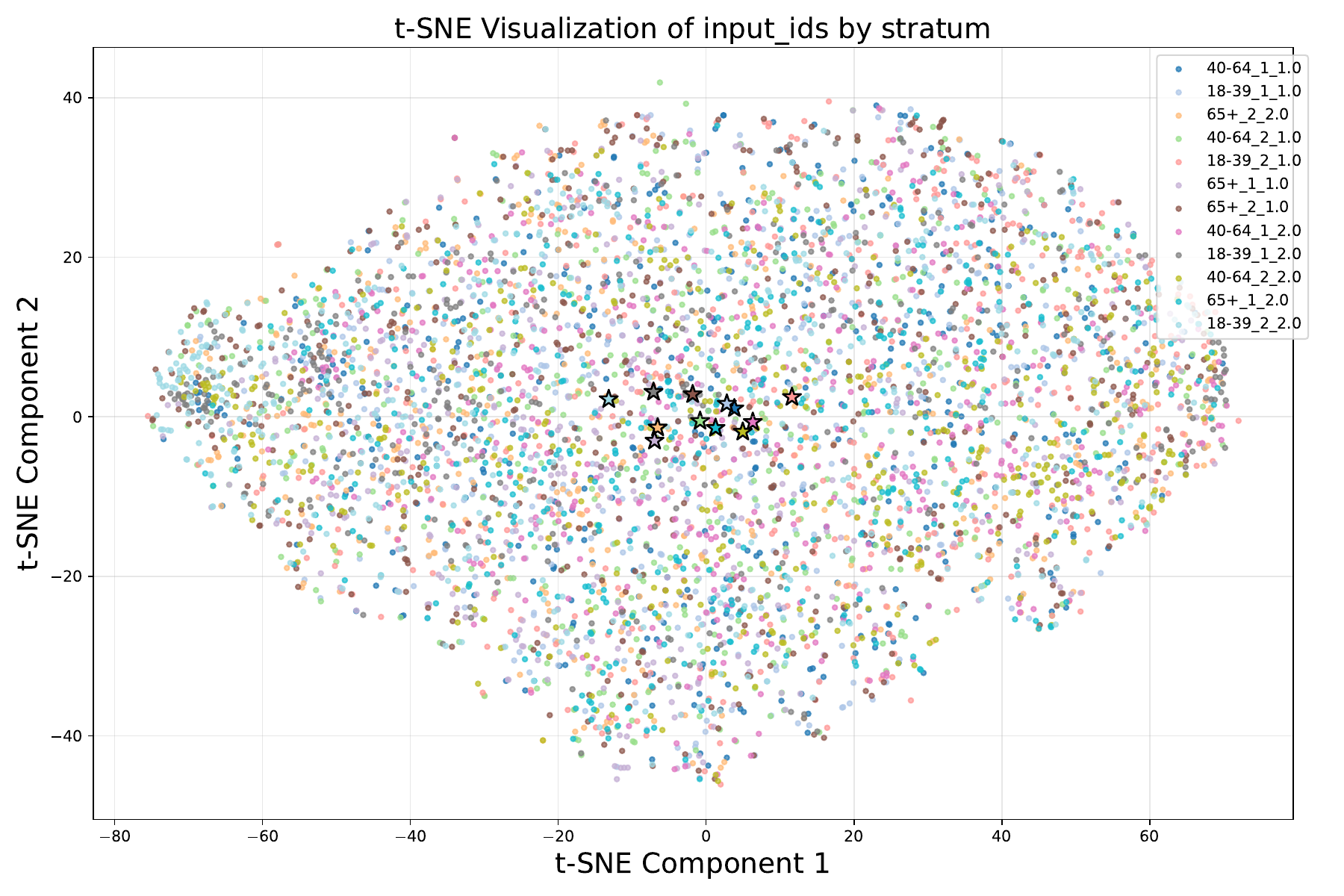}
        \caption{By \textbf{stratum} (age $\times$ gender $\times$ type)}
    \end{subfigure}

    \vspace{0.5em}

    \begin{subfigure}[t]{0.32\linewidth}
        \centering
        \includegraphics[width=\linewidth]{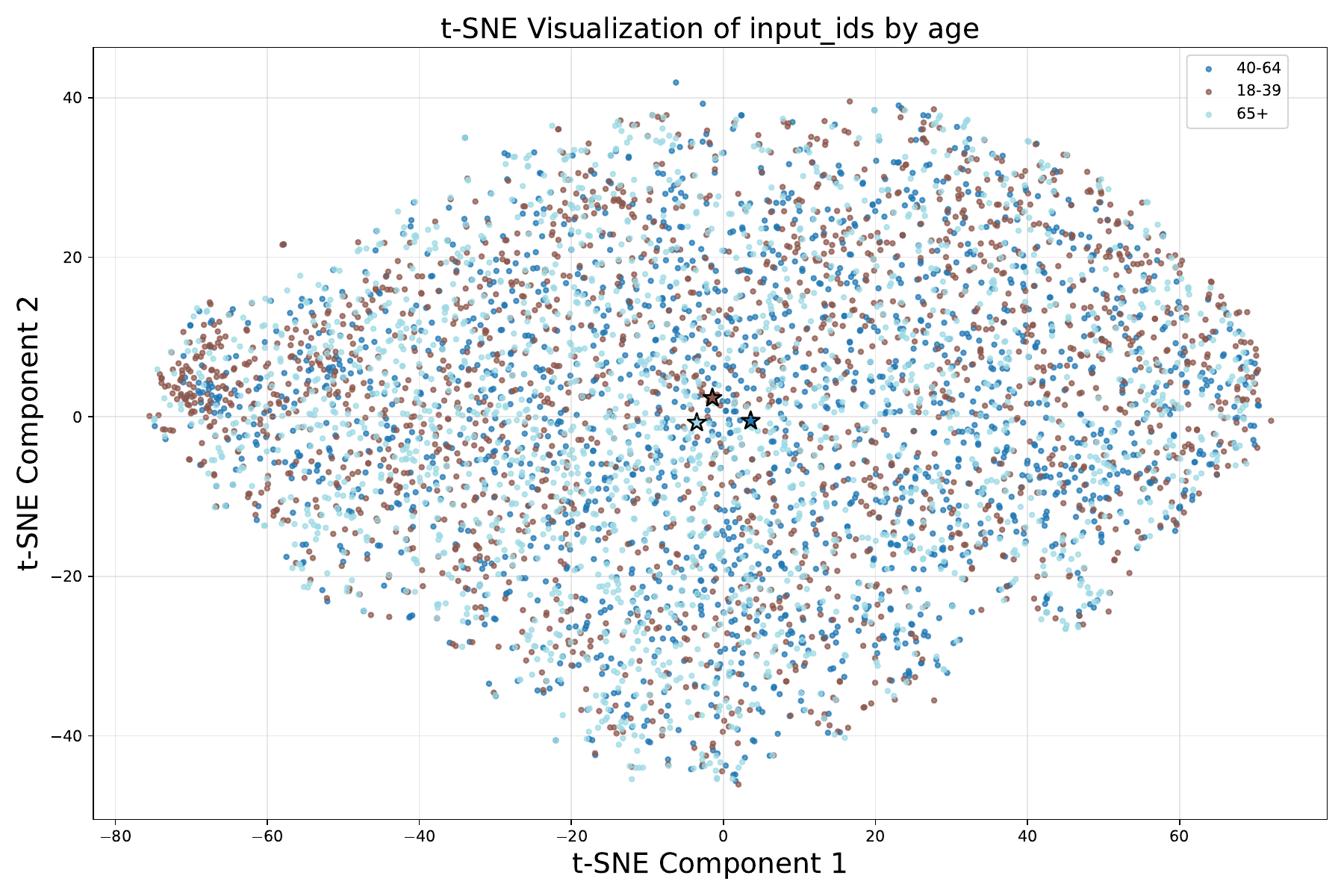}
        \caption{By \textbf{age}}
    \end{subfigure}
    \hfill
    \begin{subfigure}[t]{0.32\linewidth}
        \centering
        \includegraphics[width=\linewidth]{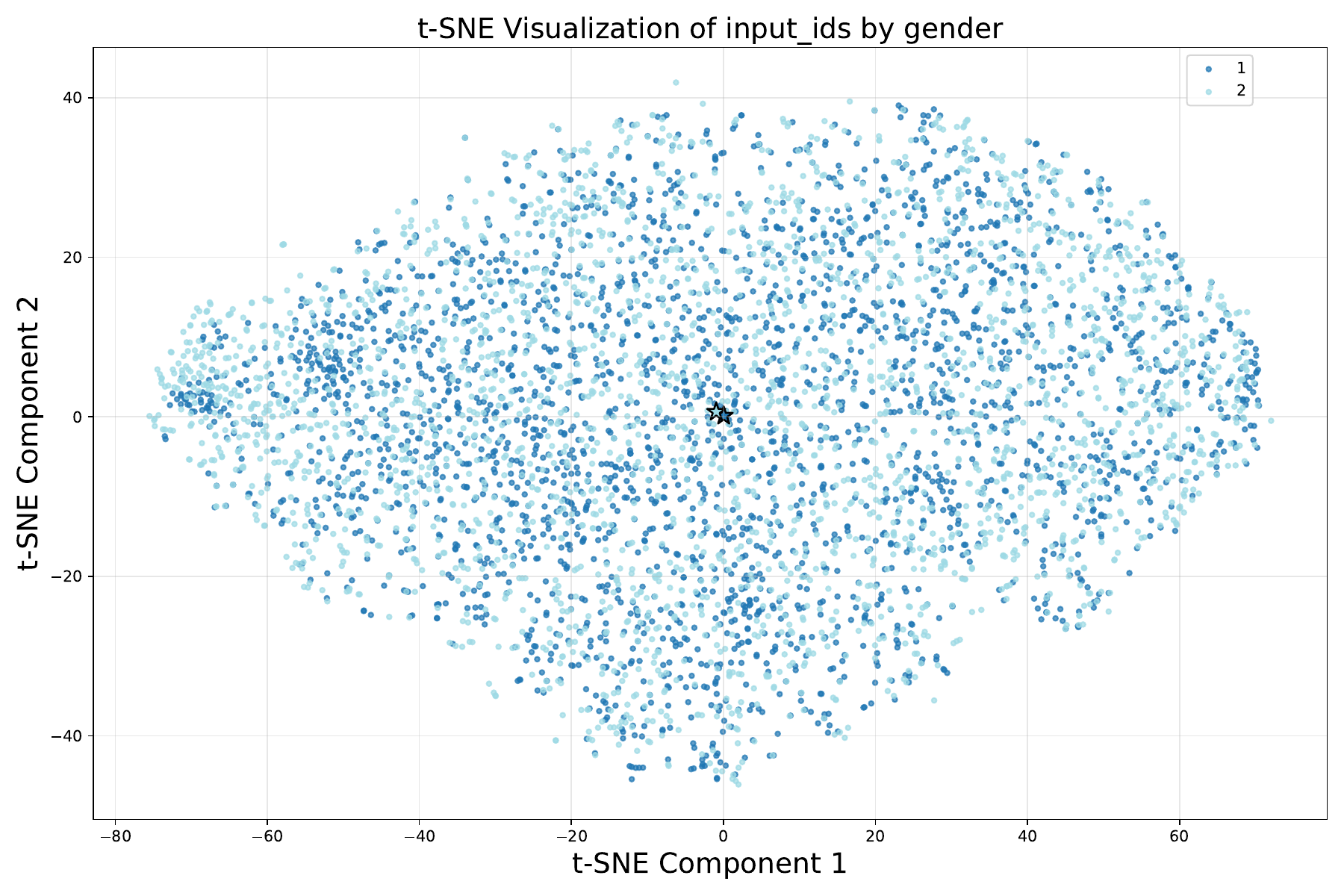}
        \caption{By \textbf{gender}}
    \end{subfigure}
    \hfill
    \begin{subfigure}[t]{0.32\linewidth}
        \centering
        \includegraphics[width=\linewidth]{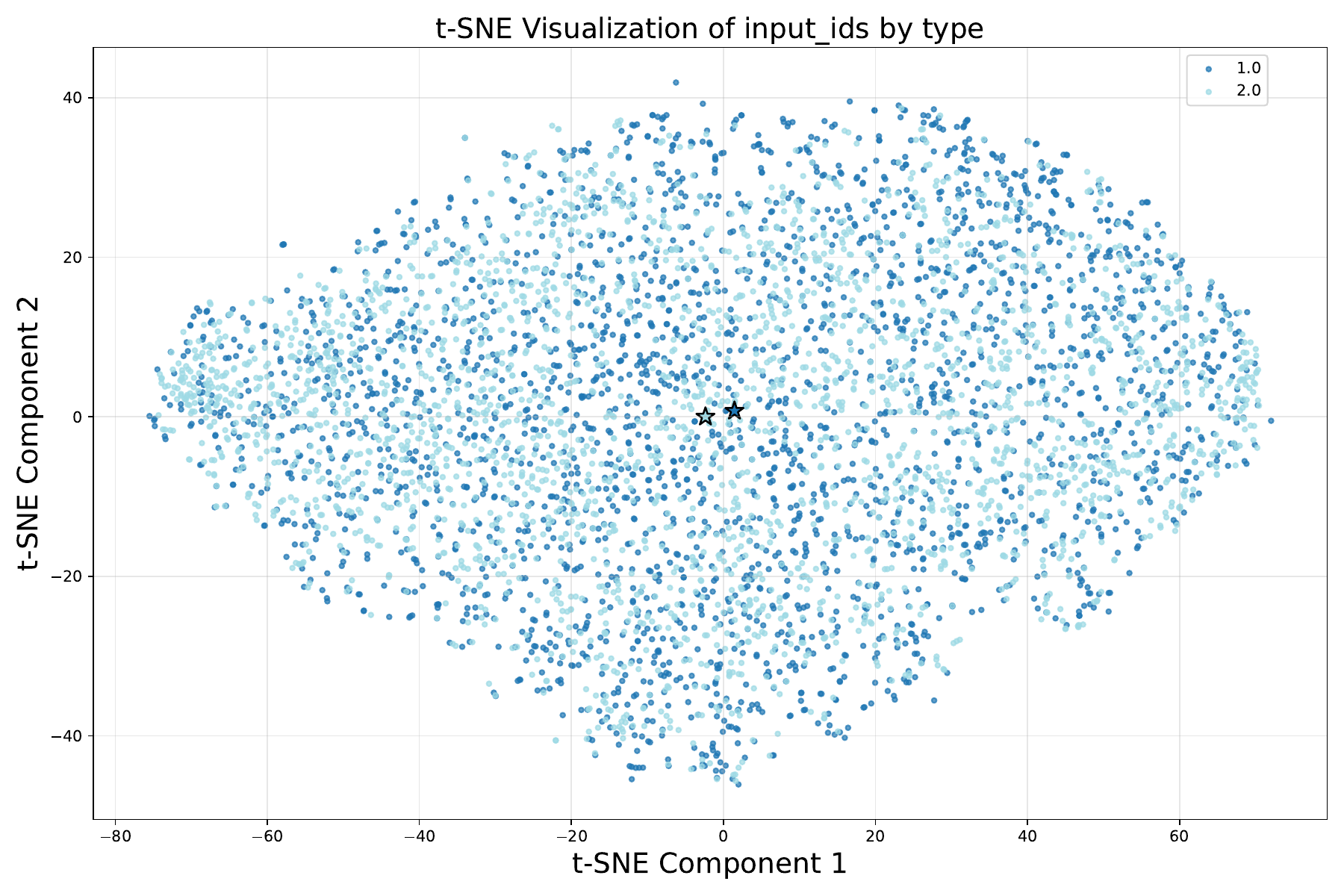}
        \caption{By \textbf{diabetes type}}
    \end{subfigure}

    \caption{t-SNE embeddings of 24-hour CGM patient-day traces (288 5-minute steps), colored by (top) composite stratum and (bottom) univariate attributes. Stratum-coloring produces the most separable clusters; univariate colorings overlap substantially. Star markers denote subgroup centroids.}
    \label{fig:tsne-visualization-grid}
\end{figure}

To verify that the 12 demographic strata constitute distinguishable subpopulations in the input space (rather than balanced groups that nevertheless overlap in CGM dynamics), we visualized 2-D t-SNE embeddings of 288-step 24-hour patient-day CGM traces under five subgroup-coloring schemes (Figure~\ref{fig:tsne-visualization-grid}). The composite \emph{stratum}-coloring (age $\times$ gender $\times$ diabetes type) produces the most pronounced cluster separation; univariate colorings by age, gender, or diabetes type alone show substantially overlapping clusters. This confirms that the cohort's multi-dimensional stratification captures structure that univariate stratification would miss, and supports the interpretation that downstream model-class-invariant disparities reflect genuine inter-subgroup differences in the prediction task rather than nominal labeling.

\subsection{Visualizing Feature Correlations and Subgroup Differences in CGM Data}

To investigate the distributional characteristics of CGM sequence features across different patient subgroups, we computed a set of descriptive statistics for each 24-hour sequence. These include: mean glucose, standard deviation (std), slope (reflecting directional change), number of peaks (n\_peaks), and clinically meaningful measures such as Time-In-Range (TIR), Time-Below-Range (TBR), and Time-Above-Range (TAR). These features were extracted using a custom feature engineering function (\texttt{advanced\_sequence\_features}) applied uniformly across all sequences.

We visualized the relationships among these features using \texttt{seaborn.pairplot} with \texttt{corner=True}, stratified by various patient attributes such as stratum, age group, gender, diabetes type, and dataset split. Diagonal plots show kernel density estimates (KDEs) for feature distributions, while off-diagonal plots display pairwise scatter plots, allowing for the examination of feature correlations and cluster separability across groups.

Our exploratory analysis reveals that certain features are strongly correlated. Notably, mean glucose exhibits a strong positive correlation with TAR and an inverse relationship with TIR, reflecting the direct impact of average glucose levels on clinical range metrics. The number of peaks and slope—features that capture temporal dynamics—exhibit weaker correlations with mean or range-based metrics but still capture unique behavioral signatures of glucose traces.

Importantly, we observe substantial differences in the distributional shapes and feature interactions across strata. For example, younger age groups (18–39) tend to show wider dispersion in both mean glucose and TIR, indicating greater physiological variability or lifestyle-driven fluctuations. In contrast, some strata, such as older patients with T2D, demonstrate narrower distributions, possibly reflecting more stable or medically regulated glucose profiles. These cross-stratum differences highlight the importance of subgroup-sensitive modeling.

From a modeling perspective, these patterns have direct implications. High intra-stratum variability can challenge prediction consistency, while inter-stratum distributional shifts may introduce bias if not properly addressed. Such differences motivate the design of fairness-aware models that explicitly account for heterogeneous feature landscapes.
Moreover, the scatter plots also reveal outlier patterns and nonlinear dependencies not easily captured by global metrics. For instance, in some strata, the relationship between TIR and n\_peaks is non-monotonic, suggesting that overly simplistic feature assumptions may fail in real-world deployment.

\subsection{Resource Availability and Use}

FairGlucose is provided through two complementary access mechanisms designed to balance research utility with patient-data confidentiality.

\textbf{Method 1: Controlled access via Data Use Agreement (DUA).}~Researchers may request access to the de-identified CGM cohort by submitting a written research proposal, institutional affiliation, planned analyses, and IRB approval or exemption. Approved requests proceed to a formal DUA between the requesting institution and Welldoc, Inc.\ that governs scope of use, data security, publication policy, and redistribution restrictions. Access is provisioned within a secure computational environment that enables interactive analysis while preventing unauthorized export of patient-level data.

\textbf{Method 2: API-based evaluation server.}~For investigators who require benchmarking without raw-data access, we provide an API endpoint that accepts trained forecasting models and returns standardized metrics (rMSE, MAE, Clarke Error Grid zone counts, and subgroup-disaggregated variants) on the held-out test set. Each submitted model runs in a sandboxed environment and is never exposed to user-controlled execution paths; results are returned in a uniform machine-readable report. This model-to-data pattern preserves data privacy while enabling reproducible third-party comparison against the reference 33-model leaderboard.

Cohort metadata, demographic strata definitions, evaluation splits, and the 33-model reference leaderboard will be made openly available at \url{https://github.com/JHU-CDHAI/FairGlucose} upon acceptance. Both access mechanisms ensure that FairGlucose can support the broadest possible community of investigators while protecting the patients whose data made the benchmark possible.

%% file: 0-sections/B_model_evaluation.tex
\section{Model Evaluation Details}

\subsection{Expanded Model Evaluation Across Forecast Horizons}

This section presents accuracy comparisons across different forecast horizons (30 minutes, 1 hour, and 2 hours) using three metrics: MAE, rMSE, and TIR Absolute Error \cite{armandpour2021deep, battelino2019clinical}.

To complement the main results, we present detailed performance metrics for all 33 models across three prediction horizons, 30 minutes, 1 hour, and 2 hours, evaluated on both in-distribution (Test-ID) and out-of-distribution (Test-OD) sets. Each horizon-specific evaluation reports root mean squared error (rMSE), mean absolute error (MAE), and Time-in-Range Absolute Error (TIR-AE), capturing both numerical and clinically relevant dimensions of model performance. The corresponding results are reported in the tables below.

At the 30-minute horizon, attention-based neural models consistently achieve the strongest performance across all three metrics and test splits. NS-Transformer and PatchTST \cite{nie2022time} rank among the top performers, with rMSE and MAE among the lowest of all models. N-HiTS performs similarly well, especially in the MAE metric.
TFT \cite{lim2021temporal} also delivers competitive results at this horizon, though with slightly higher TIR error. Among classical models,
ARIMA \cite{yang2018arima} shows surprisingly strong performance, often outperforming machine learning methods like XGBoost \cite{chen2016xgboost} and CatBoost \cite{prokhorenkova2018catboost} in TIR error, particularly on the Test-OD set.
LightGBM \cite{ke2017lightgbm} and CatBoost, while strong in rMSE and MAE, show more variability in TIR prediction across subgroups. Foundation models, including GPT-5.1 \cite{openai_gpt51_model_docs}, Gemini 3 Flash \cite{google_gemini3_dev_guide}, and TimeGPT \cite{garza2023timegpt}, consistently underperform across all 30-minute evaluations. Their rMSE and TIR error are notably higher than neural and classical models, often close to or worse than the naive baseline. These results indicate that large-scale pretraining alone is insufficient for accurate short-term physiological forecasting.

Moving to the 1-hour horizon, the performance gap between top and bottom models becomes more pronounced. Attention-based architectures (NS-Transformer, PatchTST, TimeXer) maintain their lead in both error-based and TIR-based metrics. N-HiTS and TFT continue to perform competitively, though their relative rankings fluctuate slightly depending on the metric and test split. As the prediction window increases, statistical and machine learning models begin to fall further behind. ARIMA's relative strength diminishes, particularly on the Test-OD set, where its TIR error increases more sharply than the neural baselines. The degradation in performance is more severe for foundation models, which show clear limitations in both MAE and TIR prediction. These results highlight the challenges of mid-horizon glucose prediction and reaffirm the superiority of specialized neural time-series models for this task.

At the 2-hour horizon, performance deteriorates further across all models, as expected given the increased forecasting difficulty. Despite this, NS-Transformer maintains the best overall rMSE (25.6\,mg/dL), followed closely by Transformer (25.7), TimeXer and PatchTST (25.8). TFT remains strong but exhibits slightly higher TIR error, suggesting increased difficulty in capturing clinical relevance at extended horizons. In contrast, all classical models, including ARIMA, linear regression, and tree-based methods, exhibit larger declines in performance, particularly in TIR error, indicating limited capacity to generalize over long-term physiological dynamics. Foundation models continue to underperform substantially. Their rMSE and TIR metrics are consistently worse than even the simplest baselines, underscoring their inadequacy for long-horizon CGM prediction without targeted adaptation or fine-tuning. The widening performance gap reinforces the importance of task-specific architecture and domain-aware training for reliable long-range forecasting.

Across all horizons, comparison between Test-ID and Test-OD performance reveals remarkable stability. For nearly all models, the performance degradation under distribution shift is minimal, with OD/ID error ratios typically close to 1.0. This consistency suggests that our dataset construction and evaluation framework provide a robust test bed for assessing real-world generalization. Top-performing models, particularly NS-Transformer, PatchTST, and TFT, demonstrate strong resilience to distributional variation, maintaining their relative rankings across both splits. In contrast, models with high Test-ID errors tend to generalize poorly to the Test-OD set as well, reinforcing the observation that pretraining alone does not confer robustness in this domain.

Overall, the extended results affirm the main findings of the benchmark. Attention-based neural architectures (NS-Transformer, PatchTST, TimeXer) are the most accurate and robust models across horizons and evaluation settings. N-HiTS and TFT also perform strongly, especially under longer horizons and distribution shifts.
Classical models offer solid baselines, particularly at short horizons, while foundation models fail to generalize effectively without adaptation. These observations emphasize the value of rigorous benchmark design and highlight the importance of domain-specific modeling approaches for fair and reliable glucose prediction.

Expanded results from Figures~\ref{fig:accuracy_30m}, \ref{fig:accuracy_1h}, and \ref{fig:accuracy_2h} also present the similar pattern. Moreover, for each time horizon, the figures illustrate the performance variation among different data splits (Test-ID and Test-OD).
Shorter horizons (30 minutes) exhibit lower errors across all metrics, while errors increase as the prediction window extends to 1 hour and 2 hours, reflecting the expected decline in forecast accuracy with longer lead times.
These visualizations highlight consistent patterns across metrics.

\subsection{Fairness Evaluation: Subgroup Heterogeneity}

\begin{figure}[p]
    \centering
    \includegraphics[width=0.90\textwidth]{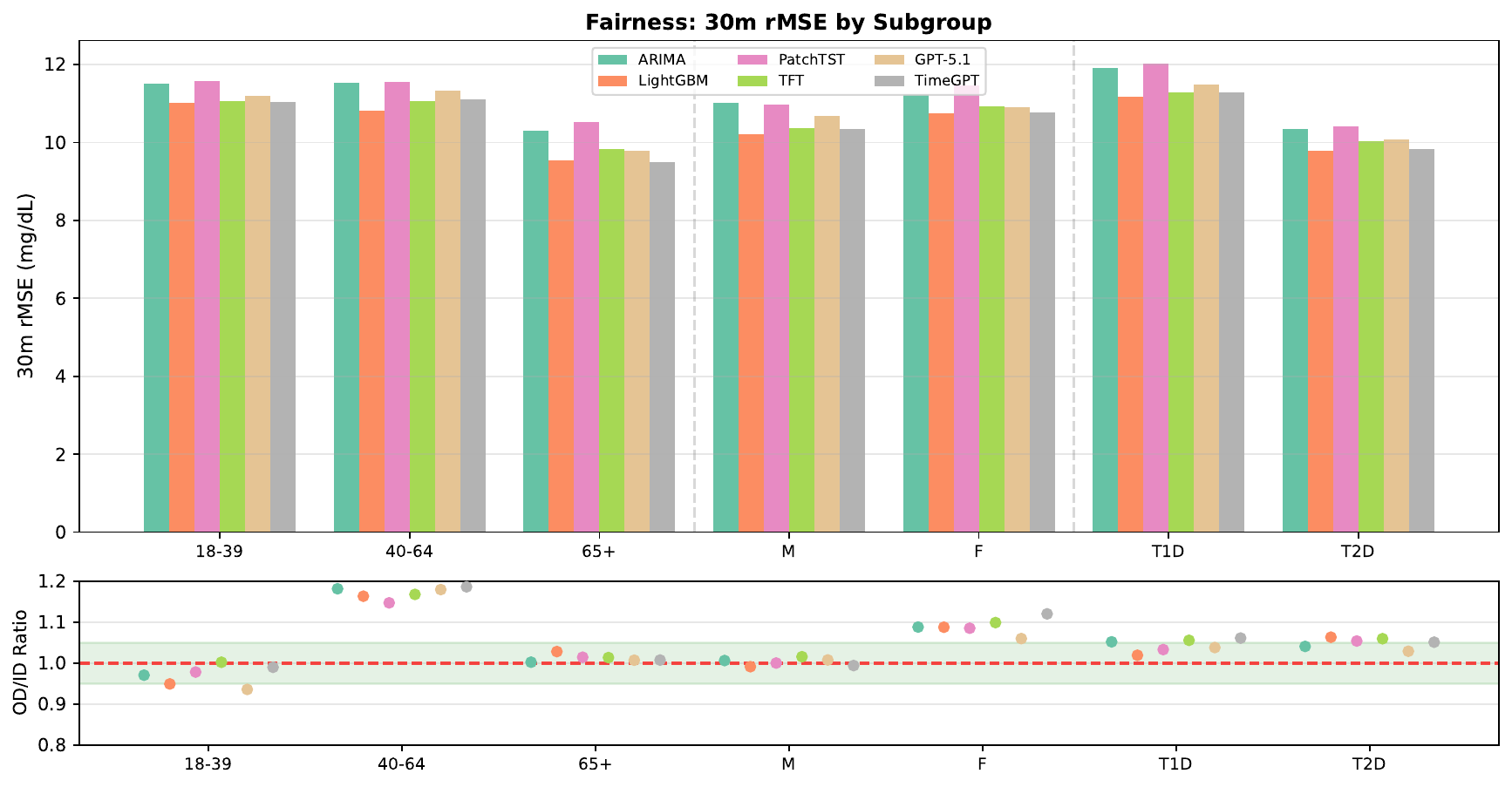}\\[-0.4em]
    {\small (a) 30-minute rMSE (mg/dL)}\\[0.4em]
    \includegraphics[width=0.90\textwidth]{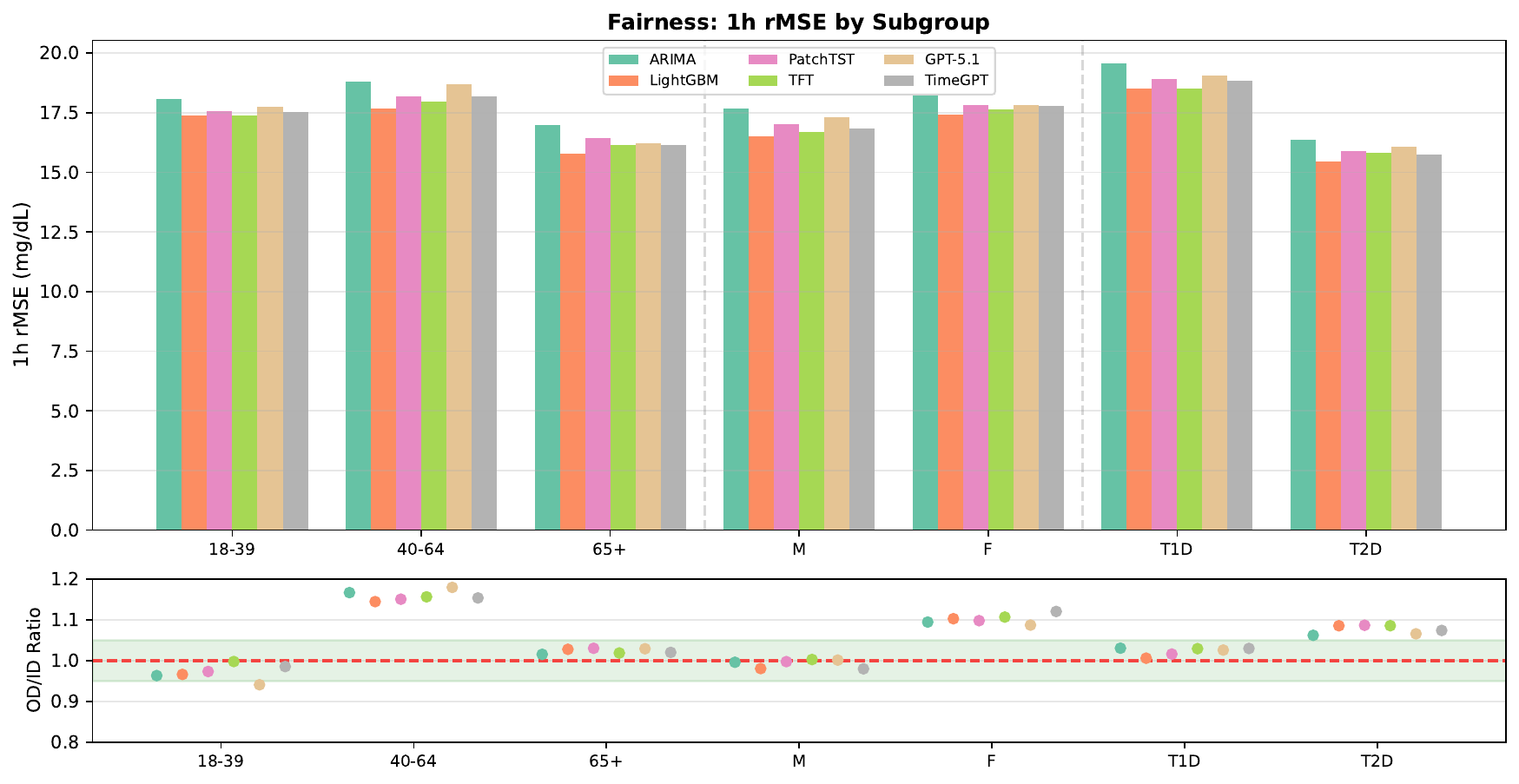}\\[-0.4em]
    {\small (b) 1-hour rMSE (mg/dL)}\\[0.4em]
    \includegraphics[width=0.90\textwidth]{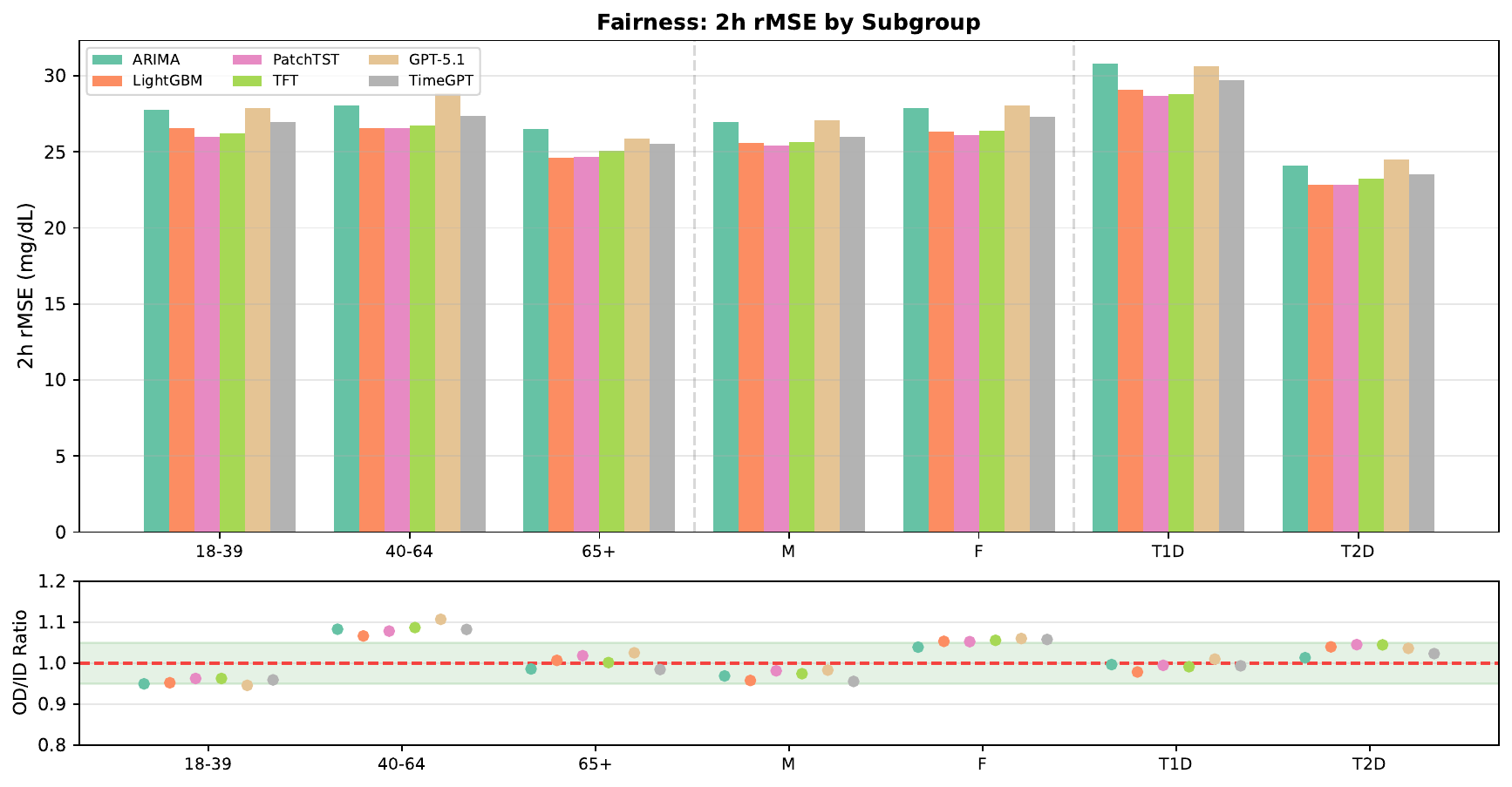}\\[-0.4em]
    {\small (c) 2-hour rMSE (mg/dL)}
    \caption{Fairness comparison across demographic subgroups (age, gender, diabetes type) for all prediction horizons. Top panel in each subfigure shows Test-ID performance; bottom panel shows OD/ID generalization ratio.}
    \label{fig:fairness_all_horizons}
\end{figure}

Figure~\ref{fig:fairness_all_horizons} presents a comprehensive comparison of model performance across demographic and clinical subgroups for all three prediction horizons (30 minutes, 1 hour, and 2 hours). Each subfigure displays rMSE performance across seven subgroups organized by dimension: three age groups (18--39, 40--64, 65+), two gender categories (male, female), and two diabetes types (T1D, T2D). Six representative models are shown spanning statistical (ARIMA), machine learning (LightGBM), neural time-series (PatchTST, TFT), and frontier LLM (GPT-5.1, TimeGPT) approaches.

The visualization employs a two-panel design to assess both accuracy and generalization. The top panel shows absolute Test-ID rMSE performance with consistent y-axis ranges across horizons (30m: 5–15 mg/dL, 1h: 10–25 mg/dL, 2h: 20–35 mg/dL), enabling direct cross-horizon comparison of error magnitudes. Grouped bar charts allow within-subgroup model comparison, with vertical separators delineating the three demographic dimensions. The bottom panel uses scatter plots to visualize OD/ID error ratios, where values near 1.0 indicate robust generalization and deviations reveal subgroup-specific distribution shift vulnerabilities. A green shaded band (0.95–1.05) highlights the region of near-perfect generalization.

Consistent with the patterns discussed in the main text, several intersectional disparities emerge. Male patients generally exhibit better accuracy than female patients in the 18--39 age group, while the reverse is observed in the 65+ group, indicating age- and gender-specific interactions in prediction difficulty. T1D patients consistently show higher rMSE than T2D patients across horizons, reaffirming that T1D populations are more challenging to forecast despite their structured insulin regimens. Neural time-series models (NS-Transformer, PatchTST, TFT) demonstrate more consistent performance across subgroups compared to frontier LLMs, which exhibit greater variability, particularly for T1D and younger age groups.

\begin{figure}[p]
    \centering
    \includegraphics[width=0.95\textwidth]{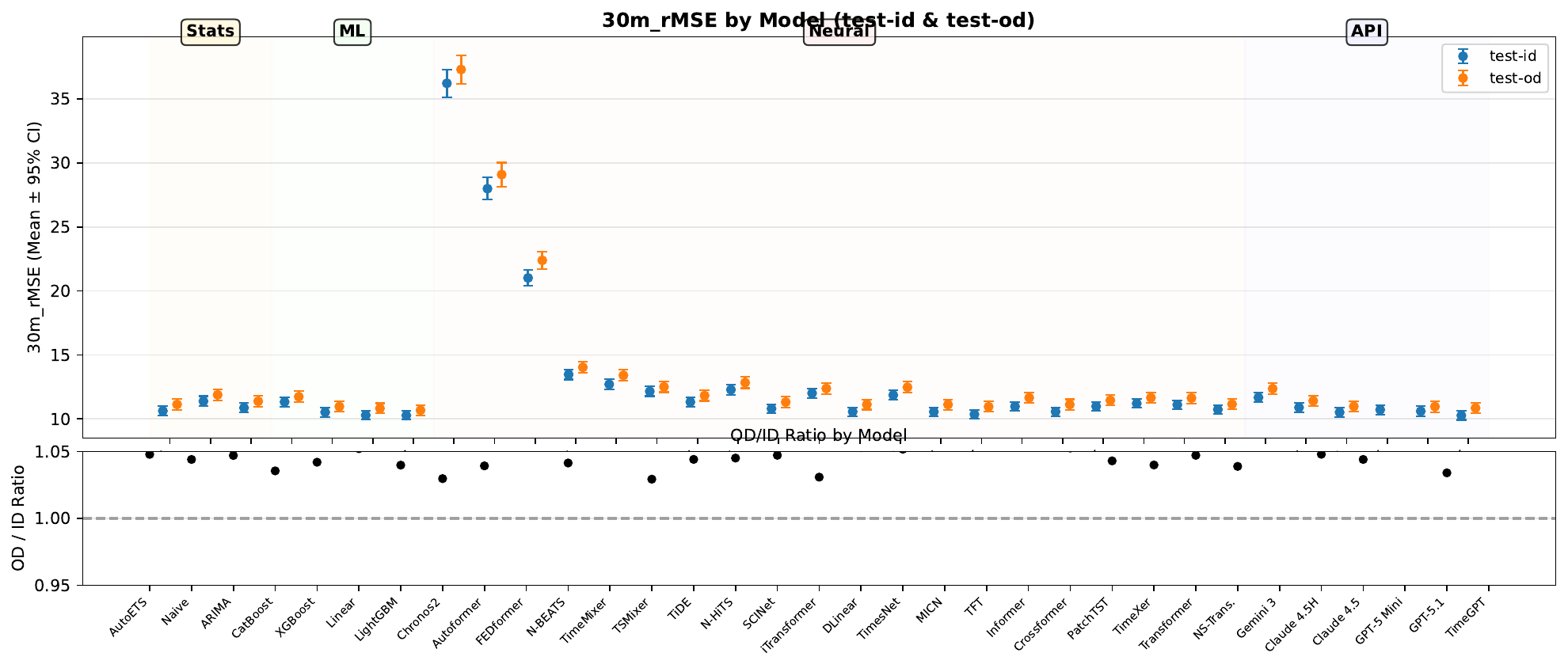}\\[-0.3em]
    {\small (a) rMSE (mg/dL)}\\[0.5em]
    \includegraphics[width=0.95\textwidth]{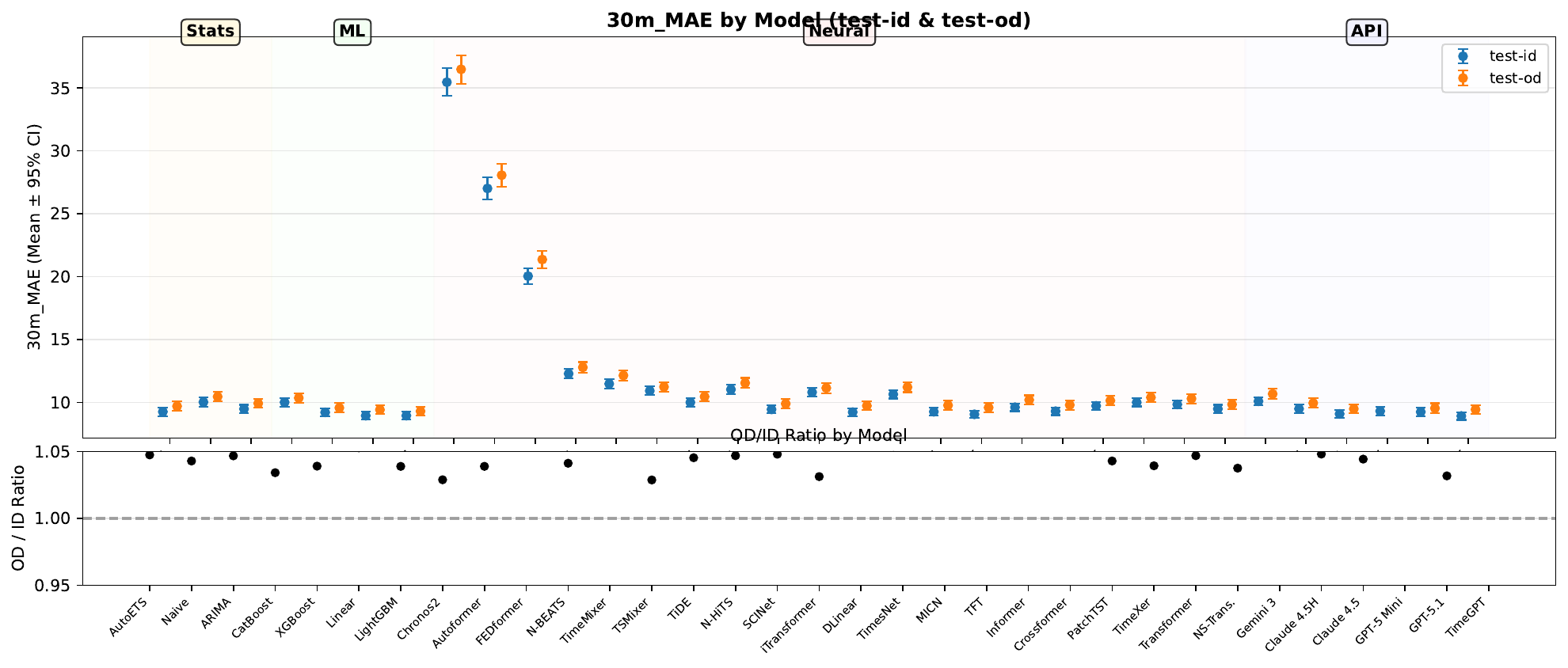}\\[-0.3em]
    {\small (b) MAE (mg/dL)}\\[0.5em]
    \includegraphics[width=0.95\textwidth]{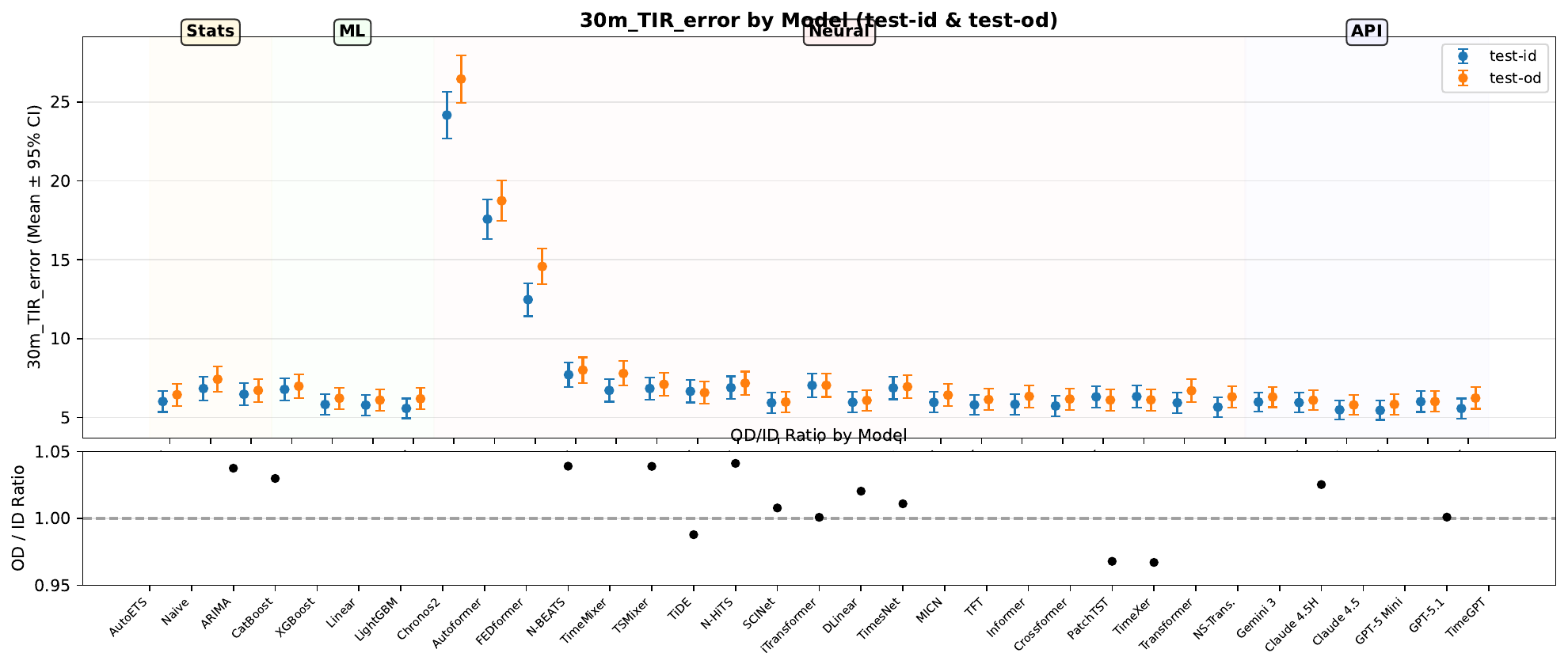}\\[-0.3em]
    {\small (c) TIR Error (\%)}
    \caption{30-minute prediction accuracy: Model comparison between Test-ID and Test-OD splits. Top panel in each subfigure shows metric values with 95\% CI; bottom panel shows OD/ID generalization ratio.}
    \label{fig:accuracy_30m}
\end{figure}

\begin{figure}[p]
    \centering
    \includegraphics[width=0.95\textwidth]{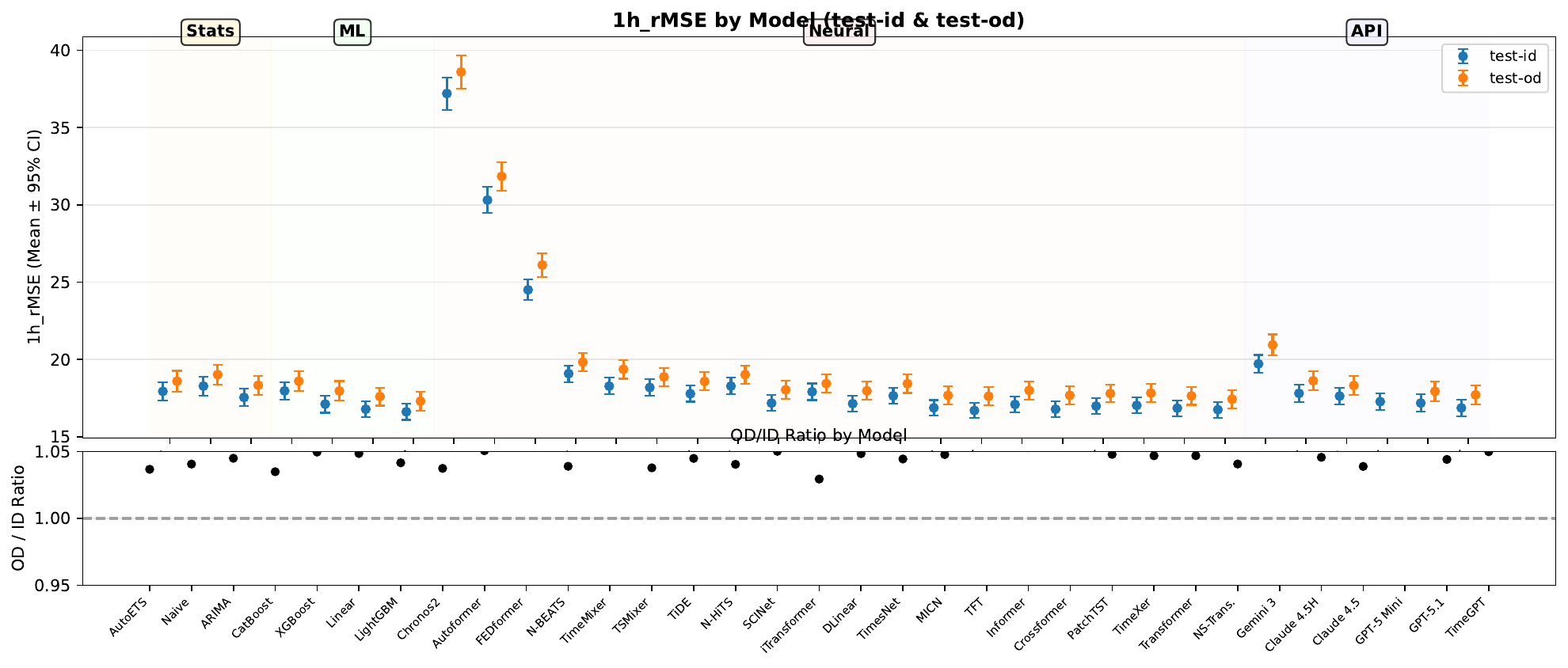}\\[-0.3em]
    {\small (a) rMSE (mg/dL)}\\[0.5em]
    \includegraphics[width=0.95\textwidth]{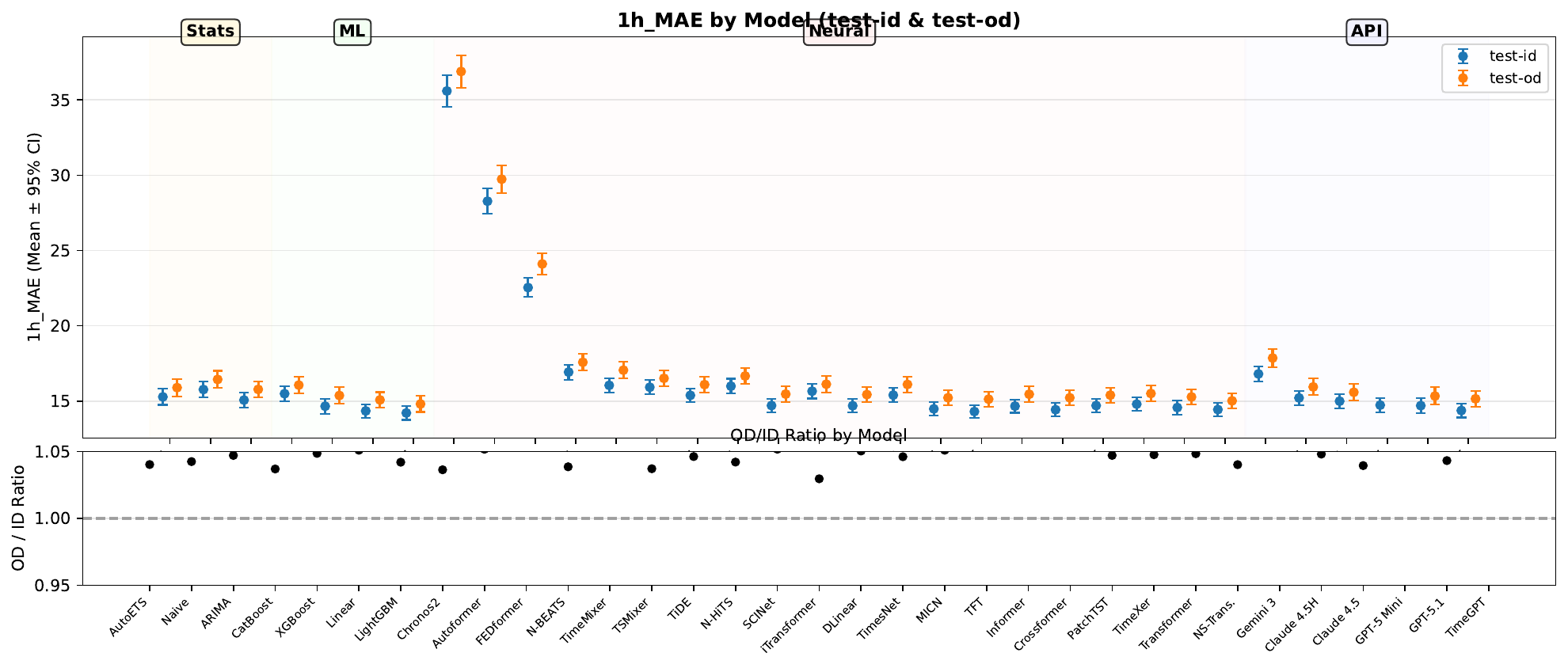}\\[-0.3em]
    {\small (b) MAE (mg/dL)}\\[0.5em]
    \includegraphics[width=0.95\textwidth]{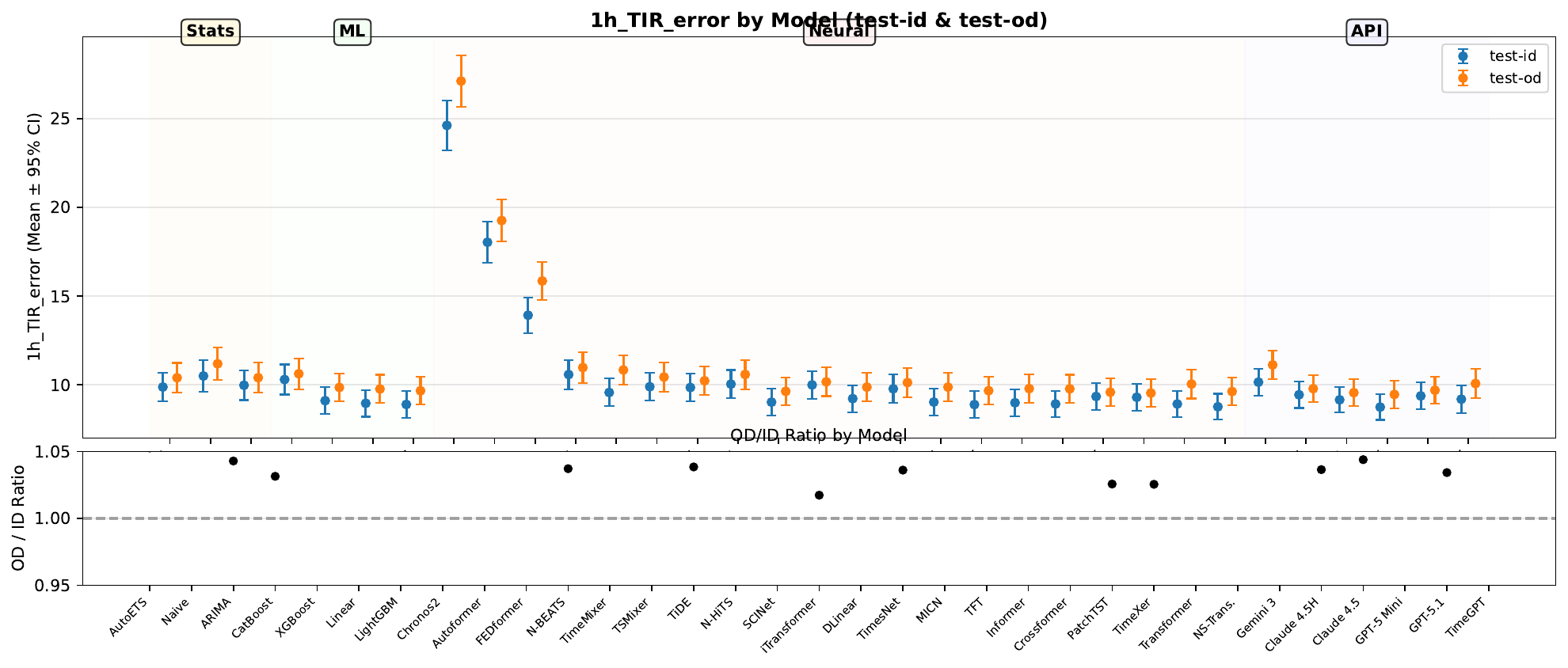}\\[-0.3em]
    {\small (c) TIR Error (\%)}
    \caption{1-hour prediction accuracy: Model comparison between Test-ID and Test-OD splits.}
    \label{fig:accuracy_1h}
\end{figure}

\begin{figure}[p]
    \centering
    \includegraphics[width=0.95\textwidth]{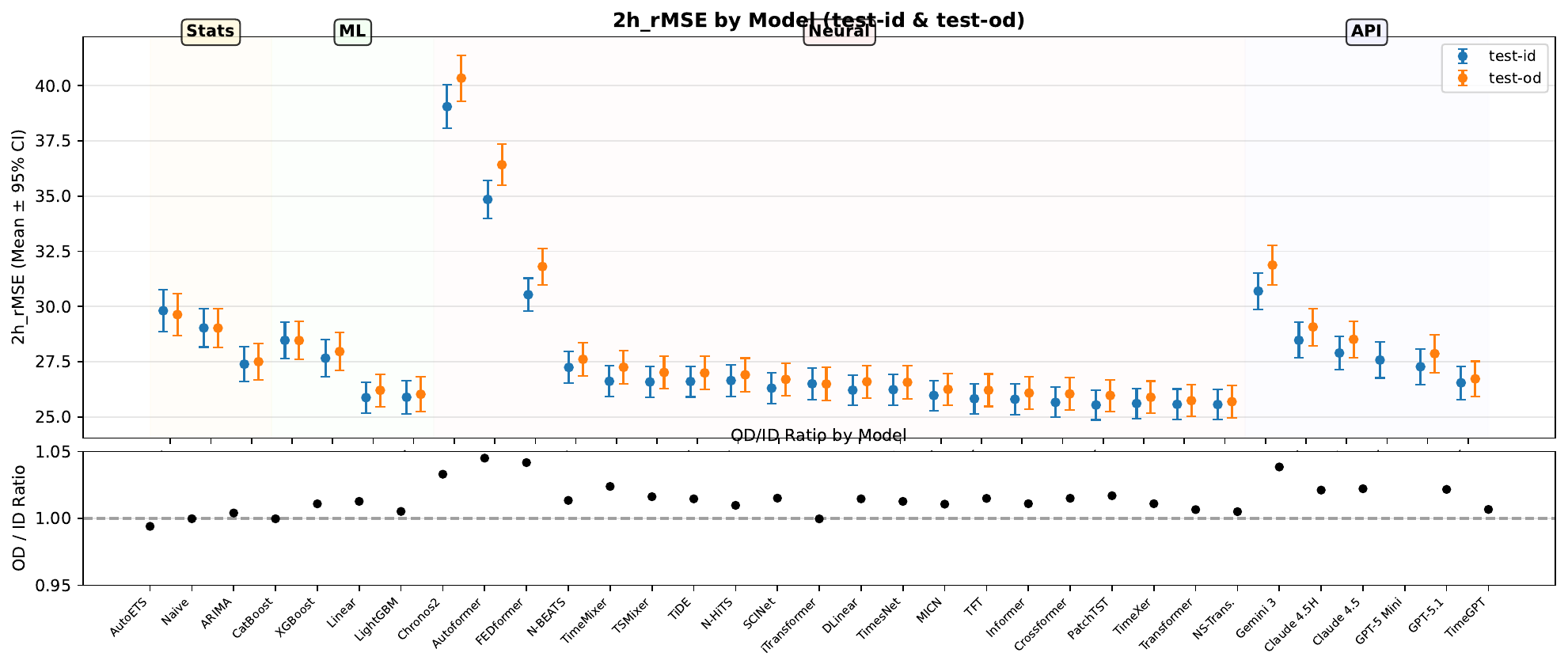}\\[-0.3em]
    {\small (a) rMSE (mg/dL)}\\[0.5em]
    \includegraphics[width=0.95\textwidth]{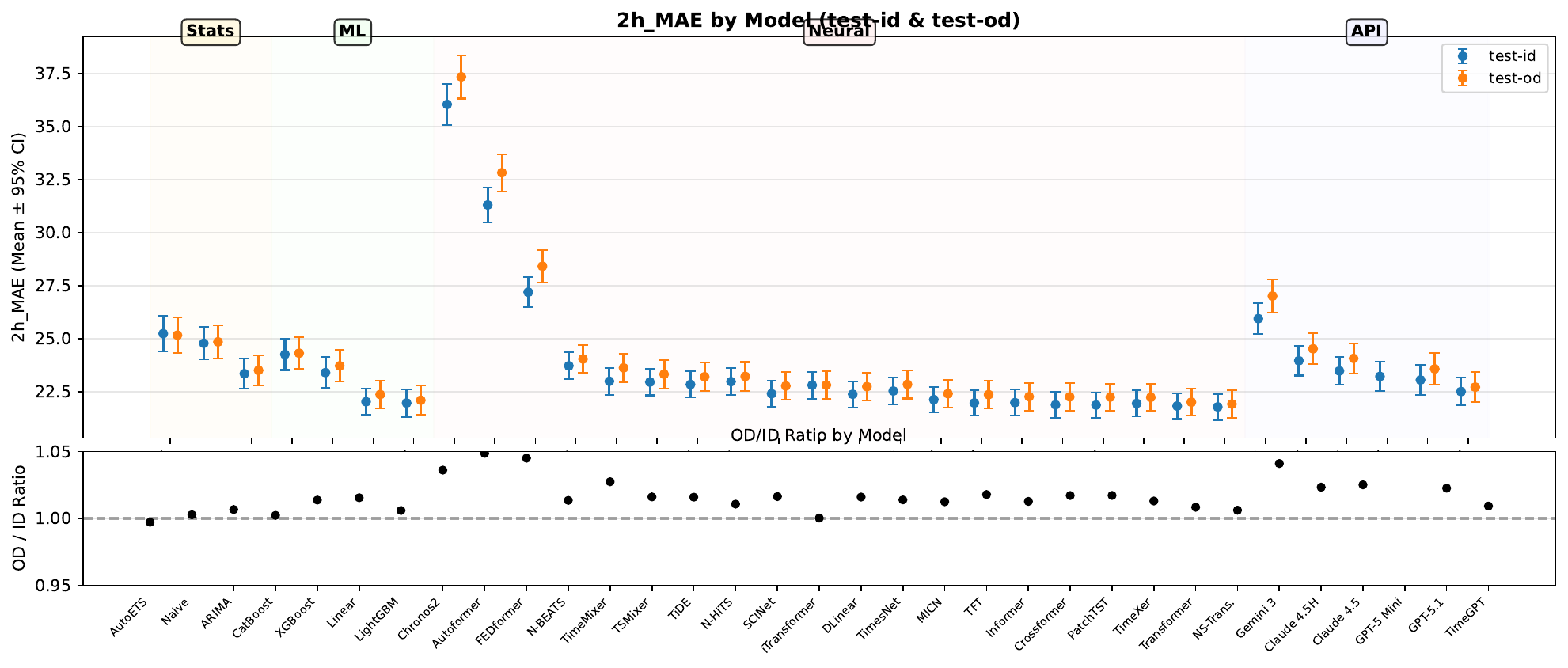}\\[-0.3em]
    {\small (b) MAE (mg/dL)}\\[0.5em]
    \includegraphics[width=0.95\textwidth]{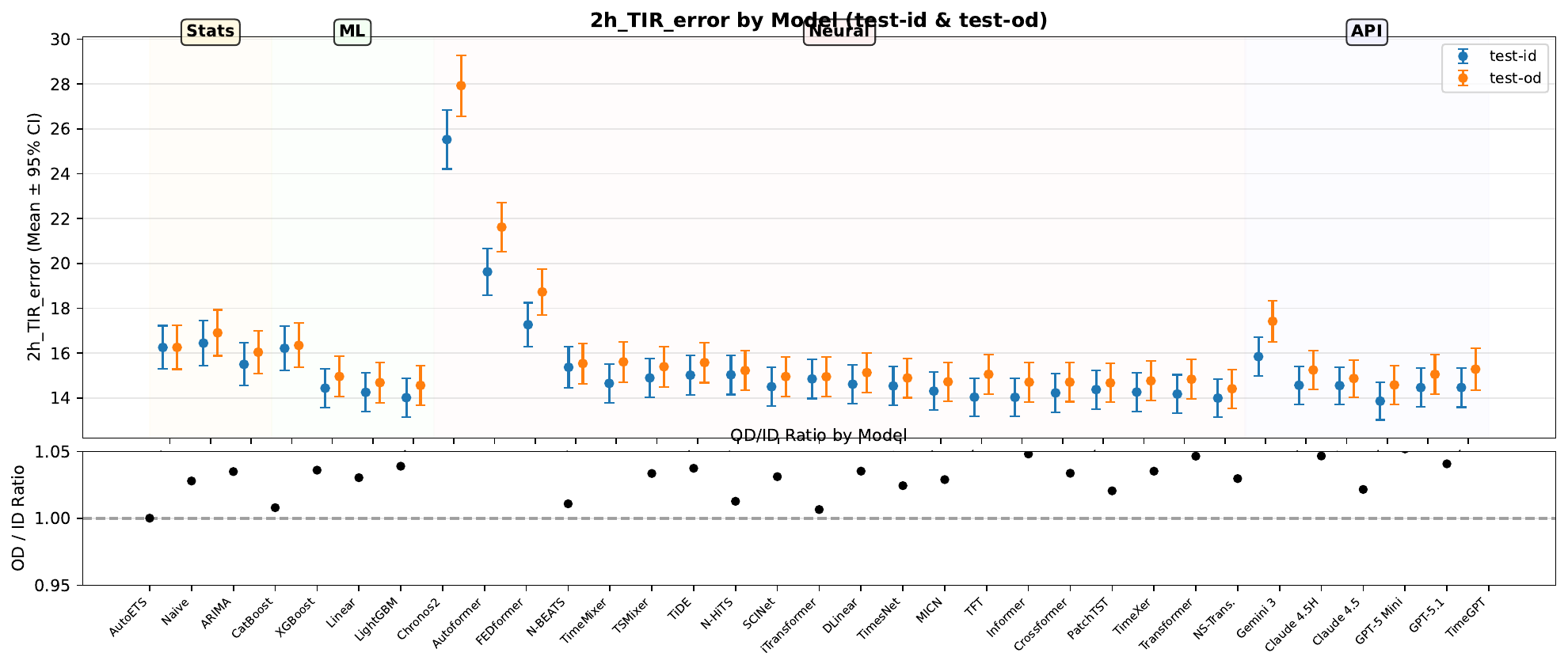}\\[-0.3em]
    {\small (c) TIR Error (\%)}
    \caption{2-hour prediction accuracy: Model comparison between Test-ID and Test-OD splits.}
    \label{fig:accuracy_2h}
\end{figure}

The OD/ID ratios in the lower panels further expose subgroup-specific vulnerabilities. While the overall generalization performance appears stable in population-level evaluations (e.g., Figures~\ref{fig:accuracy_30m}--\ref{fig:accuracy_2h}), subgroup-level ratios span a broader range, consistent with the 0.8--1.4 range reported in the main text. Notably, most models maintain ratios near 1.0 for T2D patients across all age groups, indicating robust generalization for this population. In contrast, T1D patients, particularly females aged 40--64, show more pronounced distribution shift effects. Frontier LLMs exhibit more variable generalization patterns depending on subgroup, suggesting limited robustness to demographic shifts.

As prediction horizons extend from 30 minutes to 2 hours, absolute errors increase predictably across all subgroups (note the increasing y-axis ranges), but relative subgroup disparities remain largely stable. This consistency suggests that subgroup-specific challenges are inherent to the prediction task rather than artifacts of model capacity or training. However, the generalization patterns (bottom panels) show slight deterioration at longer horizons, with OD/ID ratio variance increasing from 0.05 at 30 minutes to 0.08 at 2 hours, indicating that distribution shift effects compound with forecast difficulty.

Together, these appendix results substantiate the main text's findings: generalization robustness and predictive accuracy vary significantly across demographic and clinical subgroups. Fairness-aware evaluations should therefore extend beyond marginal attributes to capture intersectional effects, ensuring models perform reliably across all subpopulations in real-world deployments. The consistent y-axis scaling and scatter plot visualization facilitate rapid identification of vulnerable subgroups and poorly generalizing models, supporting targeted model improvement and fairness-aware deployment decisions.

\subsection{Fairness Evaluation: Model Accuracy vs Fairness}

\begin{figure}[t]
    \centering
    \begin{subfigure}[t]{0.48\textwidth}
        \includegraphics[width=\linewidth]{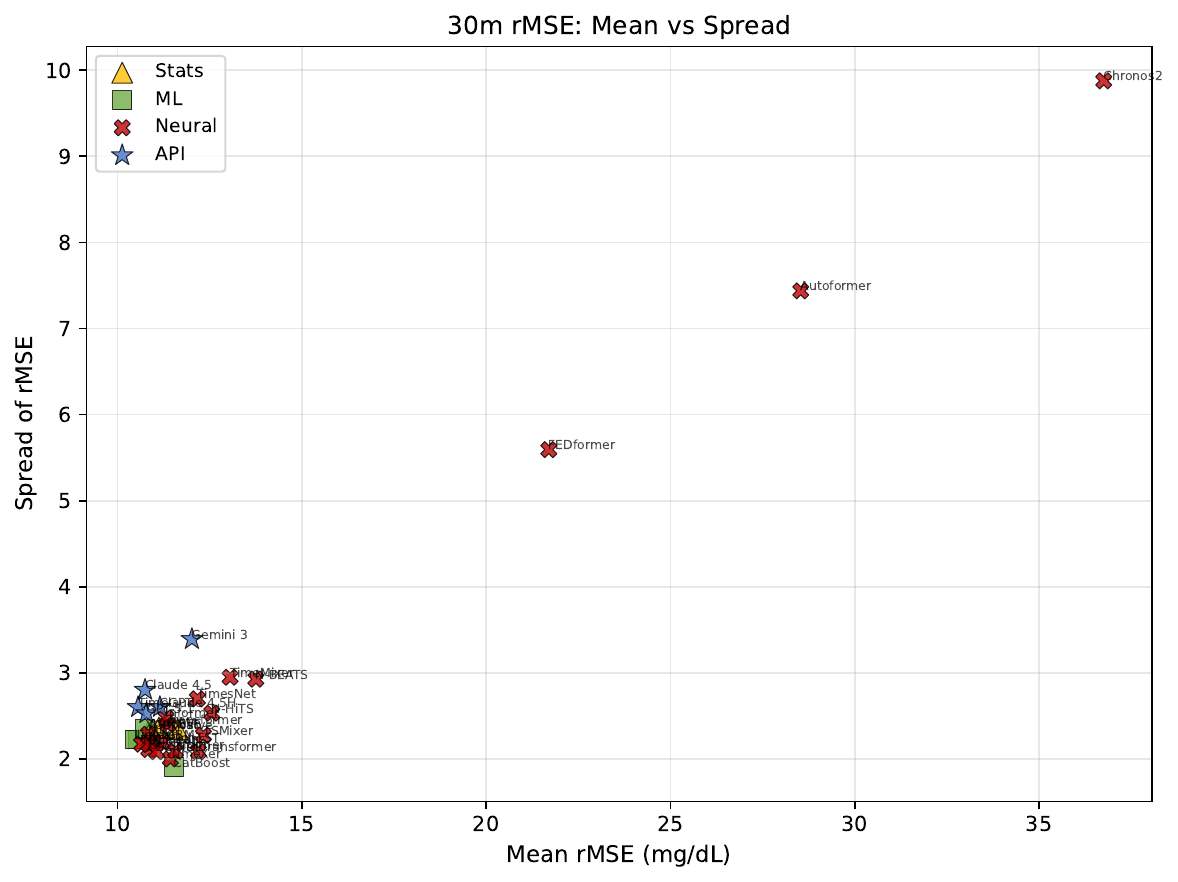}
        \caption{30m rMSE}
    \end{subfigure}\hfill
    \begin{subfigure}[t]{0.48\textwidth}
        \includegraphics[width=\linewidth]{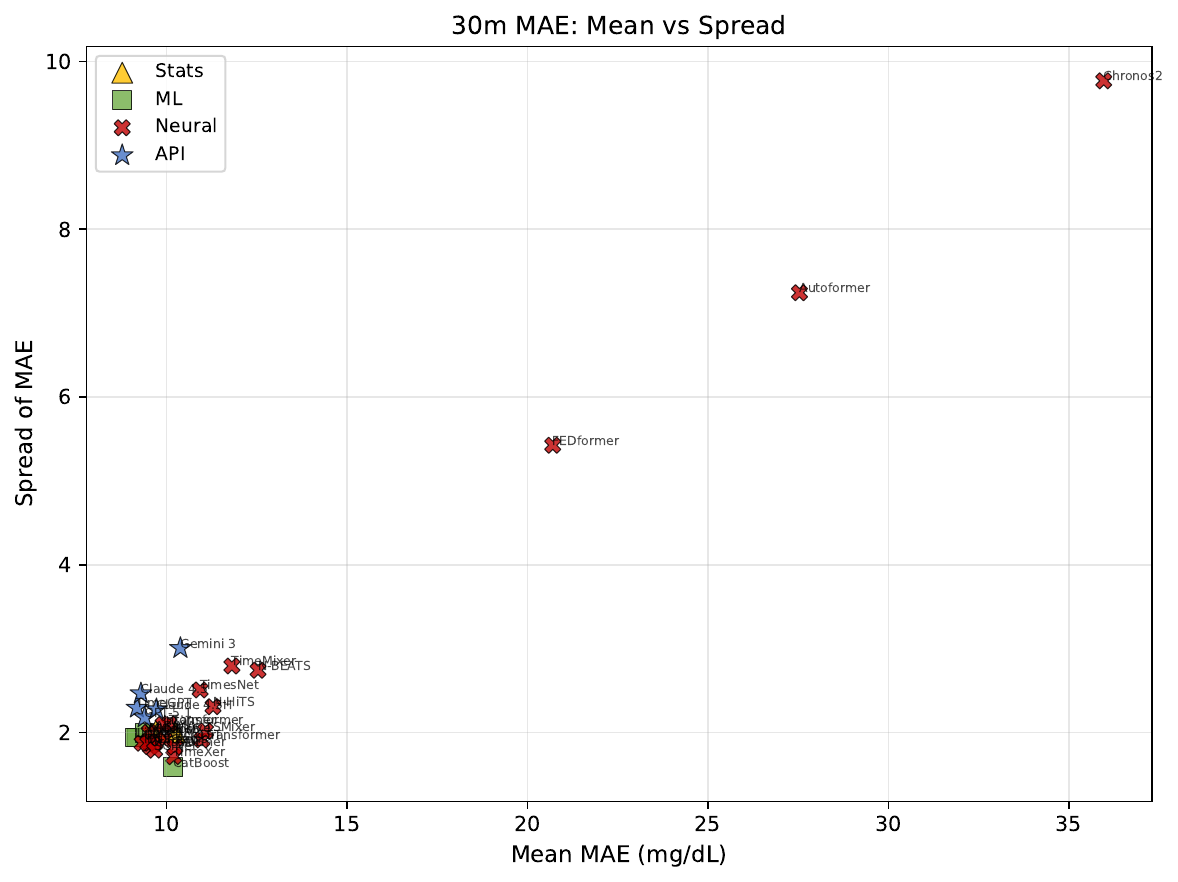}
        \caption{30m MAE}
    \end{subfigure}

    \vspace{0.8em}

    \begin{subfigure}[t]{0.48\textwidth}
        \includegraphics[width=\linewidth]{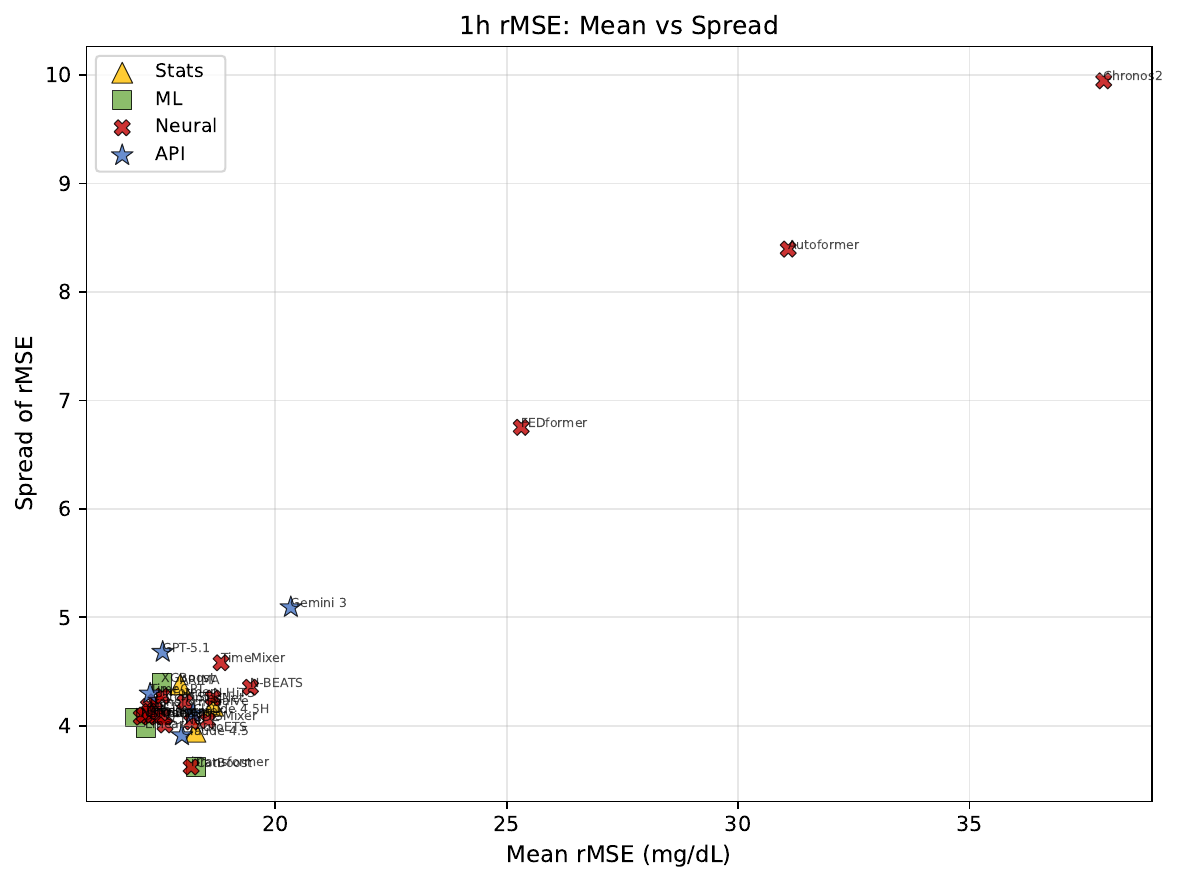}
        \caption{1h rMSE}
    \end{subfigure}\hfill
    \begin{subfigure}[t]{0.48\textwidth}
        \includegraphics[width=\linewidth]{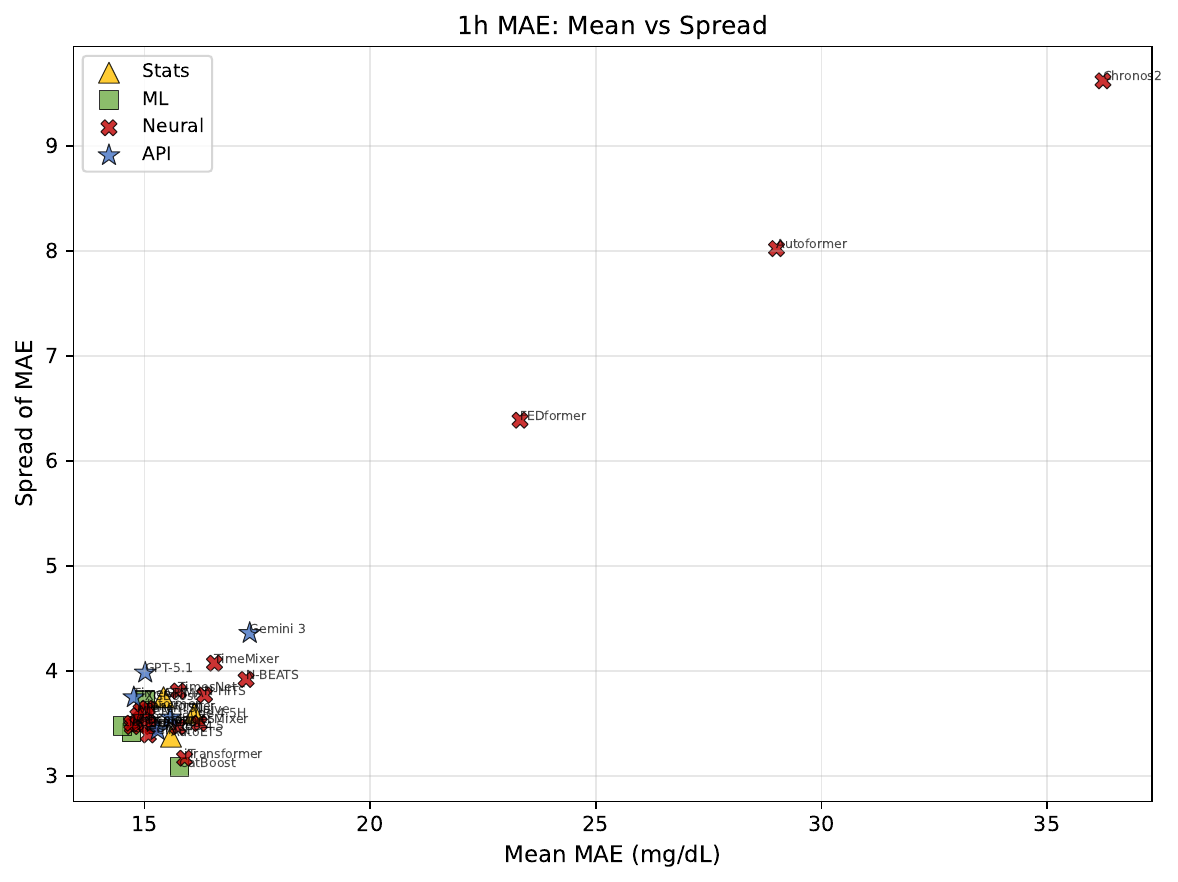}
        \caption{1h MAE}
    \end{subfigure}

    \vspace{0.8em}

    \begin{subfigure}[t]{0.48\textwidth}
        \includegraphics[width=\linewidth]{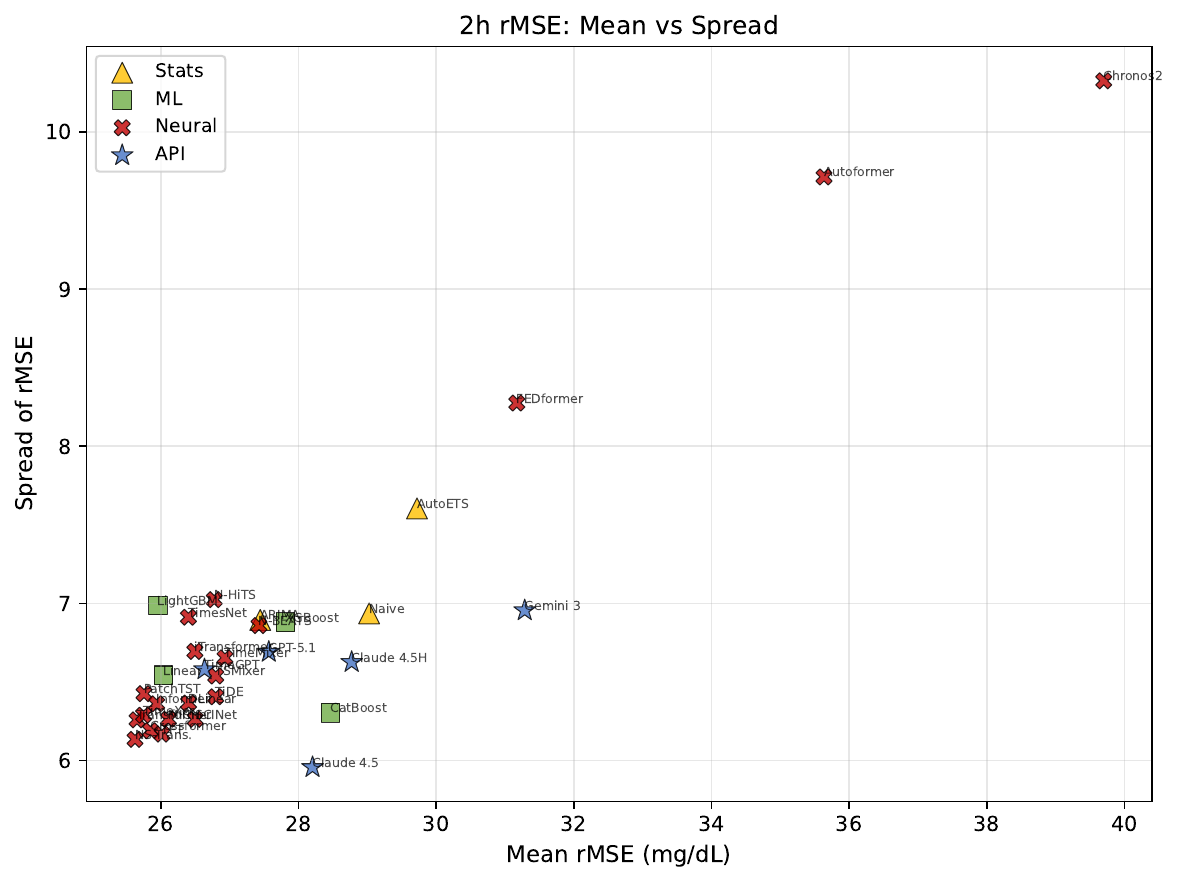}
        \caption{2h rMSE}
    \end{subfigure}\hfill
    \begin{subfigure}[t]{0.48\textwidth}
        \includegraphics[width=\linewidth]{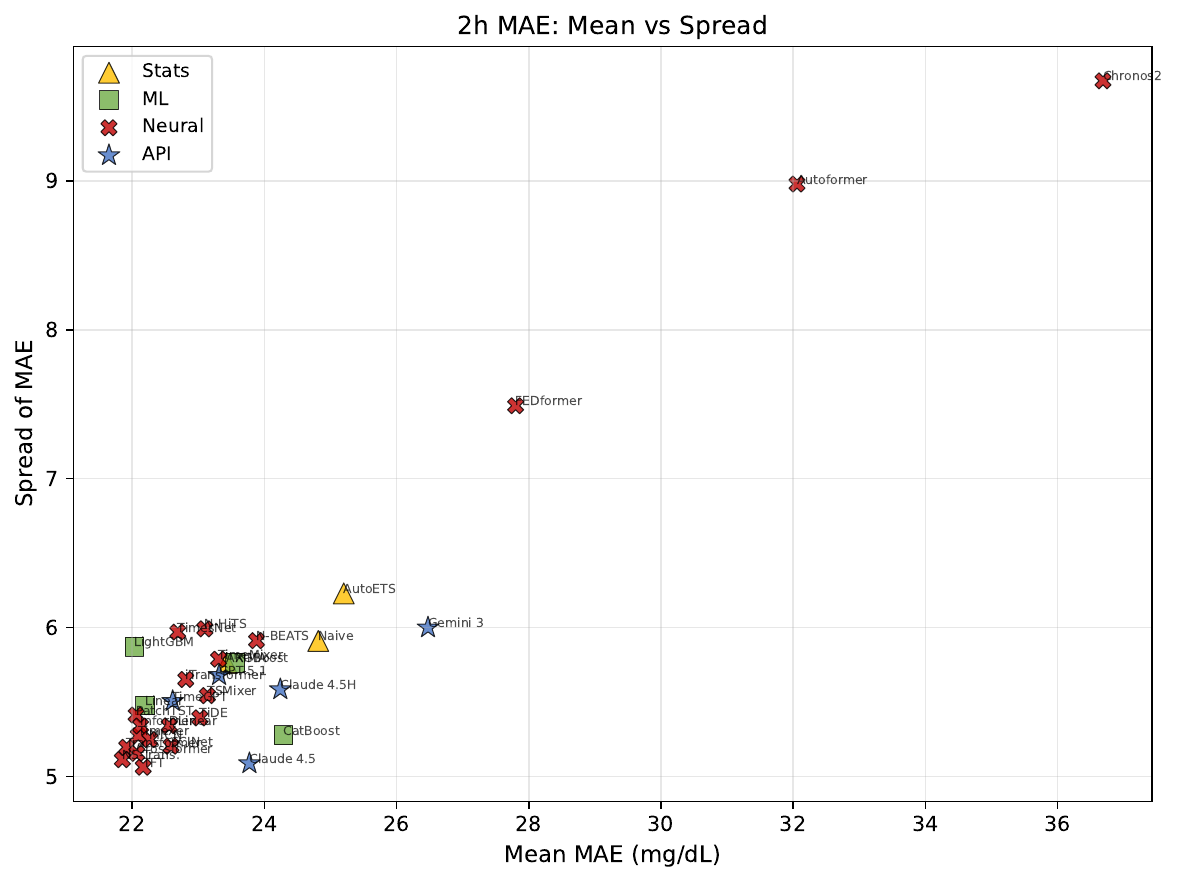}
        \caption{2h MAE}
    \end{subfigure}

    \caption{Accuracy vs. fairness tradeoff across prediction horizons (Spread measure). Columns show rMSE and MAE metrics; rows show 30-minute, 1-hour, and 2-hour predictions. Each point represents a model, colored by family. Lower values on both axes indicate better performance.}
    \label{fig:fairness-spread-3x2}
\end{figure}

\begin{figure}[t]
    \centering
    \begin{subfigure}[t]{0.48\textwidth}
        \includegraphics[width=\linewidth]{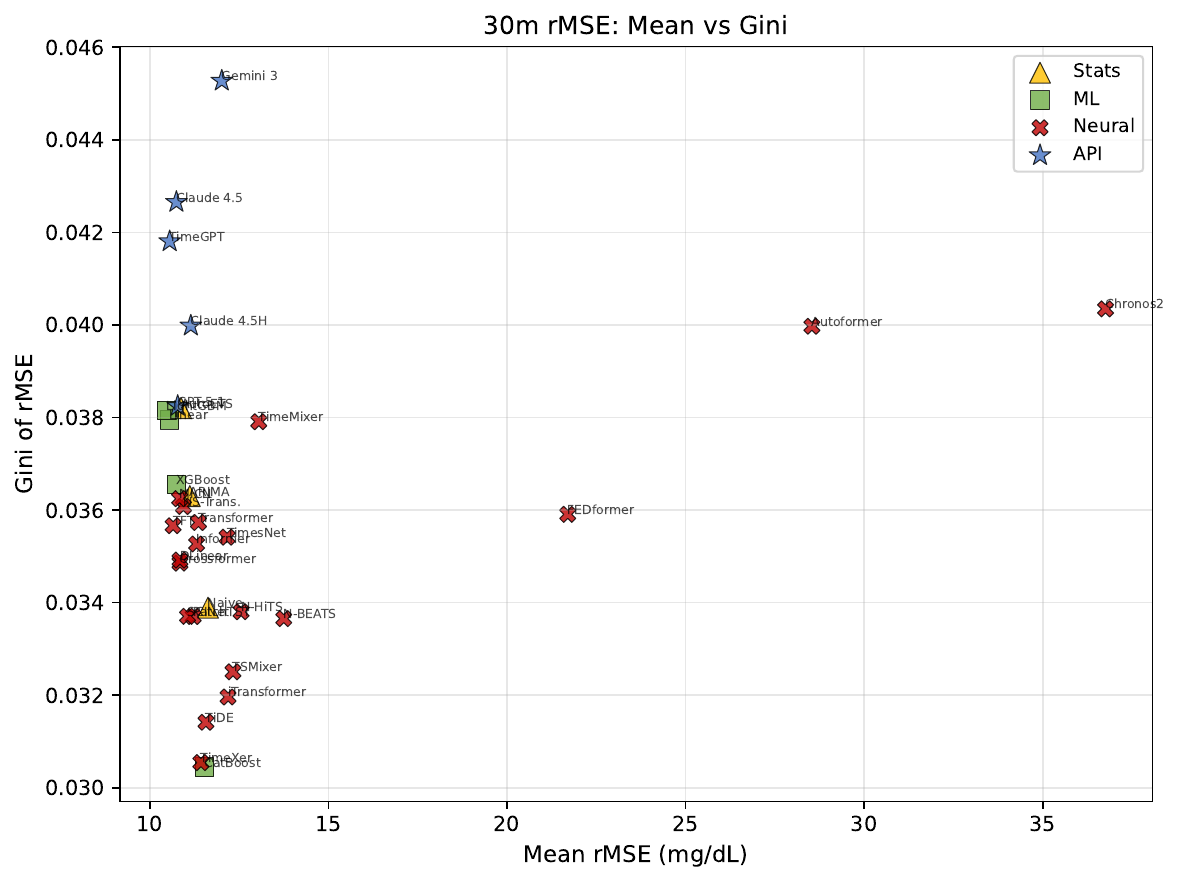}
        \caption{30m rMSE}
    \end{subfigure}\hfill
    \begin{subfigure}[t]{0.48\textwidth}
        \includegraphics[width=\linewidth]{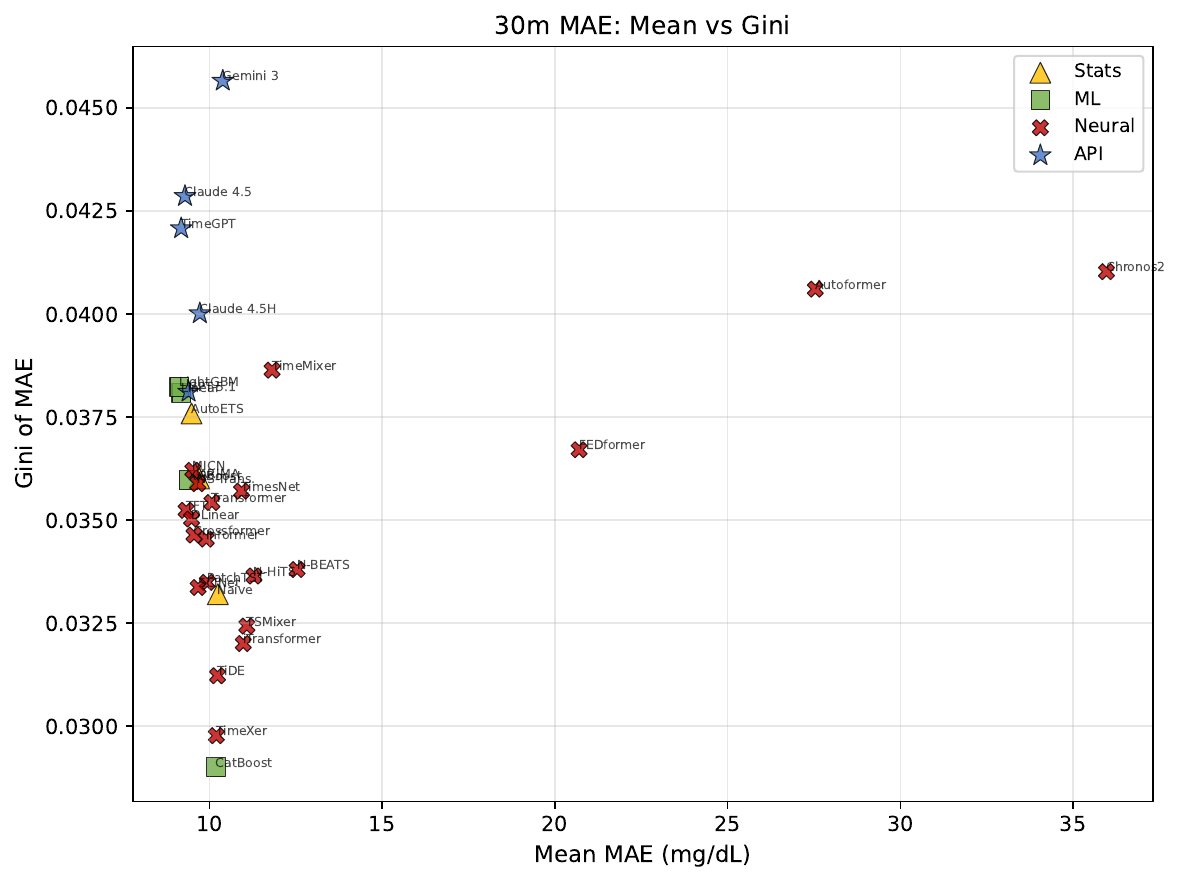}
        \caption{30m MAE}
    \end{subfigure}

    \vspace{0.8em}

    \begin{subfigure}[t]{0.48\textwidth}
        \includegraphics[width=\linewidth]{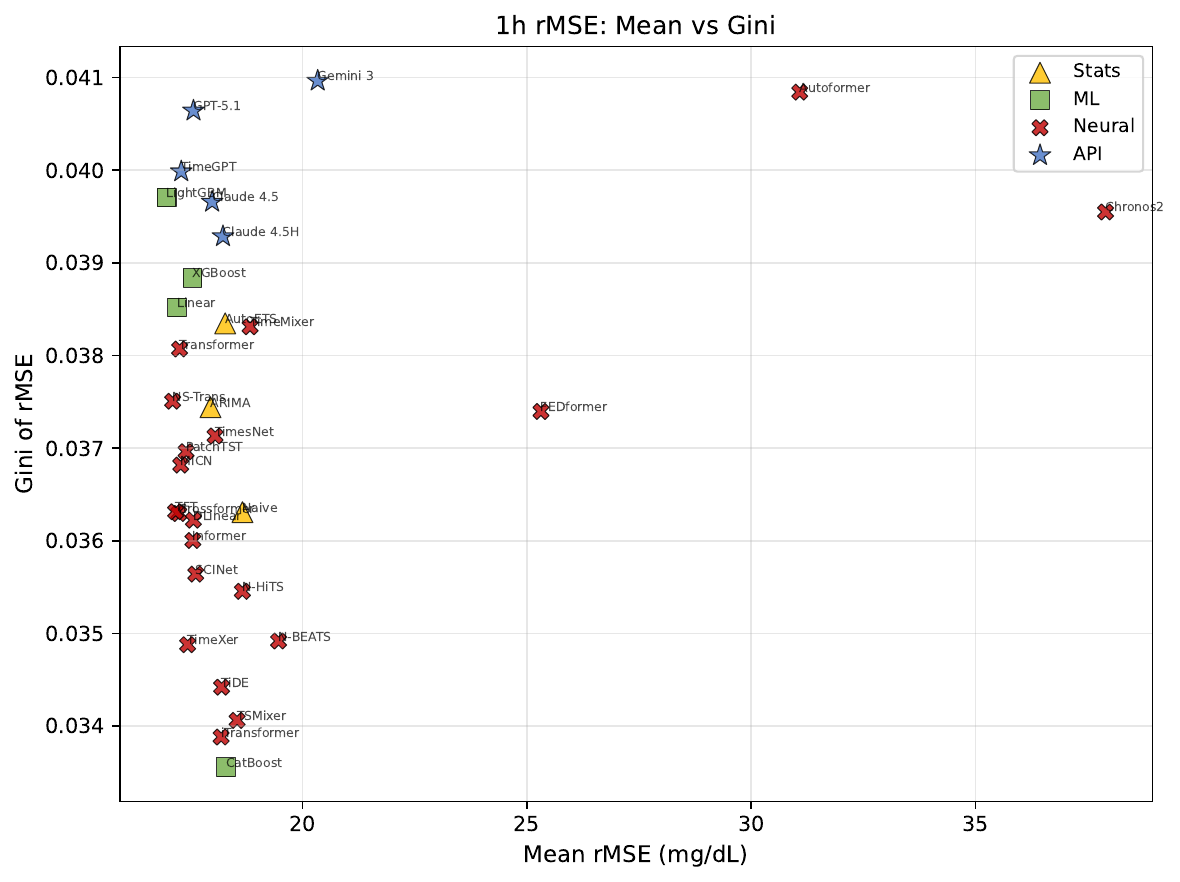}
        \caption{1h rMSE}
    \end{subfigure}\hfill
    \begin{subfigure}[t]{0.48\textwidth}
        \includegraphics[width=\linewidth]{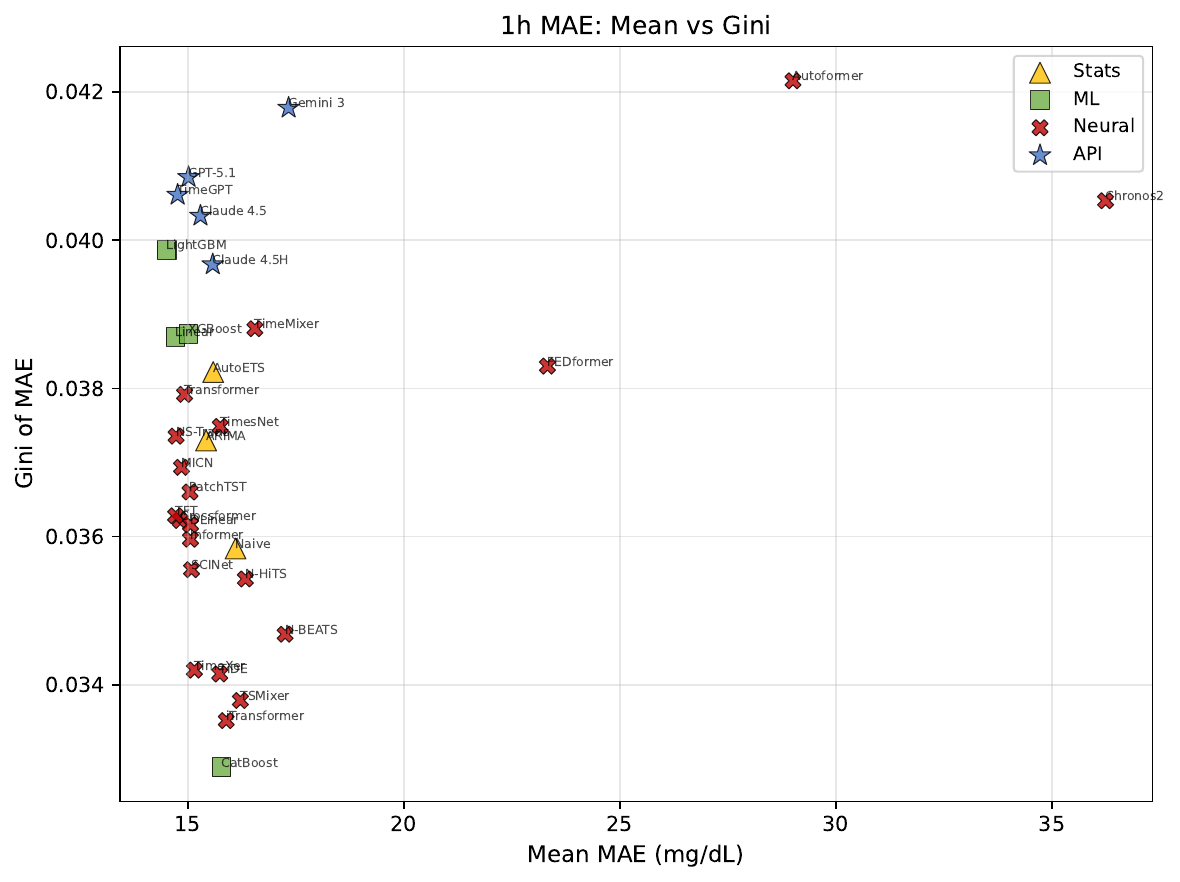}
        \caption{1h MAE}
    \end{subfigure}

    \vspace{0.8em}

    \begin{subfigure}[t]{0.48\textwidth}
        \includegraphics[width=\linewidth]{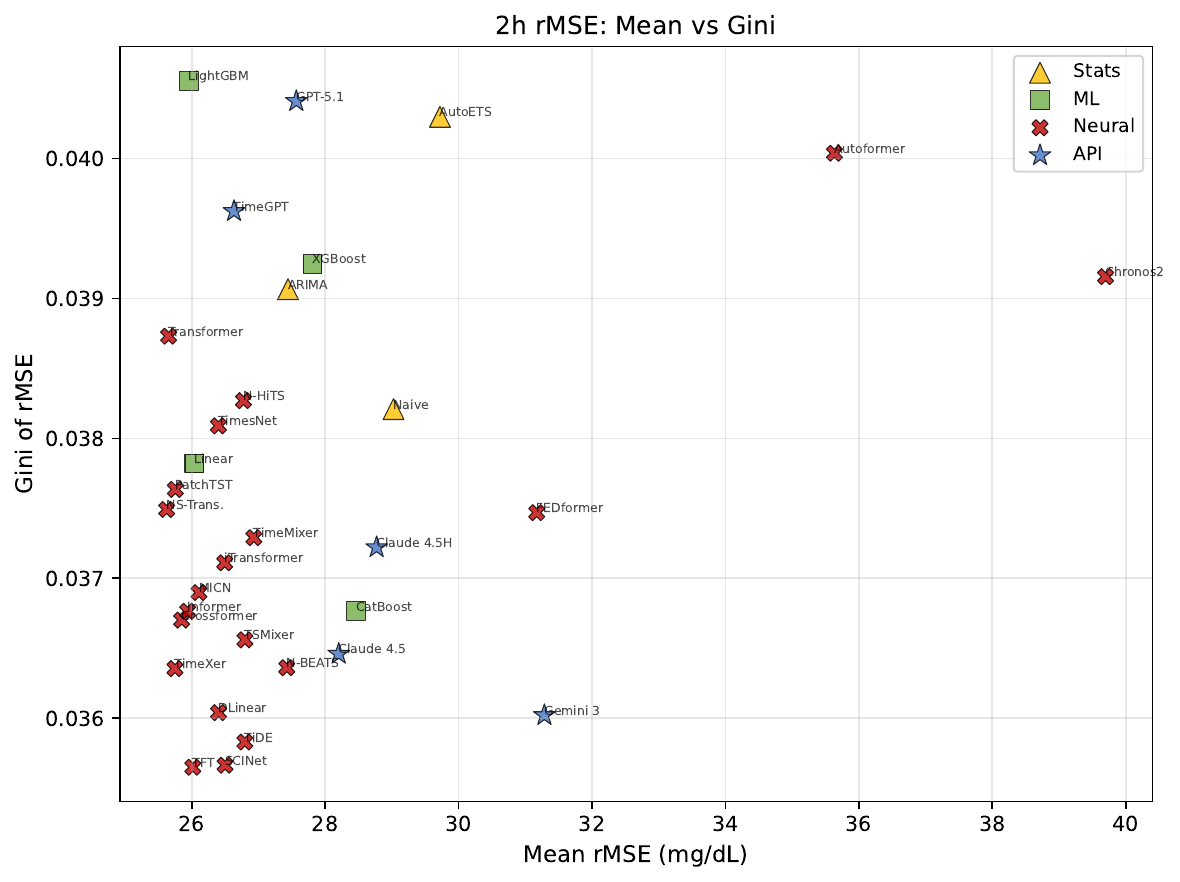}
        \caption{2h rMSE}
    \end{subfigure}\hfill
    \begin{subfigure}[t]{0.48\textwidth}
        \includegraphics[width=\linewidth]{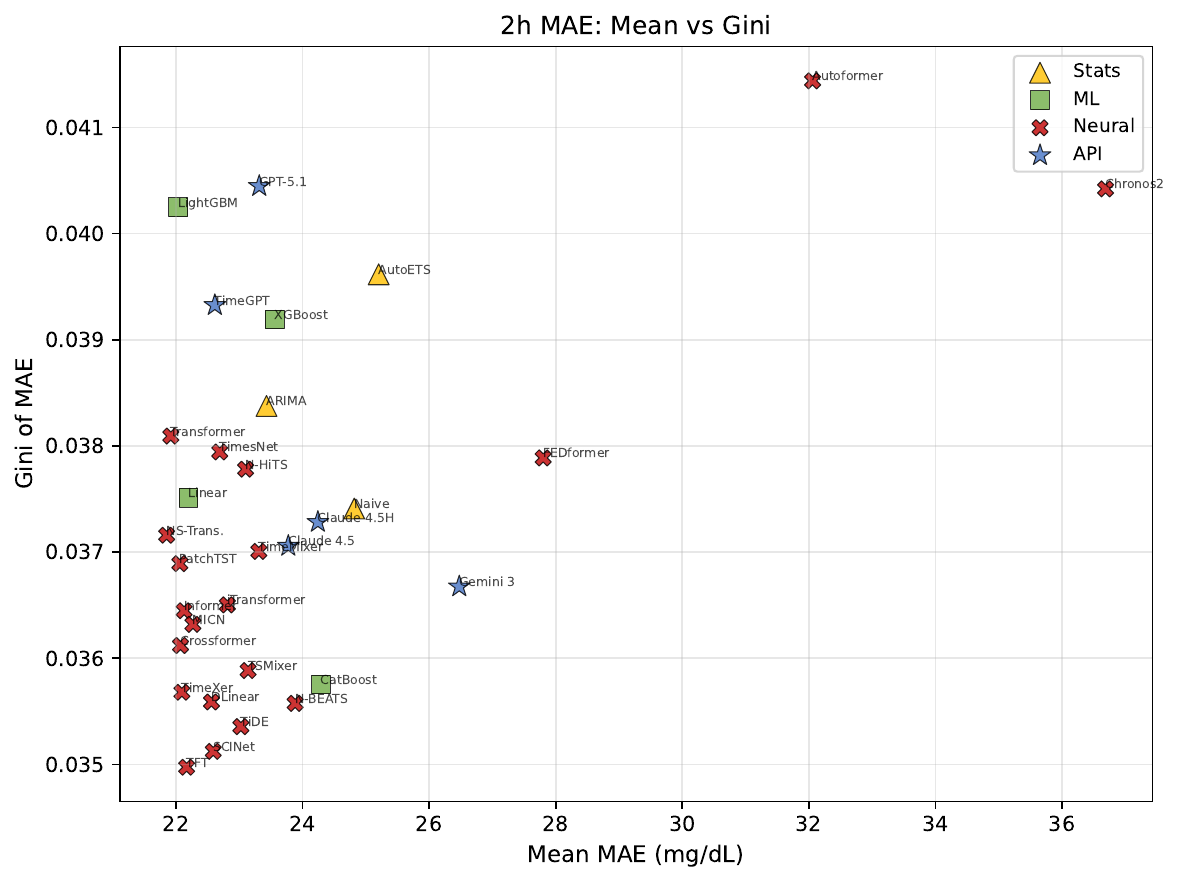}
        \caption{2h MAE}
    \end{subfigure}

    \caption{Accuracy vs. fairness tradeoff across prediction horizons (Gini coefficient measure). Columns show rMSE and MAE metrics; rows show 30-minute, 1-hour, and 2-hour predictions. Each point represents a model, colored by family. Lower values on both axes indicate better performance.}
    \label{fig:fairness-gini-3x2}
\end{figure}

To evaluate subgroup fairness across models, we computed two measures of variability: Prediction Error Spread (the difference between maximum and minimum subgroup errors, shown in Figure~\ref{fig:fairness-spread-3x2}) and Gini coefficient (shown in Figure~\ref{fig:fairness-gini-3x2}).
For each model, we first calculated the root mean squared error (rMSE) across all subgroups, followed by the fairness metric that captures how evenly errors are distributed across these subgroups.
A lower Spread or Gini metric indicates greater fairness (less disparity).

We generated separate plots for 30-minute, 1-hour, and 2-hour horizons, using both spread and Gini fairness measures. Each point in the scatter plots represents a model, with the x-axis showing average predictive error (lower is better) and the y-axis representing fairness disparity (lower is fairer). Models were grouped by model family.

The fairness analysis reveals notable differences among model families. Neural time-series models such as PatchTST and TFT occupy favorable regions with low error and low subgroup disparity, consistent with the main-text accuracy--fairness analysis.

Conversely, several lower-performing neural and frontier LLM models exhibit higher error and/or higher subgroup disparity, indicating less consistent performance across subgroups. Traditional linear models and statistical baselines occupy the middle ground with moderate fairness and error. Overall, the tradeoff analysis supports reporting both average accuracy and subgroup disparity rather than selecting models by population-level rMSE alone.

\subsection{Input Length Sensitivity: Extended Results}

The main text presents input-length sensitivity analysis for the 2-hour prediction horizon. Here we provide extended results across all three horizons (30 minutes, 1 hour, and 2 hours) using three metrics: rMSE, MAE, and TIR Absolute Error.

\begin{figure}[htbp]
    \centering
    \begin{subfigure}[t]{0.48\textwidth}
        \centering
        \includegraphics[width=\linewidth]{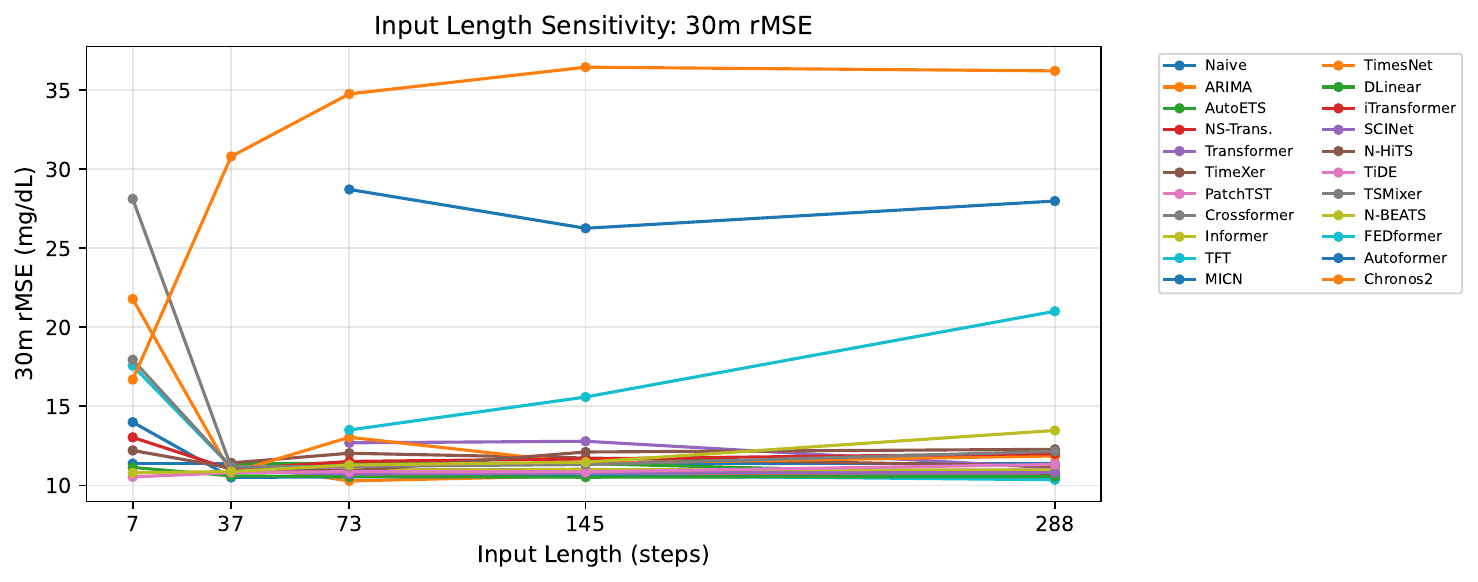}
        \caption{30-minute rMSE}
        \label{fig:difflength_30m_rmse}
    \end{subfigure}
    \hfill
    \begin{subfigure}[t]{0.48\textwidth}
        \centering
        \includegraphics[width=\linewidth]{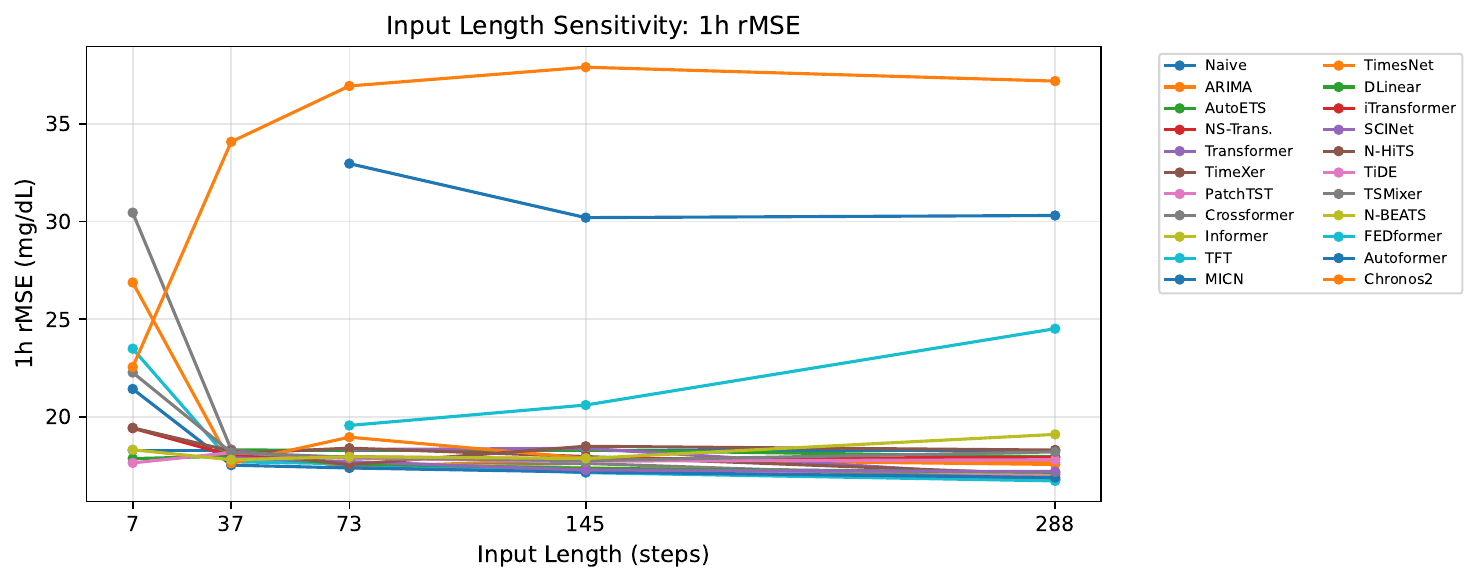}
        \caption{1-hour rMSE}
        \label{fig:difflength_1h_rmse}
    \end{subfigure}

    \vspace{1em}

    \begin{subfigure}[t]{0.48\textwidth}
        \centering
        \includegraphics[width=\linewidth]{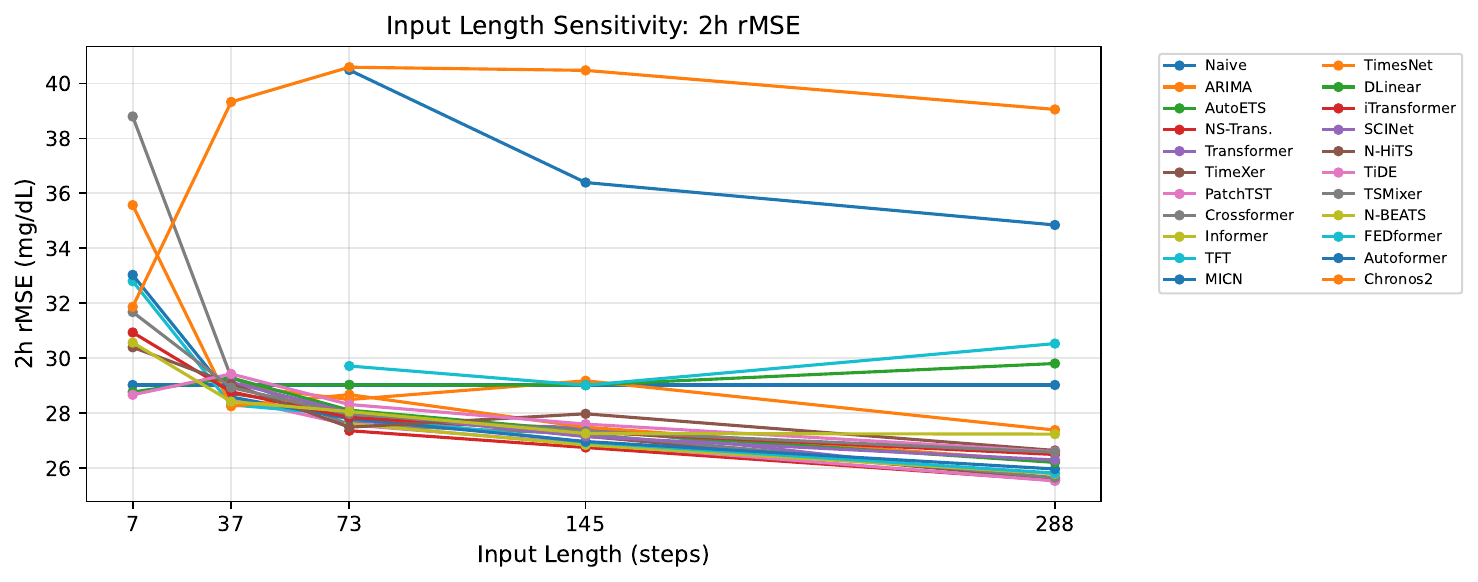}
        \caption{2-hour rMSE}
        \label{fig:difflength_2h_rmse}
    \end{subfigure}
    \hfill
    \begin{subfigure}[t]{0.48\textwidth}
        \centering
        \includegraphics[width=\linewidth]{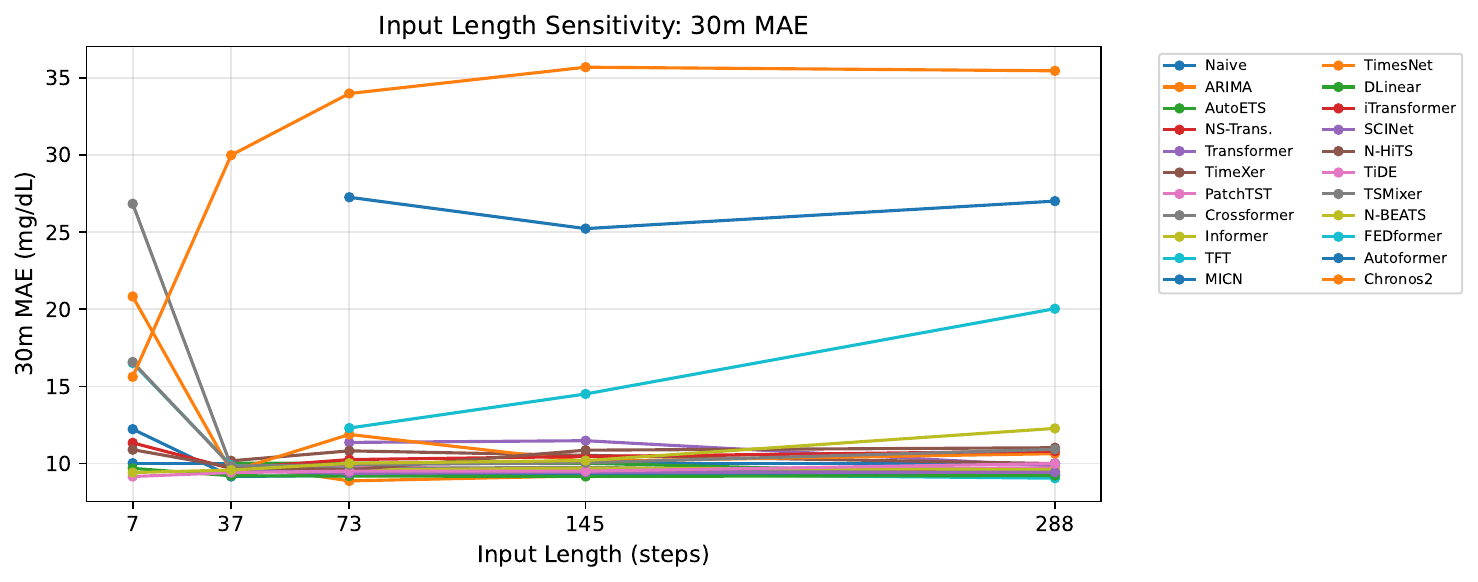}
        \caption{30-minute MAE}
        \label{fig:difflength_30m_mae}
    \end{subfigure}

    \vspace{1em}

    \begin{subfigure}[t]{0.48\textwidth}
        \centering
        \includegraphics[width=\linewidth]{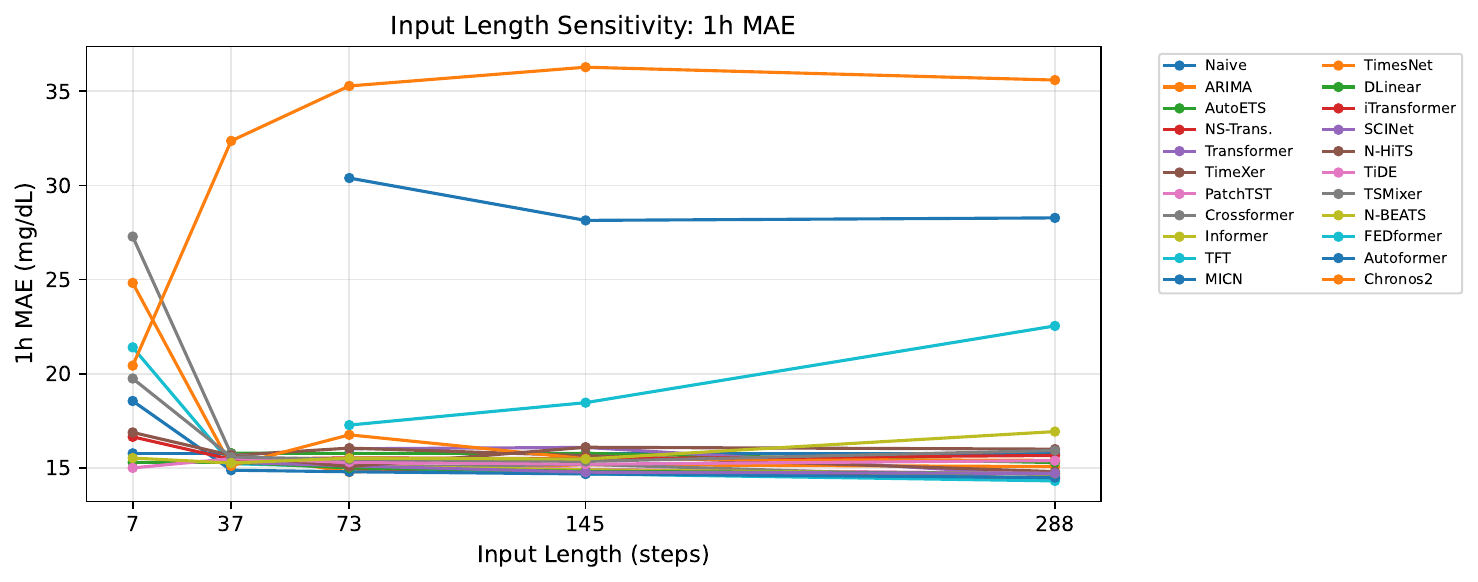}
        \caption{1-hour MAE}
        \label{fig:difflength_1h_mae}
    \end{subfigure}
    \hfill
    \begin{subfigure}[t]{0.48\textwidth}
        \centering
        \includegraphics[width=\linewidth]{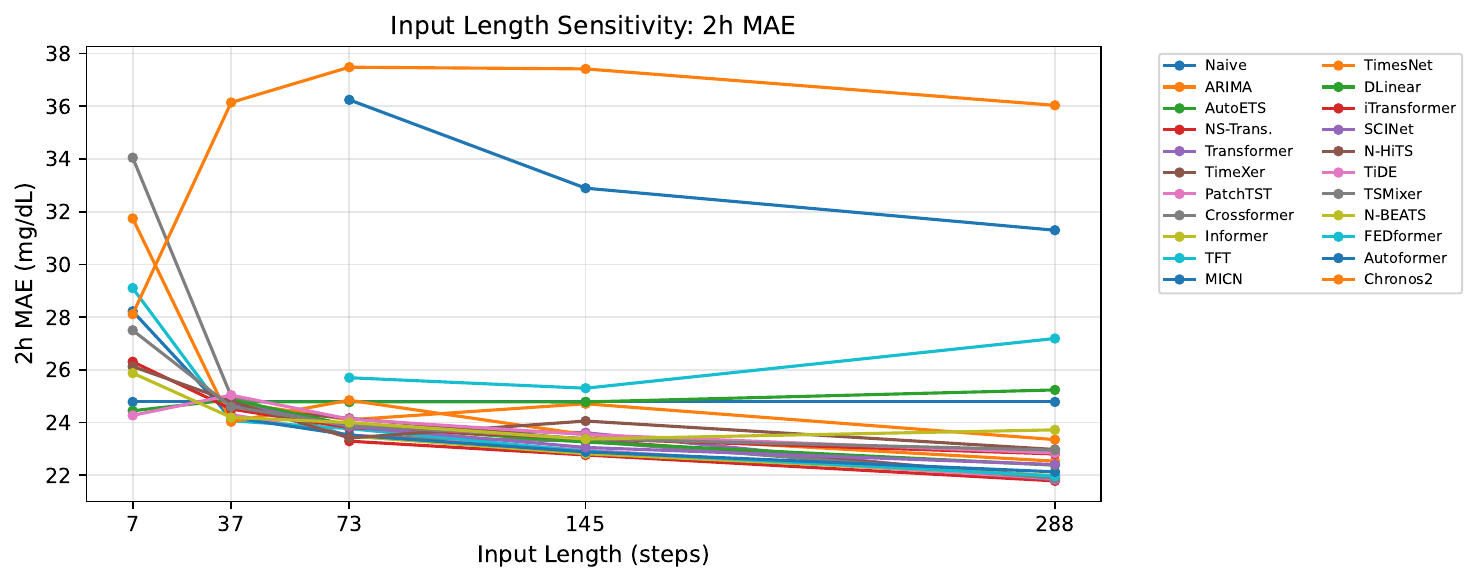}
        \caption{2-hour MAE}
        \label{fig:difflength_2h_mae}
    \end{subfigure}

    \caption{Input length sensitivity across prediction horizons and metrics. Each panel shows model performance (y-axis) at varying input lengths (x-axis) for a specific horizon-metric combination.}
    \label{fig:difflength_all}
\end{figure}

\begin{figure}[htbp]
    \centering
    \begin{subfigure}[t]{0.32\textwidth}
        \centering
        \includegraphics[width=\linewidth]{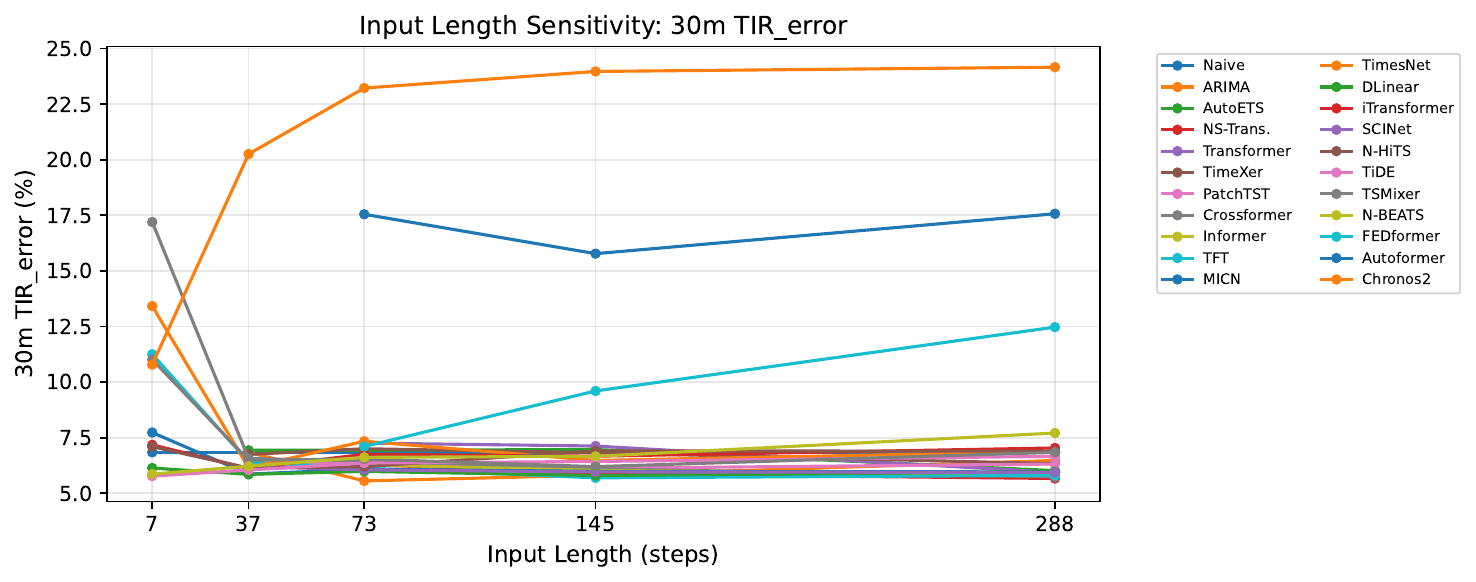}
        \caption{30-minute TIR-AE}
        \label{fig:difflength_30m_tir}
    \end{subfigure}
    \hfill
    \begin{subfigure}[t]{0.32\textwidth}
        \centering
        \includegraphics[width=\linewidth]{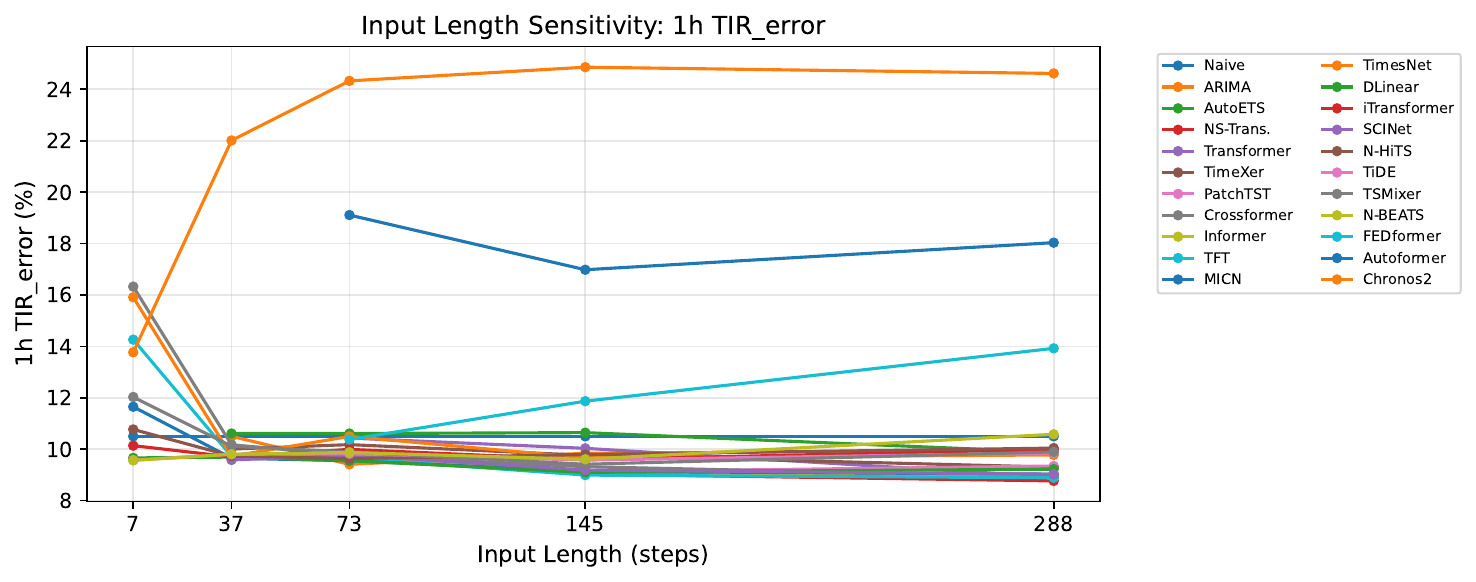}
        \caption{1-hour TIR-AE}
        \label{fig:difflength_1h_tir}
    \end{subfigure}
    \hfill
    \begin{subfigure}[t]{0.32\textwidth}
        \centering
        \includegraphics[width=\linewidth]{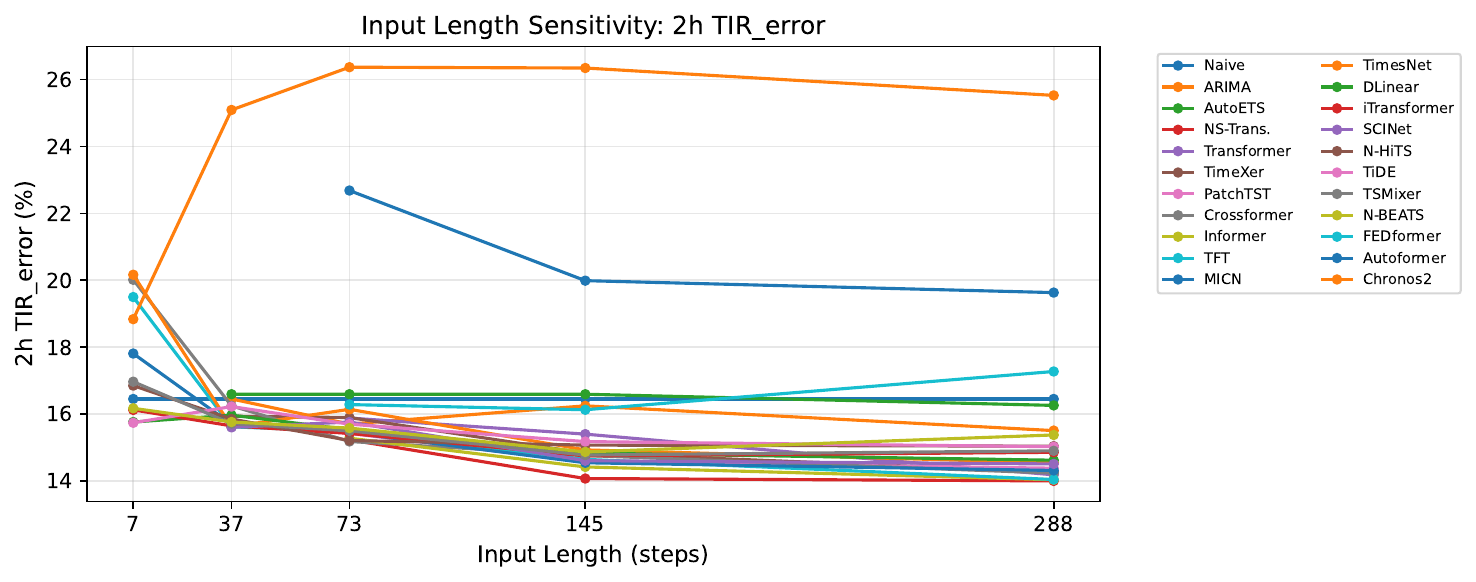}
        \caption{2-hour TIR-AE}
        \label{fig:difflength_2h_tir}
    \end{subfigure}
    \caption{Input length sensitivity for TIR Absolute Error across prediction horizons.}
    \label{fig:difflength_tir}
\end{figure}

Figure~\ref{fig:difflength_all}, panels (a)--(f), covers rMSE and MAE at all three horizons, and Figure~\ref{fig:difflength_tir}, panels (a)--(c), covers TIR Absolute Error. Together they show how model performance varies across input lengths for each horizon-metric combination. Across all settings, the same subgroup-dependent pattern holds: younger T1D patients are relatively insensitive to input-length reduction, while older T2D patients show larger degradation with shorter histories. Neural models (PatchTST, N-HiTS, TFT) consistently benefit most from longer input windows, while statistical and foundation models show more modest gains.

\subsection{Instance-Level Difficulty: Extended Results}

The main text presents instance-level difficulty analysis for the 2-hour horizon. Here we provide extended difficulty-stratified results across all three horizons.

\begin{figure}[htbp]
    \centering
    \begin{subfigure}[t]{0.48\textwidth}
        \centering
        \includegraphics[width=\linewidth]{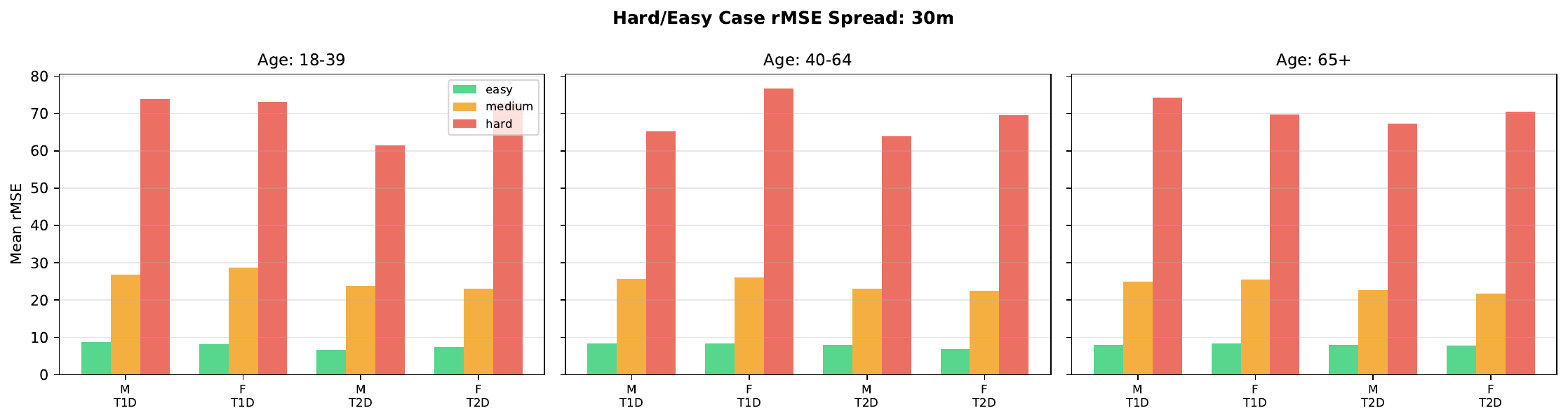}
        \caption{30-minute rMSE}
        \label{fig:hardeasy_spread_30m_rmse}
    \end{subfigure}
    \hfill
    \begin{subfigure}[t]{0.48\textwidth}
        \centering
        \includegraphics[width=\linewidth]{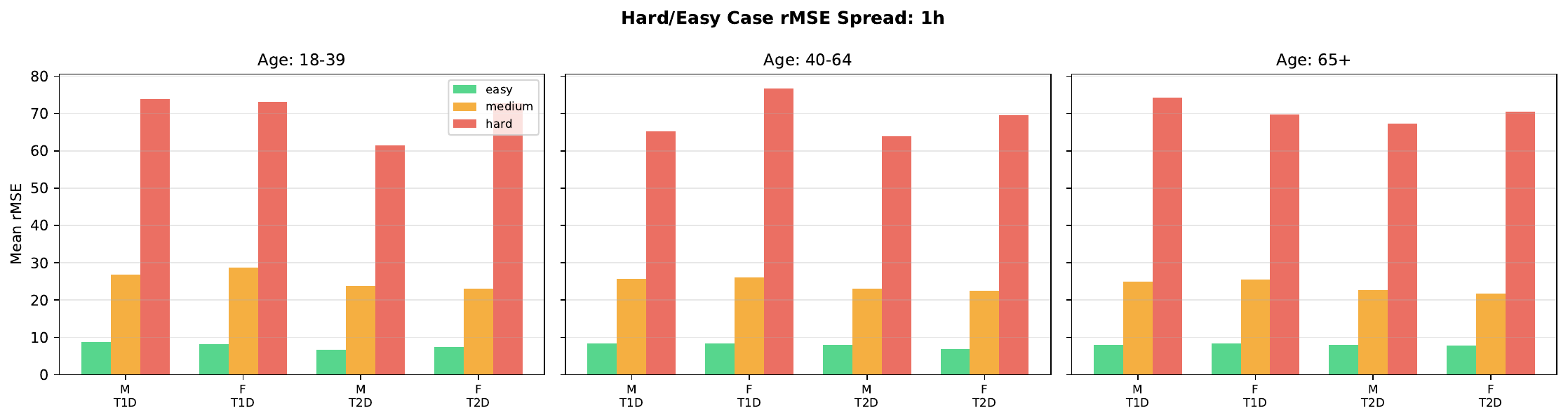}
        \caption{1-hour rMSE}
        \label{fig:hardeasy_spread_1h_rmse}
    \end{subfigure}

    \vspace{1em}

    \begin{subfigure}[t]{0.48\textwidth}
        \centering
        \includegraphics[width=\linewidth]{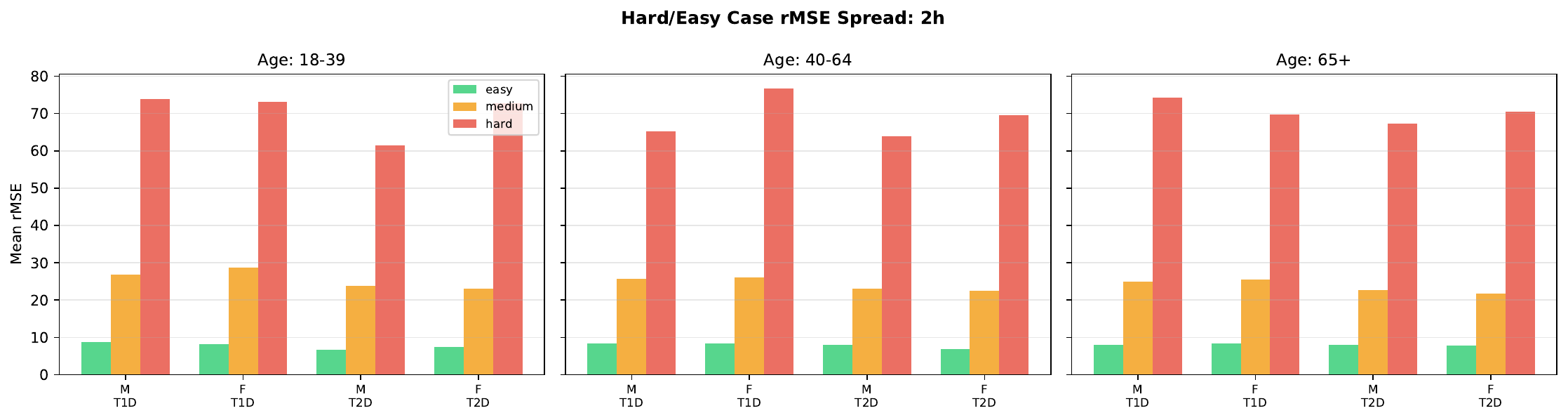}
        \caption{2-hour rMSE}
        \label{fig:hardeasy_spread_2h_rmse}
    \end{subfigure}
    \hfill
    \begin{subfigure}[t]{0.48\textwidth}
        \centering
        \includegraphics[width=\linewidth]{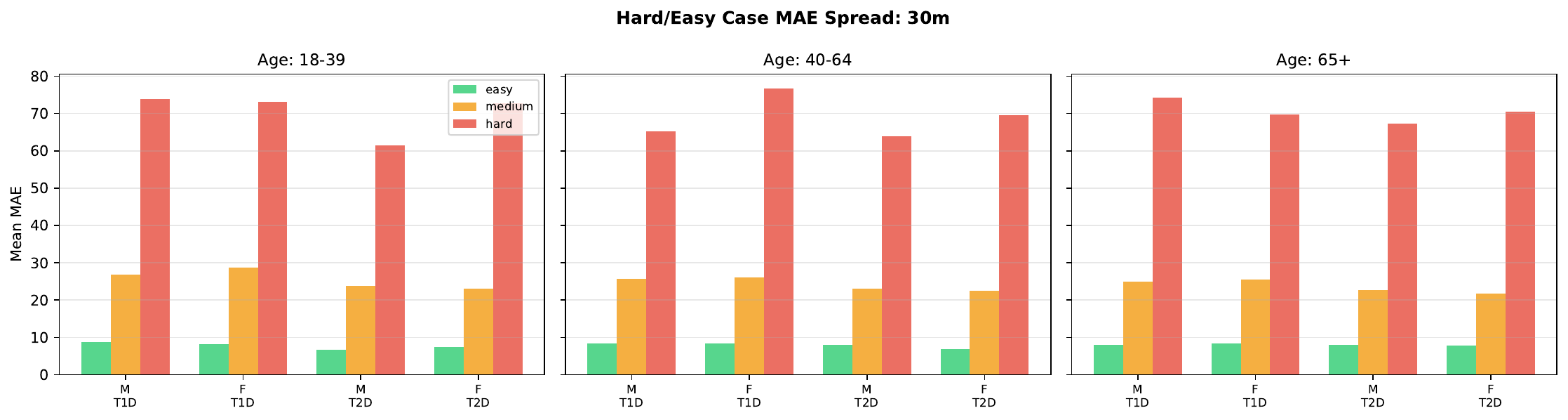}
        \caption{30-minute MAE}
        \label{fig:hardeasy_spread_30m_mae}
    \end{subfigure}

    \vspace{1em}

    \begin{subfigure}[t]{0.48\textwidth}
        \centering
        \includegraphics[width=\linewidth]{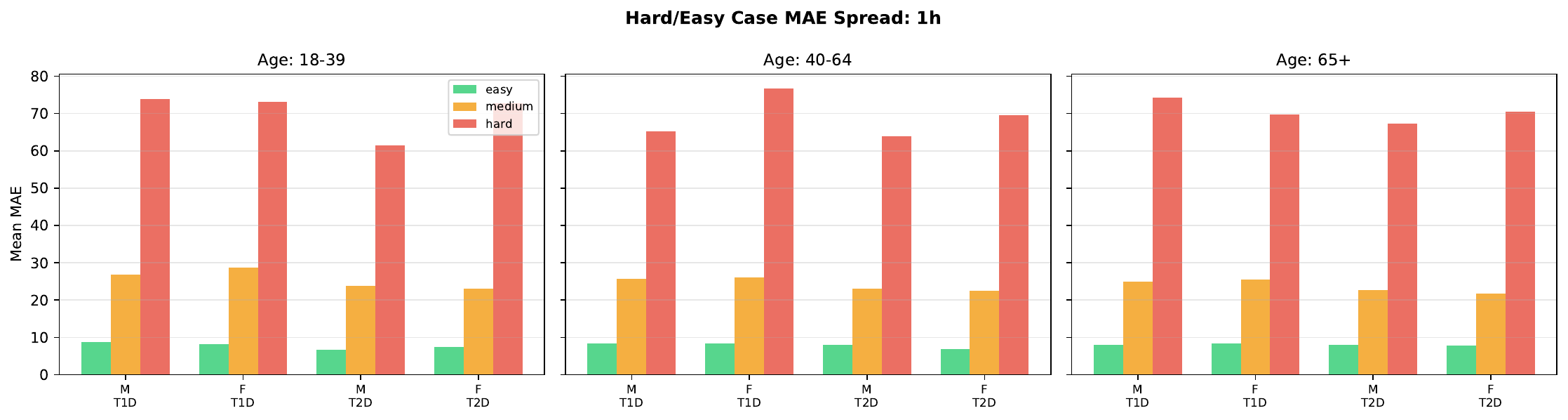}
        \caption{1-hour MAE}
        \label{fig:hardeasy_spread_1h_mae}
    \end{subfigure}
    \hfill
    \begin{subfigure}[t]{0.48\textwidth}
        \centering
        \includegraphics[width=\linewidth]{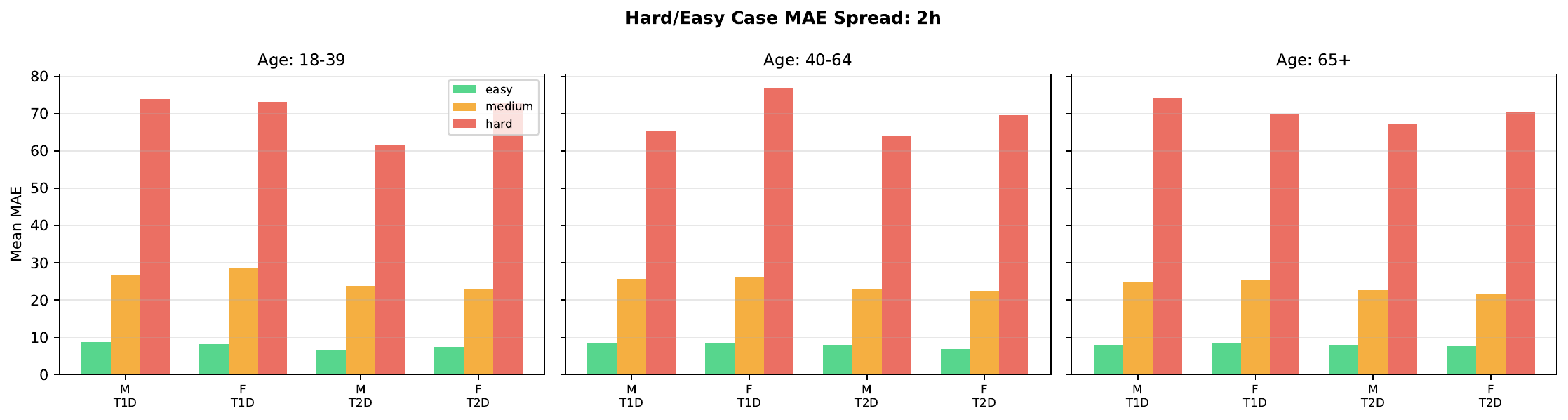}
        \caption{2-hour MAE}
        \label{fig:hardeasy_spread_2h_mae}
    \end{subfigure}

    \caption{Instance-level difficulty analysis: subgroup-level hard/easy case distribution and rMSE/MAE spread across prediction horizons.}
    \label{fig:hardeasy_all}
\end{figure}

\begin{figure}[htbp]
    \centering
    \begin{subfigure}[t]{0.32\textwidth}
        \centering
        \includegraphics[width=\linewidth]{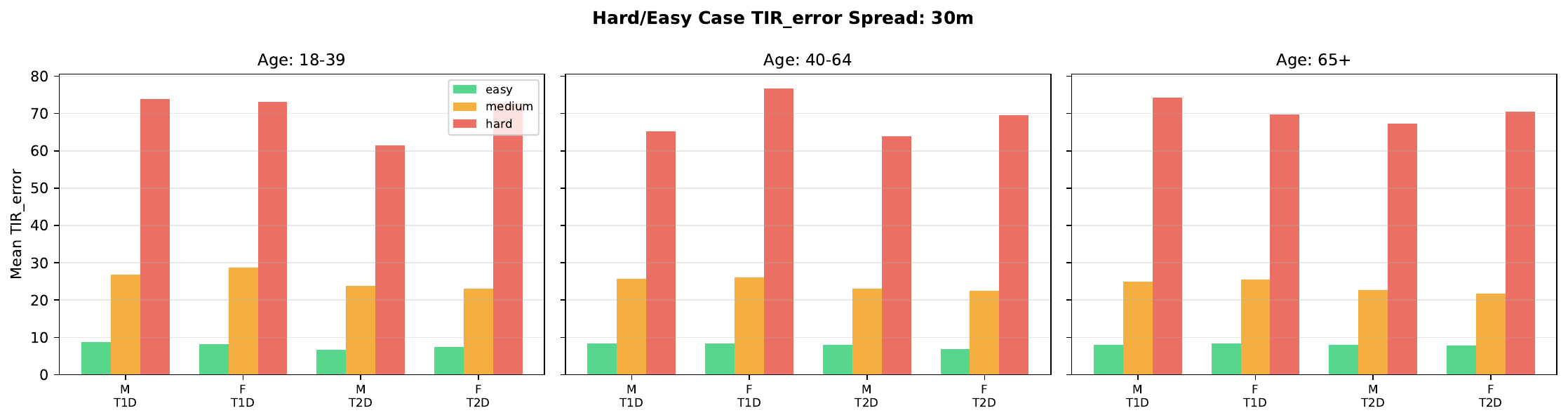}
        \caption{30-minute TIR-AE}
        \label{fig:hardeasy_spread_30m_tir}
    \end{subfigure}
    \hfill
    \begin{subfigure}[t]{0.32\textwidth}
        \centering
        \includegraphics[width=\linewidth]{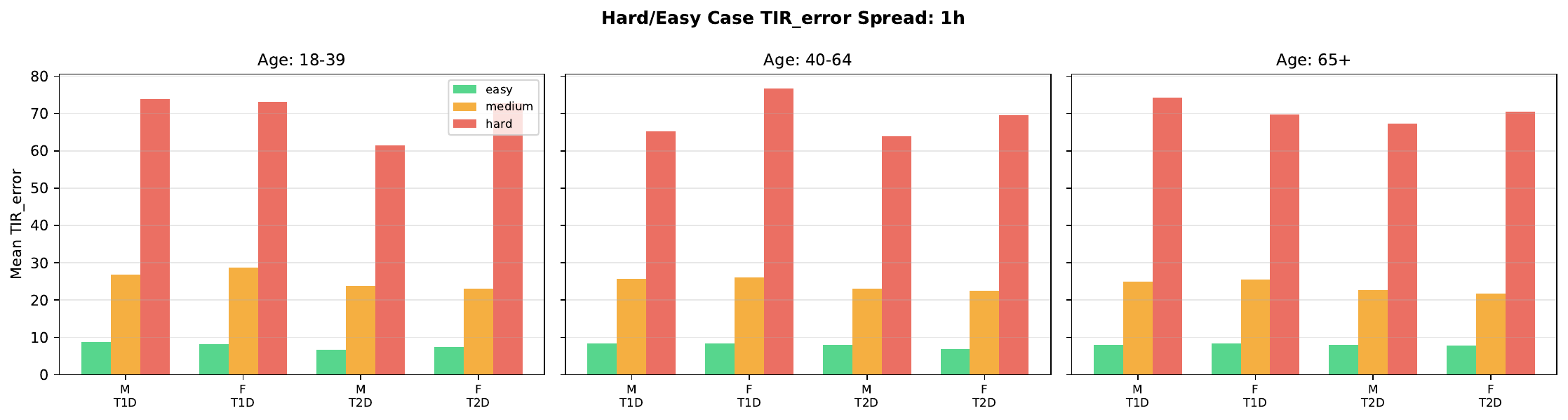}
        \caption{1-hour TIR-AE}
        \label{fig:hardeasy_spread_1h_tir}
    \end{subfigure}
    \hfill
    \begin{subfigure}[t]{0.32\textwidth}
        \centering
        \includegraphics[width=\linewidth]{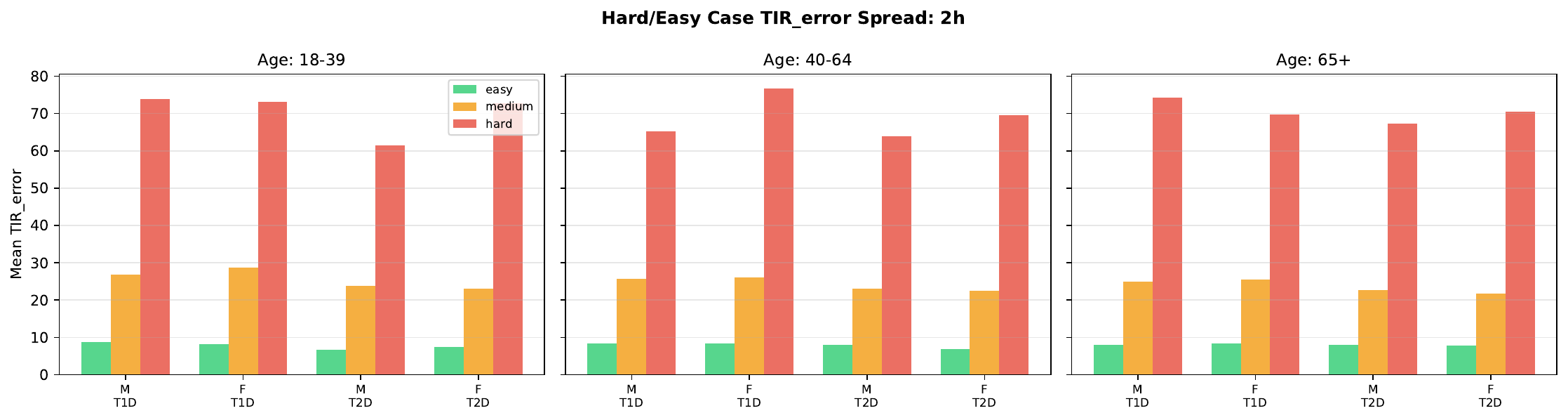}
        \caption{2-hour TIR-AE}
        \label{fig:hardeasy_spread_2h_tir}
    \end{subfigure}
    \caption{Instance-level difficulty analysis for TIR Absolute Error across prediction horizons.}
    \label{fig:hardeasy_tir}
\end{figure}

Figure~\ref{fig:hardeasy_all}, panels (a)--(f), covers rMSE and MAE at all three horizons, and Figure~\ref{fig:hardeasy_tir}, panels (a)--(c), covers TIR Absolute Error. Together they show the subgroup-level distribution of hard and easy cases and the corresponding rMSE spread across demographic strata for each horizon-metric combination. The key finding from the main text---that subgroup disparities align with the proportion of hard cases---holds consistently across all horizons. Young male T1D and middle-aged female T1D groups maintain the highest hard-case proportions regardless of the prediction window.

\subsection{Subgroup-Level Performance Tables}

Tables~\ref{tab:appendix_2h_rmse_id}--\ref{tab:appendix_2h_mae_od} present the full subgroup-disaggregated performance for all 33 models at the 2-hour prediction horizon, evaluated on both in-distribution (Test-ID) and out-of-distribution (Test-OD) splits. Models are grouped by family (Statistical, ML, Neural, API) and sorted by overall rMSE within each family. These tables complement the main results (Table~1 of the main text) by providing per-subgroup breakdowns by diabetes type (T1D, T2D), gender (Female, Male), and age group (18--39, 40--64, 65+).

\begin{table}[!htbp]
    \centering
    \caption{2-hour rMSE (mg/dL) across subgroups -- Test-ID. Lower values indicate better prediction accuracy. T1D patients consistently show higher rMSE than T2D across all models, reflecting greater glucose variability.}
    \label{tab:appendix_2h_rmse_id}
    \small
    \input{0-display/AppendixTable/appendix_table_2h_rMSE_test-id}
\end{table}

\begin{table}[!htbp]
    \centering
    \caption{2-hour rMSE (mg/dL) across subgroups -- Test-OD. Out-of-distribution evaluation on held-out patients. OD/ID ratios near 1.0 indicate robust generalization; subgroup-level variation reveals hidden heterogeneity.}
    \label{tab:appendix_2h_rmse_od}
    \small
    \input{0-display/AppendixTable/appendix_table_2h_rMSE_test-od}
\end{table}

\clearpage

\begin{table}[!htbp]
    \centering
    \caption{2-hour MAE (mg/dL) across subgroups -- Test-ID. Mean absolute error provides a complementary view to rMSE, being less sensitive to large outlier predictions.}
    \label{tab:appendix_2h_mae_id}
    \small
    \input{0-display/AppendixTable/appendix_table_2h_MAE_test-id}
\end{table}

\begin{table}[!htbp]
    \centering
    \caption{2-hour MAE (mg/dL) across subgroups -- Test-OD. The same subgroup disparity patterns observed in rMSE persist in MAE, confirming the robustness of the findings across metrics.}
    \label{tab:appendix_2h_mae_od}
    \small
    \input{0-display/AppendixTable/appendix_table_2h_MAE_test-od}
\end{table}

\clearpage


\clearpage

%% file: 0-display/AppendixTable/appendix_table_2h_rMSE_test-id.tex

\begin{tabular}{lcccccccc}
\toprule
\textbf{Model} & \textbf{All} & \textbf{T1D} & \textbf{T2D} & \textbf{Female} & \textbf{Male} & \textbf{18-39} & \textbf{40-64} & \textbf{65+} \\
\midrule
\multicolumn{9}{l}{\textit{Statistical}} \\
ARIMA & 27.39 & 30.84 & 23.94 & 27.36 & 27.42 & 28.48 & 26.93 & 26.70 \\
Naive & 29.03 & 32.35 & 25.70 & 28.96 & 29.10 & 29.96 & 28.84 & 28.19 \\
AutoETS & 29.81 & 33.61 & 26.01 & 29.87 & 29.75 & 30.79 & 29.81 & 28.69 \\
\midrule
\multicolumn{9}{l}{\textit{ML}} \\
Linear & 25.87 & 29.15 & 22.60 & 25.79 & 25.96 & 27.13 & 25.70 & 24.65 \\
LightGBM & 25.89 & 29.38 & 22.40 & 25.65 & 26.13 & 27.24 & 25.73 & 24.55 \\
XGBoost & 27.66 & 31.10 & 24.22 & 27.84 & 27.47 & 29.02 & 27.53 & 26.27 \\
CatBoost & 28.47 & 31.51 & 25.42 & 28.48 & 28.46 & 29.43 & 28.20 & 27.69 \\
\midrule
\multicolumn{9}{l}{\textit{Neural}} \\
PatchTST & 25.54 & 28.75 & 22.33 & 25.41 & 25.67 & 26.48 & 25.56 & 24.44 \\
NS-Trans. & 25.56 & 28.63 & 22.49 & 25.49 & 25.63 & 26.68 & 25.50 & 24.35 \\
Transformer & 25.57 & 28.70 & 22.44 & 25.47 & 25.67 & 26.73 & 25.44 & 24.40 \\
TimeXer & 25.61 & 28.75 & 22.46 & 25.49 & 25.73 & 26.69 & 25.49 & 24.52 \\
Crossformer & 25.66 & 28.75 & 22.56 & 25.66 & 25.65 & 26.64 & 25.54 & 24.68 \\
Informer & 25.80 & 28.98 & 22.62 & 25.67 & 25.92 & 26.94 & 25.58 & 24.76 \\
TFT & 25.82 & 28.91 & 22.74 & 25.68 & 25.97 & 26.71 & 25.63 & 25.04 \\
MICN & 25.97 & 29.10 & 22.84 & 25.76 & 26.18 & 27.06 & 25.91 & 24.81 \\
DLinear & 26.21 & 29.39 & 23.03 & 25.95 & 26.48 & 27.40 & 25.95 & 25.17 \\
TimesNet & 26.23 & 29.69 & 22.78 & 26.18 & 26.29 & 27.58 & 26.02 & 24.95 \\
SCINet & 26.30 & 29.43 & 23.17 & 26.09 & 26.51 & 27.55 & 26.00 & 25.24 \\
iTransformer & 26.50 & 29.85 & 23.15 & 26.19 & 26.81 & 27.55 & 26.57 & 25.21 \\
TSMixer & 26.58 & 29.85 & 23.31 & 26.38 & 26.78 & 27.54 & 26.61 & 25.46 \\
TiDE & 26.60 & 29.81 & 23.40 & 26.29 & 26.91 & 27.72 & 26.32 & 25.66 \\
TimeMixer & 26.61 & 29.80 & 23.42 & 26.35 & 26.88 & 27.74 & 26.45 & 25.53 \\
N-HiTS & 26.65 & 30.16 & 23.13 & 26.21 & 27.08 & 27.57 & 26.64 & 25.61 \\
N-BEATS & 27.24 & 30.67 & 23.81 & 26.96 & 27.52 & 28.25 & 26.90 & 26.51 \\
FEDformer & 30.53 & 34.49 & 26.58 & 30.31 & 30.76 & 31.48 & 30.06 & 30.02 \\
Autoformer & 34.85 & 39.57 & 30.13 & 34.66 & 35.04 & 35.52 & 33.74 & 35.41 \\
Chronos2 & 39.05 & 44.17 & 33.94 & 38.58 & 39.52 & 40.12 & 37.63 & 39.53 \\
\midrule
\multicolumn{9}{l}{\textit{API}} \\
TimeGPT & 26.54 & 29.83 & 23.25 & 26.52 & 26.57 & 27.52 & 26.30 & 25.72 \\
GPT-5.1 & 27.27 & 30.46 & 24.08 & 27.21 & 27.33 & 28.67 & 27.36 & 25.58 \\
GPT-5 Mini & 27.57 & 30.74 & 24.41 & 27.57 & 27.58 & 28.50 & 27.64 & 26.43 \\
Claude 4.5 & 27.89 & 30.67 & 25.11 & 28.19 & 27.60 & 28.50 & 28.66 & 26.29 \\
Claude 4.5H & 28.47 & 31.48 & 25.46 & 28.50 & 28.44 & 29.62 & 28.54 & 27.08 \\
Gemini 3 & 30.70 & 33.50 & 27.89 & 30.66 & 30.74 & 32.12 & 30.53 & 29.27 \\
\bottomrule
\end{tabular}

%% file: 0-display/AppendixTable/appendix_table_2h_rMSE_test-od.tex

\begin{tabular}{lcccccccc}
\toprule
\textbf{Model} & \textbf{All} & \textbf{T1D} & \textbf{T2D} & \textbf{Female} & \textbf{Male} & \textbf{18-39} & \textbf{40-64} & \textbf{65+} \\
\midrule
\multicolumn{9}{l}{\textit{Statistical}} \\
ARIMA & 27.50 & 30.73 & 24.26 & 28.43 & 26.57 & 27.04 & 29.16 & 26.33 \\
Naive & 29.02 & 32.49 & 25.55 & 30.11 & 27.93 & 28.86 & 30.73 & 27.51 \\
AutoETS & 29.63 & 33.08 & 26.18 & 31.00 & 28.26 & 29.76 & 31.24 & 27.92 \\
\midrule
\multicolumn{9}{l}{\textit{ML}} \\
LightGBM & 26.02 & 28.76 & 23.29 & 27.02 & 25.03 & 25.94 & 27.44 & 24.72 \\
Linear & 26.20 & 28.70 & 23.70 & 27.24 & 25.17 & 26.47 & 27.31 & 24.85 \\
XGBoost & 27.96 & 30.86 & 25.06 & 28.98 & 26.93 & 27.12 & 29.94 & 26.86 \\
CatBoost & 28.46 & 31.61 & 25.31 & 29.64 & 27.28 & 28.77 & 29.93 & 26.69 \\
\midrule
\multicolumn{9}{l}{\textit{Neural}} \\
NS-Trans. & 25.69 & 28.29 & 23.08 & 26.56 & 24.82 & 25.29 & 27.28 & 24.53 \\
Transformer & 25.74 & 28.43 & 23.05 & 26.58 & 24.89 & 25.47 & 27.48 & 24.29 \\
TimeXer & 25.89 & 28.31 & 23.46 & 26.76 & 25.02 & 25.25 & 27.45 & 25.00 \\
PatchTST & 25.97 & 28.60 & 23.34 & 26.75 & 25.19 & 25.49 & 27.57 & 24.89 \\
Crossformer & 26.04 & 28.65 & 23.44 & 26.92 & 25.16 & 25.47 & 27.66 & 25.03 \\
Informer & 26.08 & 28.60 & 23.56 & 26.91 & 25.25 & 25.30 & 27.72 & 25.27 \\
TFT & 26.21 & 28.66 & 23.76 & 27.11 & 25.30 & 25.71 & 27.86 & 25.09 \\
MICN & 26.25 & 28.82 & 23.68 & 27.14 & 25.35 & 25.67 & 27.85 & 25.26 \\
iTransformer & 26.49 & 29.14 & 23.84 & 27.28 & 25.70 & 26.02 & 27.81 & 25.67 \\
TimesNet & 26.57 & 29.21 & 23.92 & 27.32 & 25.81 & 25.77 & 28.01 & 25.97 \\
DLinear & 26.59 & 29.18 & 24.00 & 27.35 & 25.84 & 25.87 & 28.11 & 25.85 \\
SCINet & 26.70 & 29.21 & 24.18 & 27.56 & 25.84 & 26.06 & 28.19 & 25.89 \\
N-HiTS & 26.91 & 29.97 & 23.84 & 27.70 & 26.11 & 26.61 & 27.97 & 26.17 \\
TiDE & 26.99 & 29.61 & 24.37 & 27.82 & 26.17 & 26.13 & 28.52 & 26.37 \\
TSMixer & 27.01 & 29.85 & 24.18 & 27.70 & 26.33 & 26.33 & 28.46 & 26.29 \\
TimeMixer & 27.25 & 30.08 & 24.42 & 28.08 & 26.42 & 26.68 & 28.84 & 26.26 \\
N-BEATS & 27.61 & 30.46 & 24.76 & 28.35 & 26.87 & 27.05 & 29.10 & 26.72 \\
FEDformer & 31.81 & 34.85 & 28.76 & 32.86 & 30.75 & 30.56 & 33.29 & 31.62 \\
Autoformer & 36.42 & 39.85 & 32.99 & 38.02 & 34.82 & 34.19 & 38.31 & 36.82 \\
Chronos2 & 40.34 & 44.26 & 36.42 & 41.70 & 38.98 & 37.59 & 41.78 & 41.72 \\
\midrule
\multicolumn{9}{l}{\textit{API}} \\
TimeGPT & 26.72 & 29.65 & 23.80 & 28.06 & 25.39 & 26.41 & 28.47 & 25.32 \\
GPT-5.1 & 27.86 & 30.77 & 24.95 & 28.85 & 26.87 & 27.12 & 30.30 & 26.23 \\
GPT-5 Mini & -- & -- & -- & -- & 26.61 & 27.31 & -- & 26.02 \\
Claude 4.5 & 28.51 & 31.07 & 25.95 & 29.54 & 27.48 & 28.50 & 30.77 & 26.31 \\
Claude 4.5H & 29.07 & 32.08 & 26.06 & 30.12 & 28.03 & 28.68 & 31.14 & 27.44 \\
Gemini 3 & 31.88 & 34.80 & 28.95 & 33.11 & 30.64 & 30.42 & 34.84 & 30.44 \\
\bottomrule
\end{tabular}

%% file: 0-display/AppendixTable/appendix_table_2h_MAE_test-id.tex

\begin{tabular}{lcccccccc}
\toprule
\textbf{Model} & \textbf{All} & \textbf{T1D} & \textbf{T2D} & \textbf{Female} & \textbf{Male} & \textbf{18-39} & \textbf{40-64} & \textbf{65+} \\
\midrule
\multicolumn{9}{l}{\textit{Statistical}} \\
ARIMA & 23.35 & 26.22 & 20.49 & 23.31 & 23.39 & 24.27 & 22.95 & 22.78 \\
Naive & 24.78 & 27.55 & 22.01 & 24.67 & 24.90 & 25.54 & 24.61 & 24.13 \\
AutoETS & 25.24 & 28.35 & 22.12 & 25.24 & 25.23 & 26.05 & 25.26 & 24.28 \\
\midrule
\multicolumn{9}{l}{\textit{ML}} \\
Linear & 22.03 & 24.77 & 19.29 & 21.99 & 22.07 & 23.12 & 21.85 & 20.99 \\
LightGBM & 21.97 & 24.90 & 19.03 & 21.84 & 22.10 & 23.15 & 21.83 & 20.79 \\
XGBoost & 23.40 & 26.28 & 20.52 & 23.57 & 23.23 & 24.60 & 23.27 & 22.18 \\
CatBoost & 24.26 & 26.78 & 21.74 & 24.26 & 24.26 & 25.02 & 24.03 & 23.66 \\
\midrule
\multicolumn{9}{l}{\textit{Neural}} \\
PatchTST & 21.87 & 24.58 & 19.17 & 21.81 & 21.93 & 22.70 & 21.83 & 20.97 \\
NS-Trans. & 21.78 & 24.34 & 19.23 & 21.78 & 21.79 & 22.79 & 21.70 & 20.75 \\
Transformer & 21.83 & 24.42 & 19.23 & 21.78 & 21.87 & 22.85 & 21.68 & 20.84 \\
TimeXer & 21.95 & 24.59 & 19.31 & 21.90 & 22.00 & 22.88 & 21.79 & 21.07 \\
Crossformer & 21.88 & 24.45 & 19.31 & 21.93 & 21.83 & 22.74 & 21.74 & 21.08 \\
Informer & 21.99 & 24.66 & 19.32 & 21.92 & 22.05 & 23.02 & 21.74 & 21.11 \\
TFT & 21.97 & 24.50 & 19.44 & 21.91 & 22.04 & 22.73 & 21.78 & 21.34 \\
MICN & 22.13 & 24.75 & 19.50 & 22.00 & 22.26 & 23.11 & 21.99 & 21.18 \\
DLinear & 22.38 & 25.05 & 19.71 & 22.18 & 22.58 & 23.44 & 22.07 & 21.54 \\
TimesNet & 22.53 & 25.52 & 19.55 & 22.52 & 22.55 & 23.73 & 22.24 & 21.53 \\
SCINet & 22.41 & 25.01 & 19.80 & 22.27 & 22.54 & 23.53 & 22.06 & 21.54 \\
iTransformer & 22.81 & 25.63 & 19.98 & 22.56 & 23.05 & 23.76 & 22.85 & 21.68 \\
TSMixer & 22.95 & 25.70 & 20.20 & 22.84 & 23.07 & 23.82 & 22.90 & 22.03 \\
TiDE & 22.84 & 25.54 & 20.15 & 22.61 & 23.07 & 23.82 & 22.52 & 22.12 \\
TimeMixer & 22.99 & 25.70 & 20.28 & 22.80 & 23.18 & 23.98 & 22.80 & 22.09 \\
N-HiTS & 22.98 & 25.97 & 19.98 & 22.65 & 23.30 & 23.82 & 22.88 & 22.13 \\
N-BEATS & 23.72 & 26.68 & 20.77 & 23.50 & 23.94 & 24.52 & 23.39 & 23.22 \\
FEDformer & 27.19 & 30.75 & 23.62 & 27.04 & 27.34 & 27.98 & 26.74 & 26.82 \\
Autoformer & 31.30 & 35.68 & 26.92 & 31.14 & 31.46 & 31.81 & 30.16 & 32.07 \\
Chronos2 & 36.04 & 40.79 & 31.29 & 35.55 & 36.53 & 36.98 & 34.55 & 36.73 \\
\midrule
\multicolumn{9}{l}{\textit{API}} \\
TimeGPT & 22.51 & 25.26 & 19.76 & 22.49 & 22.52 & 23.36 & 22.29 & 21.81 \\
GPT-5.1 & 23.05 & 25.72 & 20.39 & 23.04 & 23.07 & 24.29 & 23.10 & 21.58 \\
GPT-5 Mini & 23.22 & 25.82 & 20.62 & 23.22 & 23.22 & 23.97 & 23.32 & 22.24 \\
Claude 4.5 & 23.48 & 25.75 & 21.20 & 23.78 & 23.18 & 23.97 & 24.20 & 22.06 \\
Claude 4.5H & 23.96 & 26.43 & 21.49 & 24.01 & 23.91 & 24.94 & 24.02 & 22.77 \\
Gemini 3 & 25.94 & 28.26 & 23.63 & 25.95 & 25.93 & 27.21 & 25.83 & 24.64 \\
\bottomrule
\end{tabular}

%% file: 0-display/AppendixTable/appendix_table_2h_MAE_test-od.tex

\begin{tabular}{lcccccccc}
\toprule
\textbf{Model} & \textbf{All} & \textbf{T1D} & \textbf{T2D} & \textbf{Female} & \textbf{Male} & \textbf{18-39} & \textbf{40-64} & \textbf{65+} \\
\midrule
\multicolumn{9}{l}{\textit{Statistical}} \\
ARIMA & 23.51 & 26.25 & 20.76 & 24.27 & 22.75 & 23.00 & 24.98 & 22.58 \\
Naive & 24.85 & 27.80 & 21.89 & 25.75 & 23.95 & 24.59 & 26.30 & 23.69 \\
AutoETS & 25.16 & 28.09 & 22.24 & 26.36 & 23.97 & 25.29 & 26.48 & 23.75 \\
\midrule
\multicolumn{9}{l}{\textit{ML}} \\
LightGBM & 22.10 & 24.42 & 19.78 & 22.95 & 21.24 & 22.00 & 23.26 & 21.06 \\
Linear & 22.37 & 24.47 & 20.26 & 23.27 & 21.46 & 22.60 & 23.29 & 21.23 \\
XGBoost & 23.72 & 26.19 & 21.25 & 24.60 & 22.84 & 22.85 & 25.37 & 22.99 \\
CatBoost & 24.32 & 26.95 & 21.68 & 25.33 & 23.30 & 24.57 & 25.50 & 22.89 \\
\midrule
\multicolumn{9}{l}{\textit{Neural}} \\
NS-Trans. & 21.92 & 24.13 & 19.71 & 22.68 & 21.16 & 21.57 & 23.28 & 20.94 \\
Transformer & 22.01 & 24.27 & 19.75 & 22.72 & 21.30 & 21.73 & 23.52 & 20.80 \\
TimeXer & 22.23 & 24.28 & 20.19 & 23.01 & 21.46 & 21.64 & 23.54 & 21.55 \\
PatchTST & 22.25 & 24.47 & 20.02 & 22.91 & 21.58 & 21.80 & 23.58 & 21.39 \\
Crossformer & 22.26 & 24.47 & 20.05 & 23.03 & 21.49 & 21.75 & 23.61 & 21.45 \\
Informer & 22.27 & 24.38 & 20.15 & 22.99 & 21.55 & 21.51 & 23.66 & 21.67 \\
TFT & 22.36 & 24.42 & 20.30 & 23.17 & 21.56 & 21.93 & 23.77 & 21.42 \\
MICN & 22.40 & 24.56 & 20.25 & 23.17 & 21.64 & 21.86 & 23.77 & 21.61 \\
iTransformer & 22.81 & 25.06 & 20.56 & 23.50 & 22.12 & 22.37 & 23.89 & 22.20 \\
TimesNet & 22.85 & 25.11 & 20.58 & 23.50 & 22.19 & 22.10 & 24.01 & 22.45 \\
DLinear & 22.74 & 24.92 & 20.56 & 23.38 & 22.10 & 22.06 & 24.02 & 22.16 \\
SCINet & 22.77 & 24.88 & 20.67 & 23.53 & 22.02 & 22.17 & 24.02 & 22.15 \\
N-HiTS & 23.22 & 25.87 & 20.57 & 23.91 & 22.54 & 22.91 & 24.08 & 22.70 \\
TiDE & 23.21 & 25.43 & 20.99 & 23.92 & 22.50 & 22.38 & 24.50 & 22.77 \\
TSMixer & 23.32 & 25.75 & 20.90 & 23.87 & 22.77 & 22.64 & 24.58 & 22.77 \\
TimeMixer & 23.62 & 26.07 & 21.18 & 24.33 & 22.91 & 23.08 & 25.00 & 22.82 \\
N-BEATS & 24.04 & 26.53 & 21.56 & 24.67 & 23.42 & 23.53 & 25.28 & 23.35 \\
FEDformer & 28.41 & 31.11 & 25.71 & 29.33 & 27.50 & 27.14 & 29.64 & 28.50 \\
Autoformer & 32.82 & 35.90 & 29.74 & 34.28 & 31.35 & 30.61 & 34.46 & 33.45 \\
Chronos2 & 37.34 & 40.96 & 33.72 & 38.57 & 36.10 & 34.47 & 38.55 & 39.06 \\
\midrule
\multicolumn{9}{l}{\textit{API}} \\
TimeGPT & 22.72 & 25.20 & 20.24 & 23.84 & 21.60 & 22.40 & 24.22 & 21.56 \\
GPT-5.1 & 23.58 & 26.07 & 21.08 & 24.38 & 22.77 & 22.83 & 25.67 & 22.27 \\
GPT-5 Mini & -- & -- & -- & -- & 22.42 & 23.01 & -- & 21.94 \\
Claude 4.5 & 24.07 & 26.29 & 21.85 & 24.97 & 23.17 & 24.10 & 25.98 & 22.16 \\
Claude 4.5H & 24.52 & 27.08 & 21.96 & 25.43 & 23.61 & 24.17 & 26.30 & 23.13 \\
Gemini 3 & 27.01 & 29.52 & 24.50 & 28.10 & 25.91 & 25.71 & 29.63 & 25.75 \\
\bottomrule
\end{tabular}

%% file: 0-sections/C_experiment_details.tex
\section{Experiment Implementation Details}

\subsection{Compute Resources}

All experiments were conducted on a single compute node equipped with 2 NVIDIA RTX 4090 (24GB) GPUs and 128GB RAM. We used \texttt{Optuna} for hyperparameter tuning and retained the best-performing configurations based on validation performance.
Model training was implemented using the \texttt{Nixtla} forecasting library (StatsForecast and MLForecast) for statistical and machine learning models, and the Time-Series-Library (TSLib) framework for 20 neural time-series architectures with fixed hyperparameters and 20-epoch training with early stopping (patience 5). For language model-based experiments, we used API-based inference with deterministic settings.

Training time varied depending on model complexity and dataset size. The shallow models, including ARIMA and linear regression, completed training within 1 hour on the FairGlucose splits. Among the deep learning models, N-HiTS was among the most efficient, requiring less than 2 hours in the largest FairGlucose training runs. Transformer-based models such as Transformer and Autoformer were the most computationally intensive, with training times reaching up to 24 hours.
All models were trained with early stopping and batch-level validation to ensure computational efficiency without sacrificing performance.

\subsection{Hyperparameters}

\subsubsection{Statistical Models}
We implemented statistical time series forecasting models using the \texttt{StatsForecast} library \cite{garza2022statsforecast}  and optimized their hyperparameters with \texttt{Optuna} \cite{akiba2019optuna}. The optimization aimed to minimize the root mean square error (RMSE) on a validation set by performing a direct hyperparameter search rather than nested auto-model optimization, reducing computational cost substantially. For example, the direct ARIMA optimization takes approximately 30 seconds per trial, compared to 2--5 minutes for \texttt{ARIMA}. 

The following models were considered in our statistical forecasting framework.
\textbf{ARIMA} \cite{yang2018arima} involved direct optimization of the $(p, d, q)$ and $(P, D, Q)$ parameters.
\textbf{Naive} served as a baseline model without tunable hyperparameters.

The corresponding hyperparameter search spaces and optimal parameters were defined as follows:
\begin{itemize}
\item \textbf{ARIMA}: We searched over $p \in [0, 5]$, $d \in [0, 2]$, $q \in [0, 5]$.
Seasonal parameters were disabled ($P = D = Q = 0$) to focus on non-seasonal patterns.
The optimal configuration was: $p = 2$, $d = 1$, $q = 1$.
\item \textbf{Naive}: This model had no tunable hyperparameters.
\end{itemize}

\subsubsection{Machine Learning Forecasting}
This study uses the Nixtla MLForecast framework \cite{olivares2022library_neuralforecast} to implement machine-learning models for time-series forecasting. MLForecast is a scalable, production-grade library that automatically converts time-series data into supervised learning features (lags, rolling statistics, date features), then fits any estimator following the \texttt{scikit-learn} API (having \texttt{fit}/\texttt{predict} methods).
The library handles recursive forecasting and feature updates over the prediction horizon, and supports horizontal scaling via \texttt{pandas}, \texttt{polars}, \texttt{Dask}, \texttt{Spark}, or \texttt{Ray} execution \cite{olivares2022library_neuralforecast}.
\texttt{MLForecast} automates much of the feature engineering process: it generates lag features, applies lag-based transformations, and extracts relevant date-based features. 
The user only needs to specify the desired regressors and configuration options, while the library manages updates and iterations over time steps internally.

The models evaluated in this work were selected from standard regression algorithms compatible with MLForecast:  
\texttt{LinearRegression} (ordinary least squares) \cite{scikit-learn},  
\texttt{lightgbm.LGBMRegressor} \cite{ke2017lightgbm}, 
\texttt{xgboost.XGBRegressor} \cite{chen2016xgboost}, and 
\texttt{catboost.CatBoostRegressor} \cite{prokhorenkova2018catboost}.  
These are instantiated via MLForecast by passing a list of regressor instances. MLForecast then wraps them into a forecasting pipeline with lag-feature and date-feature generation.

For each model, we defined a model‑specific Optuna hyperparameter search space (see the code listing), including, e.g.,  
\texttt{n\_estimators}, \texttt{max\_depth}, \texttt{learning\_rate} for boosting;  
\texttt{l2\_leaf\_reg}, \texttt{border\_count} for CatBoost;  
and \texttt{fit\_intercept} for LinearRegression.  
The objective function trains MLForecast with the current trial’s parameters on the training split (using the full set of lag features up to \texttt{INPUT\_LENGTH}) and evaluates on the validation split using recursive multi‑step forecasts and average root‑mean‑squared error (rMSE).  A fixed lag vector of \texttt{lags = [1, 2, ..., INPUT\_LENGTH]} is used without additional lag transforms.

Hyperparameter search spaces and optimal parameters used in this work:
\begin{itemize}
  \item \textbf{LinearRegression}: \texttt{fit\_intercept}~$\in \{\text{True}, \text{False}\}$ and \texttt{n\_jobs} = $-1$. 
The optimal configuration was: \texttt{fit\_intercept} = \text{True}, \texttt{n\_jobs} = $-1$.\\
  \item \textbf{LightGBM}: \texttt{n\_estimators}~$\in [50, 500]$ (step 50); \texttt{max\_depth}~$\in [3, 15]$; 
\texttt{learning\_rate}~$\in [0.01, 0.3]$; \texttt{num\_leaves}~$\in [10, 300]$ (step 10); 
\texttt{min\_child\_samples}~$\in [5, 100]$ (step 5); \texttt{subsample}~$\in [0.5, 1.0]$; 
\texttt{colsample\_bytree}~$\in [0.5, 1.0]$; \texttt{reg\_alpha}~$\in [0.0, 10.0]$; 
\texttt{reg\_lambda}~$\in [0.0, 10.0]$. \\
The optimal configuration was: \texttt{n\_estimators} = 300, \texttt{max\_depth} = 10,
\texttt{learning\_rate} = 0.04, \texttt{num\_leaves} = 180,
\texttt{min\_child\_samples} = 75, \texttt{subsample} = 0.7,
\texttt{colsample\_bytree} = 0.5, \texttt{reg\_alpha} = 6.5, and
\texttt{reg\_lambda} = 4.5.\\
  \item \textbf{XGBoost}: same as LightGBM plus \texttt{gamma}~$\in [0.0, 5.0]$ and \texttt{min\_child\_weight}~$\in [1, 10]$. 
The optimal configuration was: \texttt{n\_estimators} = 50, \texttt{max\_depth} = 5, \texttt{learning\_rate} = 0.2,
\texttt{subsample} = 0.9, \texttt{colsample\_bytree} = 0.7, \texttt{reg\_alpha} = 4.25, \texttt{reg\_lambda} = 3.5,
\texttt{gamma} = 0.6, and \texttt{min\_child\_weight} = 7.\\
  \item \textbf{CatBoost}: \texttt{iterations}~$\in [50, 500]$ (step 50); 
\texttt{depth}~$\in [3, 10]$ (step 1); 
\texttt{learning\_rate}~$\in [0.01, 0.3]$ (step 0.01); 
\texttt{l2\_leaf\_reg}~$\in [0.1, 10.0]$ (step 0.5); 
\texttt{border\_count}~$\in [32, 255]$ (step 32). 
The optimal configuration was: \texttt{iterations} = 350,
\texttt{depth} = 3,
\texttt{learning\_rate} = 0.28,
\texttt{l2\_leaf\_reg} = 8.1, and
\texttt{border\_count} = 32.
\end{itemize}

All models were trained using recursive forecasts over the horizon (\texttt{OUTPUT\_LENGTH}), matching the temporal frequency (\texttt{freq=5min}). Optuna optimization was run with a fixed number of 50 trials, using the validation rMSE as the objective. The final model for each best trial was refit using MLForecast with the optimal parameter set, yielding a MLForecast model object ready for out-of-sample forecasting and evaluation.

\subsubsection{Neural Time-Series Models (TSLib)}

We trained 20 neural time-series architectures using the Time-Series-Library (TSLib) framework, which provides a unified training pipeline across diverse architectures. All models use fixed hyperparameters with 20-epoch training and early stopping (patience 5), input length 288 (24 hours), and prediction length 24 for the main 2-hour evaluation. The architectures include attention-based models (PatchTST, iTransformer, TimeXer, Crossformer, NS-Transformer, Transformer, Informer, Autoformer, FEDformer, TFT), MLP/linear models (DLinear, TiDE, TSMixer, TimeMixer, N-BEATS, N-HiTS, MICN), convolutional models (TimesNet, SCINet), and the zero-shot foundation model Chronos2. Below we describe the hyperparameter configurations for representative architectures.

\paragraph{PatchTST (\texttt{PatchTST})} \cite{nie2022time}:
An efficient channel-independent transformer that uses sequence patching for scalable long-horizon forecasts \cite{nie2022time}. The Optuna search space included: \texttt{patch\_len}~$\in [8,32]$ (step 4); \texttt{stride}~$\in [4,16]$ (step 4); \texttt{hidden\_size}~$\in [64,512]$ (step 32); \texttt{n\_heads}~$\in \{2,4,8,16,32\}$; \texttt{dropout}~$\in [0.00,0.30]$ (step 0.05); \texttt{learning\_rate}~$\in \{1\text{e}^{-6}, 5\text{e}^{-6}, 1\text{e}^{-5}, \dots, 1\text{e}^{-2}\}$; \texttt{batch\_size}~$\in \{32,64,128,256\}$; and \texttt{max\_steps} derived from 10--25 epochs. 
The optimal configuration was: \texttt{patch\_len} = 8, \texttt{stride} = 8, \texttt{hidden\_size} = 320, \texttt{n\_heads} = 8, \texttt{dropout} = 0.1, \texttt{learning\_rate} = $2\text{e}^{-4}$, \texttt{batch\_size} = 256.

\paragraph{DLinear (\texttt{DLinear})} \cite{zeng2023transformers}:
A linear decomposition model with separate trend and seasonality paths, enabling quick convergence for long sequences \cite{zeng2023transformers}. 
The Optuna search space included: \texttt{learning\_rate}~$\in \{1\text{e}^{-6}, 5\text{e}^{-6}, 1\text{e}^{-5}, \dots, 1\text{e}^{-2}\}$; 
\texttt{batch\_size}~$\in \{32,64,128,256\}$; and \texttt{max\_steps} derived from 3--10 epochs. 
The optimal configuration was: \texttt{learning\_rate} = $1\text{e}^{-3}$, \texttt{batch\_size} = 64.

\paragraph{Transformer (\texttt{VanillaTransformer})} \cite{vaswani2017attention}: Standard encoder–decoder transformer with multi-head attention and an MLP decoder, serving as a baseline for modeling long-range dependencies \cite{vaswani2017attention}. The Optuna search space included: \texttt{hidden\_size} $\in [32,128]$ with a step of $32$; \texttt{n\_head} $\in \{4,6\}$; \texttt{encoder\_layers} $\in \{2,3,4\}$; \texttt{decoder\_layers} $\in \{2,3,4\}$; \texttt{dropout} $\in [0.00,0.30]$ with a step of $0.05$; \texttt{learning\_rate} $\in \{1e^{-6}, \dots, 1e^{-2}\}$; \texttt{batch\_size} $\in \{32, 64\}$; and \texttt{max\_steps} derived from 5–10 epochs. The optimal configuration was: \texttt{hidden\_size} = 128, \texttt{n\_head} = 4, \texttt{encoder\_layers} = 4, \texttt{decoder\_layers} = 2, \texttt{dropout} = 0.15, \texttt{learning\_rate} = $5\text{e}^{-4}$, \texttt{batch\_size} = 64, \texttt{activation} = \texttt{gelu}.

\paragraph{NHITS (\texttt{NHITS})} \cite{challu2023nhits}:
A hierarchical interpolation model that divides forecasting into frequency‑specialized submodels to improve accuracy and efficiency \cite{challu2023nhits}.
Optuna search space: \texttt{n\_pool\_kernel\_size} $\in$ \{\texttt{[2,2,2]}, \texttt{[4,4,4]}\};
\texttt{n\_freq\_downsample} $\in$ \{\texttt{[8,4,1]}, \texttt{[16,8,1]}\};
\texttt{interpolation\_mode} $\in$ \{\texttt{linear}, \texttt{nearest}, \texttt{cubic}\};
\texttt{n\_blocks} $\in$ \{\texttt{[1,1,1]}, \texttt{[2,2,2]}\};
\texttt{mlp\_units} $\in$ \{\texttt{[[512,512],...]}, \texttt{[[256,256],...]}\};
\texttt{activation} $\in$ \{\texttt{ReLU}, \texttt{Softplus}, \texttt{Tanh}\};
along with \texttt{learning\_rate}, \texttt{batch\_size}, and \texttt{max\_steps} derived from 8–20 epochs.
The optimal configuration was: \texttt{n\_pool\_kernel\_size} = \texttt{[2,2,2]}, \texttt{n\_freq\_downsample} = \texttt{[16,8,1]}, \texttt{interpolation\_mode} = \texttt{nearest}, \texttt{n\_blocks} = \texttt{[2,2,2]}, \texttt{mlp\_units} = \texttt{[[512,512],[512,512],[512,512]]}, \texttt{activation} = \texttt{ReLU}, \texttt{learning\_rate} = $1\text{e}^{-5}$, \texttt{batch\_size} = 256.

\paragraph{TFT (\texttt{TFT})} \cite{lim2021temporal}: Temporal Fusion Transformer, which incorporates gating mechanisms, recurrent encoders, and a multi-head attention decoder for interpretable forecasting and variable selection. The Optuna search space included: \texttt{hidden\_size} $\in {32, 64, 128, 256}$; \texttt{n\_rnn\_layers} $\in {1, 2, 3}$; \texttt{dropout} $\in [0.00, 0.30]$ with a step of $0.05$; along with \texttt{learning\_rate}, \texttt{batch\_size}, and \texttt{max\_steps} corresponding to 5–15 epochs. The optimal configuration was: \texttt{hidden\_size} = 64, \texttt{n\_rnn\_layers} = 3, \texttt{dropout} = 0.1, \texttt{learning\_rate} = $1\text{e}^{-4}$, and \texttt{batch\_size} = 64.

\paragraph{Autoformer (\texttt{Autoformer})} \cite{wu2021autoformer}: A decomposition transformer that integrates progressive trend/seasonal splitting and auto-correlation mechanisms for efficient periodic pattern modeling. Optuna search space: \texttt{encoder\_layers} $\in$ \{1, 2, 3, 4, 5, 6\}; \texttt{decoder\_layers} $\in$ \{1, 2\}; \texttt{hidden\_size} $\in$ \{32, 64, 128, 256\}; \texttt{n\_head} $\in$ \{2, 4, 8\}; \texttt{conv\_hidden\_size} $\in$ \{16, 32, 64\}; \texttt{activation} $\in$ \{\texttt{gelu}, \texttt{relu}\}; \texttt{dropout} $\in$ [0.10, 0.40] step 0.05; plus \texttt{learning\_rate}, \texttt{batch\_size}, \texttt{max\_steps} derived from 8–20 epochs.
The optimal configuration was: \texttt{hidden\_size} = 32, \texttt{encoder\_layers} = 3, \texttt{decoder\_layers} = 2, \texttt{n\_head} = 2, \texttt{conv\_hidden\_size} = 32, \texttt{activation} = \texttt{relu}, \texttt{dropout} = 0.4, \texttt{learning\_rate} = $5\text{e}^{-3}$, \texttt{batch\_size} = 256.

\paragraph{N-BEATS (\texttt{NBEATS})} \cite{oreshkin2019nbeats}: A neural basis expansion model with interpretable forecast and backcast stacks for univariate time-series forecasting. The architecture uses doubly residual stacking of blocks to model trend and seasonality components separately. Optuna search space: \texttt{stack\_types} $\in$ \{\texttt{identity}, \texttt{trend}, \texttt{seasonality}\}; \texttt{n\_blocks} $\in$ \{1, 2, 3, 4, 5\}; \texttt{mlp\_units} constructed from two sampled integers $\in$ [64, 512] with step 64; along with \texttt{learning\_rate}, \texttt{batch\_size}, and \texttt{max\_steps} derived from 8–20 epochs.
The optimal configuration was: \texttt{stack\_types} = \texttt{[seasonality]}, \texttt{n\_blocks} = \texttt{[5]}, \texttt{mlp\_units} = \texttt{[[448, 512]]}, \texttt{learning\_rate} = $1\text{e}^{-4}$, \texttt{batch\_size} = 32.

\subsection{LLM for Glucose Forecasting}

We implement a prompt-based glucose prediction agent using large language models (LLMs) to generate short-term glucose forecasts from historical CGM sequences. 
Given a sequence of CGM values $\{x_1, x_2, \dots, x_T\}$ sampled every 5 minutes and a desired prediction horizon $H$ (defined as the number of future 5-minute intervals), the agent formats the input into a natural language prompt that encodes domain knowledge and physiological constraints. This prompt is then passed to a specified LLM backend using LangChain’s unified interface. 
The model's textual output, expected to be in a strict JSON format, is parsed to extract the predicted future glucose values $\{\hat{x}_{T+1}, \dots, \hat{x}_{T+H}\}$.

The prompt guides the model to return predictions that are physiologically plausible, including glucose values between 30–400 mg/dL, with typical rates of change bounded between $-5$ and $+5$ mg/dL per minute. It also encourages the model to account for known glucose dynamics such as glucose inertia and time-of-day effects (e.g., dawn phenomenon and meal-related spikes). Predictions that do not conform to these structural or physiological constraints are flagged for review or discarded.

The agent also includes a fallback interface to the Nixtla time-series API, which provides non-LLM-based forecasting using classical methods. 
All models follow the same prompt and output specification to ensure comparability across different model families and implementations.

\subsubsection{Prompt Template}

The prompt issued to the LLMs is shown below:

\subsubsection{Implementation Details}

We implement an end-to-end evaluation pipeline to benchmark LLM-based models for glucose forecasting using the FairGlucose dataset. The system is modular, supports multiple backends, and operates on a standardized interface for CGM input handling, model inference, and evaluation.

The pipeline is orchestrated through a Python script that accepts various command-line arguments, including the model type, model name, task index, number of samples, input length, and output horizon. 
These parameters allow for parallel and distributed execution over patient records using task batching. 
Model access credentials (e.g., OpenAI, Anthropic, Google Vertex AI) are dynamically configured via environment variables, allowing authenticated API access at runtime.
Each selected CGM record is sliced into an input window of 24 hours (288 values sampled every 5 minutes) and a prediction horizon of up to 8 hours (3--96 steps), depending on the experimental setup. The main LLM comparison uses a 24-step, 2-hour prediction horizon.

\vspace{1em}

\begin{tcolorbox}[title=Glucose Prediction Prompt,colback=gray!5,colframe=gray!80!black]

\textbf{System Prompt:}

You are a precise time-series forecasting assistant for glucose prediction.
You must return ONLY a JSON array of integers (predicted glucose values in mg/dL).

Example format: [120, 125, 130, 128, 126, 124, 122, 120, ...]

Do not include any explanatory text, markdown formatting, or additional fields.
Return ONLY the JSON array of integers.

\vspace{0.5em}

\textbf{User Prompt:}

You are a glucose prediction model for diabetes patients. Given historical CGM (Continuous Glucose Monitor) readings at 5-minute intervals, predict the next \{forecast\_horizon\} glucose values.

Historical CGM values (most recent \{input\_size\} readings, in mg/dL):

\{cgm\_history\}

Instructions:
\begin{enumerate}
\item Analyze the glucose trends, patterns, and rate of change
\item Consider typical post-meal curves and insulin effects
\item Predict the next \{forecast\_horizon\} glucose values (5-minute intervals)
\item Return ONLY a JSON array of \{forecast\_horizon\} numbers (predicted glucose values in mg/dL)
\end{enumerate}

Example output format: [120, 125, 130, 128, 126, 124, 122, 120, 118, 116, 115, 114, ...]

Your prediction (JSON array only):

\end{tcolorbox}

For each record, the system constructs a structured prompt and passes it to the selected LLM using the \texttt{GlucosePredictor} agent. The agent handles the model-specific API interaction, parsing, and prediction recovery. If the selected model is a time-series backend (e.g., Nixtla’s TimeGPT \cite{garza2023timegpt}), the system constructs a timestamped time series and passes it directly to the statistical model API. Parallel processing using a thread pool is used to accelerate batch inference.

Predictions are stored in a structured format that includes the patient ID, prediction time, input sequence, ground truth, and model prediction. After inference, the results are saved in Parquet format and evaluated using a custom sequence evaluation toolkit. This toolkit calculates root mean square error (rMSE) over multiple horizons and supports subgroup-level evaluation (e.g., by diabetes type, age interval, or prediction hour). The evaluation framework ensures consistency across all experiments and provides both neat and full-format reports for downstream analysis.

The key parameters for LLM-based glucose forecasting are: input context window of 288 time steps (24 hours at 5-minute intervals), forecast horizon of 24 time steps (2 hours ahead), and temperature set to 0.0 for deterministic generation. All LLM models (GPT-5.1, GPT-5 Mini, Claude 4.5 Sonnet, Claude 4.5 Haiku, Gemini 3 Flash) use the same prompt template and configuration to ensure fair comparison.

The system supports a wide range of model backends, including GPT-5.1, GPT-5 Mini, Claude 4.5 Sonnet, Claude 4.5 Haiku, Gemini 3 Flash, and TimeGPT \cite{garza2023timegpt}. 
Each model is evaluated under the same input-output specification and metrics to ensure fair comparison. 
The resulting predictions and performance metrics are used to assess fairness, subgroup robustness, and generalization across temporal and demographic dimensions.